\documentclass[fleqn,usenatbib]{mnras}
\usepackage{newtxtext,newtxmath}
\usepackage[T1]{fontenc}
\DeclareRobustCommand{\VAN}[3]{#2}
\let\VANthebibliography\thebibliography
\def\thebibliography{\DeclareRobustCommand{\VAN}[3]{##3}\VANthebibliography}
\usepackage{graphicx}	
\usepackage{longtable}
\usepackage{siunitx}
\usepackage{times}
\usepackage{epstopdf}
\usepackage{caption}
\usepackage{pdflscape}
\usepackage{amsmath}
\usepackage{rotating}
\usepackage{tablefootnote}
\usepackage{rotating}
\newlength{\colwidth}
\renewcommand{\footnotesize}{\scriptsize}
\usepackage{verbatim}
\usepackage{xcolor}

\title[Blazar Photometry with RINGO3 and Fermi]{Distinguishing radiation mechanisms and particle populations in blazar jets through long-term multi-band monitoring with RINGO3 and Fermi}

\author[McCall et al.]{
Callum McCall,$^{1}$\thanks{E-mail: c.mccall@2017.ljmu.ac.uk}
Helen Jermak,$^{1}$
Iain A. Steele,$^{1}$
Iv\'an Agudo,$^{2}$
Ulisses Barres de Almeida,$^{3,4}$
\newauthor
Talvikki Hovatta,$^{5,6}$
Gavin P. Lamb,$^{1}$
Elina Lindfors$^{7}$
and Carole Mundell$^{8}$
\\ \\
$^{1}$ Astrophysics Research Institute, Liverpool John Moores University, Liverpool Science Park IC2, 146 Brownlow Hill, Liverpool, UK, L3 5RF\\
$^{2}$ Instituto de Astrof\'isica de Andaluc\'ia-CSIC, Glorieta de la Astronom\'ia, E-18008, Granada, Spain\\
$^{3}$ Centro Brasileiro de Pesquisas Físicas, Rua Xavier Sigaud 150, RJ 22290-180, Rio de Janeiro, Brazil\\
$^{4}$ Instituto de Astronomia, Geofísica e Ciências Atmosféricas - Universidade de São Paulo, Cidade Universitária, R. do Matão, 1226, CEP 05508-090,\\
São Paulo, SP, Brazil\\
$^{5}$ Finnish Centre for Astronomy with ESO (FINCA), Quantum, Vesilinnantie 5, FI-20014 University of Turku, Finland\\
$^{6}$ Aalto University Mets\"ahovi Radio Observatory, Mets\"ahovintie 114, FI-02540 Kylm\"al\"a, Finland\\
$^{7}$ Tuorla Observatory, University of Turku, Väisäläntie 20, FI-21500 Piikkiö, Finland \\
$^{8}$ European Space Agency, European Space Astronomy Centre, E-28692 Villanueva de la Canñada, Madrid, Spain\\
}

\date{Accepted XXX. Received YYY; in original form ZZZ}

\pubyear{2024}

\begin{document}
\label{firstpage}
\pagerange{\pageref{firstpage}--\pageref{lastpage}}
\maketitle

\begin{abstract}
We present the results of seven years of multicolour photometric monitoring of a sample of 31 $\gamma$-ray bright blazars using the RINGO3 polarimeter on the Liverpool Telescope from 2013--2020. We explore the relationships between simultaneous observations of flux in three optical wavebands along with Fermi $\gamma$-ray data in order to explore the radiation mechanisms and particle populations in blazar jets. We find significant correlations between optical and $\gamma$-ray flux with no detectable time lag, suggesting leptonic emission processes in the jets of these sources. Furthermore, we find the spectral behaviour against optical and $\gamma$-ray flux for many sources is best fit logarithmically. This is suggestive of a transition between bluer-/redder-when-brighter into stable-when-brighter behaviour during high activity states; a behaviour that might be missed in poorly sampled data, resulting in apparent linear relationships.
\end{abstract}


\section{Introduction}\label{sec:intro}
Blazars are active galactic nuclei (AGN) orientated on the sky with jet viewing angles $\lesssim$20$^\circ$ with respect to the observer \citep{urrypadovani1995,hovatta2009,hovatta2019}. Relativistic beaming of the jet results in highly variable emission which is seen across the entire electromagnetic spectrum \citep{blandfordrees1978}. Blazars can be split into two subclasses based on the strength of emission lines present in their optical spectra. Flat spectrum radio quasars, FSRQs, are those sources originally identified to have emission line equivalent widths $\geq5$\,\AA, and BL~Lacertae-type objects, BL~Lacs, with emission line equivalent widths $<5$\,\AA, or absent altogether \citep{stickel1991}. Recent studies have suggested an additional population of transitional blazars \citep{ghisellini2011}, or masquerading BL Lacs, whose higher luminosities and accretion rates are more similar to that of FSRQs \citep{padovani2019}, but are classified as BL Lacs due to emission from the broad line region being overpowered by the jet continuum \citep{ruan2014}.

The spectral energy distributions (SEDs) of blazars take a unique double-hump shape. The lower energy peak is attributed to the powerful beamed jet emission and emission from the disk, which is most prevalent during periods of jet quiescence. It is possible to distinguish between jet and disk emission by exploring the spectral properties of blazars. Accretion disk emission is created by in-falling matter converting gravitational potential energy to luminosity, and as such is thermal emission \citep{perlman2008}. Emission from the jet will predominantly be non-thermal as relativistic charged particles create synchrotron emission as they spiral around magnetic field lines \citep{celotti1994}. 

Furthermore, blazars can be sub-classified according to the rest-frame location of this lower-energy synchrotron ($\nu_\mathrm{s}$) peak in their SEDs. This was first introduced by \cite{padovaniandgiommi1995} and has been adapted by \cite{abdo2010b} for use on large samples of \textit{Fermi} blazars. BL Lac sources are classified as low synchrotron peak (LSP; $\nu_\mathrm{s}$ < 10$^{14}$\,Hz), intermediate synchrotron peak (ISP; 10$^{14}$ < $\nu_\mathrm{s}$ < 10$^{15}$\,Hz) and high synchrotron peak (HSP; $\nu_\mathrm{s}$ >10$^{15}$\,Hz). FSRQs are all classified as LSPs, based on the location of their synchrotron peak \citep{abdo2010b}. 

The origin of the higher-energy peak, located at hard X-rays to very-high-energy (VHE) $\gamma$-rays, is still debated. Leptonic modelling of this high-energy emission suggests inverse-Compton scattering as the likely origin \citep{maraschi1992,bloom1996,bottcher2013b}. In this scenario, low-energy photons originating from within the jet \cite[synchrotron-self Compton (SSC);][]{maraschi1992} or from outside \cite[external Compton (EC);][]{dermer1993} are upscattered by interactions with the population of synchrotron electrons within the jet, thus producing the observed high-energy emission. Conversely, assuming hadronic modelling, the acceleration of protons to VHE can lead to the high-energy emission directly through proton synchrotron emission, or via interactions between the protons producing both charged and neutral pions \citep{mannheim1992,aharonian2000,bottcher2013b}. It is the decay of these charged pions that produces high-energy neutrinos, the smoking-gun signature of hadronic emission \citep{reimer2012}, that have been detected coinciding with several flaring blazars \citep{plavin2023}.

The behaviour of blazars' $\gamma$-ray and optical flux gives an insight into the locations of emitting regions and the underlying emission mechanisms occurring within the jet. Strongly correlated behaviour between the two fluxes suggests the emission may originate from linked processes within the jet, favouring leptonic models. Specifically, an increase in synchrotron photons leads to an increase in the seed photons available for inverse-Compton upscattering \citep{bottcher2010a}. On the other hand, a lack of correlated behaviour including orphan optical and $\gamma$-ray flares could favour both leptonic and hadronic models, or even a combination (lepto-hadronic) \citep{sol2022}. In the leptonic scenario, a localised enhancement of the seed photon fields further out into the jet interacts with a relativistic blob travelling along a shocked portion of the jet sheath, resulting in increased inverse-Compton scattering and $\gamma$-ray emission without an optical counterpart \citep{macdonald2015}. Conversely, the orphan flares could be the result of high-energy emission produced completely independently of any lower energy synchrotron behaviour through hadronic emission processes \citep{liodakis2019}. Additionally, one can look for a temporal separation between flaring events seen at the different frequencies. In the leptonic model, a lag between optical and $\gamma$-ray emission suggests a larger spatial separation between the synchrotron and inverse-Compton emitting regions \citep{cohen2014}. It follows that optical and $\gamma$-ray monitoring over many year-long timescales is a powerful discriminator of the dominant jet content and origin of the detected radiation, leading to the possible distinction between leptonic SSC and EC emission, and hadronic processes.


A frequent optical photometric feature of blazars is their colour evolution during various levels of jet activity. Most are `bluer-when-brighter' (BWB) i.e., their SED at optical frequencies flattens during periods of higher flux. This behaviour can be explained with a one-component synchrotron model with an injection of fresh electrons into the jet with a hard energy distribution. These electrons cool and the resulting increased radiation is bluer in colour. Two-component modelling suggests two underlying components to the observed flux, one stable and one variable \citep{fiorucci2004}. The stable component consists of thermal emission from the accretion disc and broad line regions (BLR) whereas the variable component originates from non-thermal synchrotron emission. In some sources, predominantly FSRQs \citep{zhang2015b,negi2022}, we see the opposite behaviour, that is: the source appears `redder-when-brighter' (RWB), or its SED at optical frequencies steepens during periods of higher flux. This is thought to be due to an increased amount of thermal emission from the disc, resulting in the composite spectrum being flatter in the optical region \cite[the `blue/UV bump';][]{gu2006} and subsequently steepening during periods of heightened flux. Moreover, a stable-when-brighter trend has been observed in some objects \citep{ghosh2000,zhang2015b}, where the colour of the source remains constant during flux increases.

In this paper, we present 7 years of RINGO3 multicolour photometric data and use it to explore the colour and flux behaviour of a sample of 31 $\gamma$-ray bright blazars, in particular focusing on the behaviours of the different classes of objects. The paper is organised as follows: Section \ref{sec: obs} describes the observations, data, and facilities used in this work. Section \ref{sec: var stats} describes the correlation analysis between the data including optical flux and colour, $\gamma$-ray flux, and inter-waveband lags. In Section \ref{sec: discussion} we discuss the implications of our findings and compare the results of each correlation.

\section{Observations}
\label{sec: obs}
The RINGO2 \citep{steele2006} and DIPOL blazar monitoring campaign \citep{ringo2paper}, ran from 2008 to 2012 on the Liverpool Telescope. It was designed to provide optical photo-polarimetric monitoring of 15 $\gamma$-ray flaring blazars (also monitored at high energies by the \textit{Fermi Gamma-ray Space Telescope}). The sample has slowly grown since 2008 with the introduction of new sources that have exhibited $\gamma$-ray flaring.
With the commissioning of RINGO3 \citep{arnold2012} in 2013, photo-polarimetric monitoring of the existing sample was expanded with an additional 16 blazars. RINGO3 operated on the Liverpool Telescope until 2020; we present the photometric results of this monitoring campaign here (polarimetric results will be presented in a subsequent paper).


\subsection{RINGO3 polarimeter}
\label{sec:r3}
The Liverpool Telescope (LT) is a 2.0-m, fully autonomous, robotic telescope located on the Canary Island of La Palma \citep{steele2004}. The LT's intelligent dispatch scheduler allows the telescope to operate entirely autonomously, selecting observational sequences according to weather conditions, science aims, source visibility and location on sky, along with priority gradings. The LT's autonomous operation makes it ideal for blazar monitoring. Regular blazar observations can be scheduled amongst other science objects across periods of months, as well as the possibility of more intensive periods of intra-night monitoring on occasions when weather and instrument availability allow.

RINGO3 is fitted with 2 dichroic mirrors that transmit/reflect the incoming beam of light into three optical wavebands: `blue-visible' (350-640 nm), `green' (650-760 nm) and `red' (770-1000 nm) \citep{arnold2012}. The light from the source is modulated by a rotating Polaroid (one rotation every 4 seconds), with triggered imaging at 8 rotor positions of the Polaroid. The combination of measurements at these 8 rotor positions are used to calculate the Linear Stokes parameters according to the equations in \cite{clarke&neumayer2002}.

The three optical wavebands in RINGO3 are dictated by the dichroic mirrors. The mirrors were selected at the time of construction to maximise the amount of flux detected in gamma-ray burst follow-up and as such do not correspond with standard astronomy passbands (e.g. Johnson-Cousin or $ugriz$;  \cite{arnold2012}). In this paper the different bands will be referred to as \textit{b*} (350-640 nm), \textit{g*} (650-760 nm) and \textit{r*} (770-1000 nm).

\subsection{Photometric calibration}
Similar to the RINGO2 data reduction procedure, differential photometry was used to remove the effect of variable seeing, airmass and atmospheric transparency using in-frame calibration stars. Due to the non-standard photometric bands, the magnitudes of these calibration stars for each blazar had to be defined in the RINGO3 \textit{b*g*r*} photometric system. 

To achieve this we used observations of unreddened A0 stars as they, by definition, have the same magnitude in all photometric passbands (i.e. zero colour) for a Vega-referenced magnitude system. A sample of bright, unreddended A0 stars with high-quality optical photometry in the Johnson-Cousins system was therefore observed with RINGO3 on a small number of photometric nights (non-coincident with the blazar observations). To account for the variable throughput of the camera/optical system over time due to dust accumulation and similar effects, approximately nightly observations of polarimetric standards (BD+64 106, BD+25 727, and HILT 960) were used to calibrate the rate of degradation. This rate was measured as $4.69 \pm 0.15 \times10^{-4}$ per cent per day, irrespective of the camera, relative to the initial counts measured from the object. The counts from the A0 stars were adjusted using this degradation rate to produce zeropoints calibrated to a common date. In the same way, the counts of the calibration star in the science frames were also adjusted to the same common date.  In combination with the zeropoints this allowed the calculation of the average magnitude for each calibration star per waveband in the natural RINGO3 system. 

The average uncertainty on each calibration star magnitude achieved was 0.07.
We note that this uncertainty in the calibration star magnitudes does not affect the correlation statistics presented in this work as it will offset all data blazar points by the same amount. 

Since the electron multiplying CCD gives an effective read noise of $<1e^-$, we can stack images without penalty and remain photon-limited. However, electron multiplication noise reduces the final signal-to-noise ratio by a factor of $\sqrt{2}$. This means that the final photometric uncertainty is increased by this factor compared with photometry using a single conventional CCD image \citep{robbins2003}.

\subsection{Fermi data}\label{sec:Fermi}
Fermi data were taken from the Fermi Large Area Telescope (LAT) Light Curve Repository (LCR)\footnote{https://fermi.gsfc.nasa.gov/ssc/data/access/lat/LightCurveRepository/} \citep{abdollahi2023}. This database consists of flux-calibrated light curves from over 1500 variable sources \citep{kocevski2021} with variability indexes $>21.67$. The fractional variability is described by \cite{abdollahi2023} as a proxy for the average fractional variability exhibited by an object over a one-year timescale. The threshold of 21.67 indicates that the $\gamma$-ray flux of the object has a less than one per cent chance of being steady. The data can be downloaded at different binning intervals (three days, one week, and one month). We chose to use data binned over three days to ensure sufficient data when correlated against the optical.

\subsection{Sample}
\label{sec:sample}
In our program, photopolarimetric data were obtained from 49 sources between 2013 and 2020. To ensure enough data per source to measure long-term variability characteristics, a minimum of 60 observations were required for the source to be a part of our final sample. An exception was made for any object where observations were taken with a density greater than once every ten days, given that at this observation density, any correlations might be detectable. Application of these conditions resulted in 31 objects in the sample; see Table \ref{sample_data} for details on redshift, spectral peak (emission) classification, \textit{r*}-band magnitude range and Fermi $\gamma$-ray flux range for the given observational period. This sample includes the 15 sources in the DIPOL and RINGO2 sample \citep{ringo2paper}, with additional sources selected for $\gamma$-ray activity.

The data for the majority of these sources are presented fully in this paper with the following exceptions. MRK 421 lacks usable comparison stars in its field (due in part to the presence of a bright foreground star that causes ghosting in the frame), for this reason, differential photometry is not possible. IC310 and 1ES 1426+428 do not have data in the Fermi LAT LCR as their variability indices are less than the defined threshold.

Sources are classified according to the two methods discussed in Section \ref{sec:intro}: the location of the synchrotron peak in the SED (LSP, ISP, HSP) and the size/presence of emission lines in their optical spectra (FSRQ, BL Lac).
BL Lacs span all three classes of spectral peaks: HSP, ISP or LSP, whereas all FSRQs are LSP \citep{abdo2010b}. Note that due to the variable nature of blazars some of these classifications have been known to change (ie. changing-look blazars; see
\cite{xiao2022} and references therein) and we use the most recent classification available in the literature.

\begin{table*}
\centering
\caption{The RINGO3 blazar sample with source classification, redshift, \textit{r*}-band optical magnitude range, $\gamma$-ray flux range, and observation MJD range for each source shown.}
\label{sample_data}
\resizebox{\textwidth}{!}{
\begin{tabular}{cccccc}
\hline
Name & Type & $z$ & \textit{r*} mag. range & Fermi range (erg cm$^{-2}$ s$^{-1}$) & Observational Period (MJD) \\ 
\hline
IC 310        & HSP        & 0.0189 & 13.137 - 12.873 & -                                                               & 56317.842 - 57989.231      \\
1ES 1011+496  & HSP        & 0.212  & 15.408 - 14.676 & 2.16$\times 10^{-11}$ - 3.46$\times 10^{-10}$                   & 56321.963 - 58521.050      \\
MRK 421       & HSP        & 0.03   & -  & 1.12$\times 10^{-10}$ - 1.09$\times 10^{-9}$                    & 56272.275 - 58526.096      \\
MRK 180       & HSP        & 0.045  & 14.752 - 14.241 & 9.98$\times 10^{-12}$ - 9.58$\times 10^{-11}$                   & 56321.980 - 58521.079      \\
PG 1218+304   & HSP        & 0.184  & 16.125 - 14.962 & 3.24$\times 10^{-11}$ - 3.28$\times 10^{-10}$                   & 56268.274 - 58519.227      \\
1ES 1426+428  & HSP        & 0.129  & 16.157 - 15.503 & -                                                               & 56322.127 - 58521.280      \\
PG 1553+113   & HSP        & 0.36   & 14.106 - 13.006 & 2.61$\times 10^{-11}$ - 5.78$\times 10^{-10}$                   & 56318.196 - 58521.292      \\
MRK 501       & HSP        & 0.033  & 12.817 - 12.594 & 1.60$\times 10^{-11}$ - 3.86$\times 10^{-10}$                   & 56318.203 - 58535.199      \\
1ES 1959+650  & HSP        & 0.047  & 13.843 - 13.365 & 4.34$\times 10^{-11}$ - 5.14$\times 10^{-10}$                   & 57509.122 - 57975.920      \\
3C 66A        & ISP        & 0.444  & 14.905 - 13.539 & 2.32$\times 10^{-11}$ - 3.59$\times 10^{-10}$                   & 56321.954 - 58496.939      \\
S5 0716+714   & ISP        & 0.127  & 14.444 - 11.263 & 2.45$\times 10^{-11}$ - 1.04$\times 10^{-9}$                    & 56331.918 - 58527.891      \\
ON 231        & ISP        & 0.102  & 15.460 - 13.488 & 1.89$\times 10^{-11}$ - 1.99$\times 10^{-10}$                   & 57206.950 - 58535.117      \\
A0 0235+164   & LSP        & 0.94   & 18.473 - 14.618 & 1.81$\times 10^{-11}$ - 7.39$\times 10^{-10}$                   & 57051.876 - 58394.129      \\
TXS 0506+056  & LSP        & 0.336  & 14.249 - 13.731 & 8.35$\times 10^{-11}$ - 2.48$\times 10^{-10}$                   & 58339.233 - 58360.174      \\
OJ 287        & LSP        & 0.306  & 15.096 - 12.456 & 1.59$\times 10^{-11}$ - 3.70$\times 10^{-10}$                   & 56316.033 - 58759.242      \\
S4 0954+65    & LSP        & 0.367  & 16.689 - 14.065 & 1.24$\times 10^{-11}$ - 8.54$\times 10^{-10}$                   & 57051.117 - 58535.055      \\
4C 09.57      & LSP        & 0.322  & 17.301 - 14.384 & 2.24$\times 10^{-11}$ - 1.24$\times 10^{-9}$                    & 57090.231 - 58540.284      \\
BL Lac        & LSP        & 0.069  & 13.900 - 11.967 & 3.33$\times 10^{-11}$ - 8.25$\times 10^{-10}$                   & 56407.184 - 58460.905      \\
PKS 0502+049  & LSP (FSRQ) & 0.954  & 18.206 - 15.126 & 3.41$\times 10^{-11}$ - 1.48$\times 10^{-9}$                    & 56652.997 - 57983.222      \\
PKS 0736+01   & LSP (FSRQ) & 0.189  & 16.274 - 14.384 & 2.63$\times 10^{-11}$ - 6.68$\times 10^{-10}$                   & 57007.998 - 57881.871      \\
PKS 1222+216  & LSP (FSRQ) & 0.435  & 15.223 - 13.018 & 1.68$\times 10^{-11}$ - 8.92$\times 10^{-10}$                   & 56332.163 - 58258.923      \\
3C 279        & LSP (FSRQ) & 0.536  & 15.339 - 12.818 & 3.43$\times 10^{-11}$ - 1.01$\times 10^{-8}$                    & 56322.115 - 58541.195      \\
PKS 1510-089  & LSP (FSRQ) & 0.361  & 16.023 - 13.260 & 4.60$\times 10^{-11}$ - 2.47$\times 10^{-9}$                    & 56304.292 - 58542.260      \\
OS 319        & LSP (FSRQ) & 1.399  & 17.918 - 16.381 & 1.87$\times 10^{-11}$ - 9.95$\times 10^{-11}$                   & 57110.113 - 58542.276      \\
PKS B1622-297 & LSP (FSRQ) & 0.815  & 18.629 - 16.057 & 1.94$\times 10^{-11}$ - 3.73$\times 10^{-10}$                   & 57090.217 - 58542.282      \\
4C +38.41     & LSP (FSRQ) & 1.814  & 17.531 - 15.202 & 2.45$\times 10^{-11}$ - 7.05$\times 10^{-10}$                   & 57128.155 - 58534.294      \\
3C 345        & LSP (FSRQ) & 0.593  & 17.629 - 15.750 & 1.11$\times 10^{-11}$ - 4.76$\times 10^{-10}$                   & 57083.139 - 58540.272      \\
PKS B1730-130 & LSP (FSRQ) & 0.902  & 17.643 - 16.315 & 4.15$\times 10^{-11}$ - 2.37$\times 10^{-10}$                   & 57085.216 - 58519.275      \\
3C 446        & LSP (FSRQ) & 1.404  & 18.272 - 17.361 & 1.62$\times 10^{-11}$ - 7.18$\times 10^{-11}$                   & 57175.149 - 58408.836      \\
4C 11.69      & LSP (FSRQ) & 1.037  & 16.658 - 10.560 & 4.29$\times 10^{-11}$ - 8.25$\times 10^{-9}$                    & 57143.229 - 58463.794      \\
3C 454.3      & LSP (FSRQ) & 0.859  & 15.795 - 13.346 & 8.36$\times 10^{-11}$ - 6.65$\times 10^{-9}$                    & 57143.225 - 58408.847     \\ 
\hline
\end{tabular}
}
\end{table*}

\begin{table}
\centering
\caption{Correlation strengths for Spearman rank coefficient values.}
\label{srank}
\begin{tabular}{cc}\\ \hline
Value & Correlation Degree\\ \hline
$\textit{c}$ = 0 & no correlation \\
0 $\leq$ |$\textit{c}$| < 0.2 & very weak \\
0.2 $\leq$ |$\textit{c}$| < 0.4 & weak \\
0.4 $\leq$ |$\textit{c}$| < 0.6 & moderate \\
0.6 $\leq$ |$\textit{c}$| < 0.8 & strong \\
0.8 $\leq$ |$\textit{c}$| < 1 & very strong \\
$\textit{c}$ = 1 & monotonic \\
\hline
\end{tabular}
\end{table} 


\section{Correlation analysis}
\label{sec: var stats}

We use the Spearman rank coefficient to test the strength and significance of monotonic relationships within our data in a non-parametric way. This test provides a coefficient for significance, ($p$; how likely the data are to be correlated by chance) and strength ($c$; strength of the positive or negative correlation). Table \ref{srank} shows the correlation strength coefficient values ($c$) used in this analysis with $p\leq0.05$, implying a $2\sigma$ significance level for a Gaussian distribution. 

In our analysis, we account for the differing cadence of optical and $\gamma$-ray observations by assigning the temporally closest $\gamma$-ray data point to each of the optical observations. The median difference in MJD between optical and $\gamma$-ray observations across all sources in the sample was 1.12 days. 96 per cent of observations had a corresponding Fermi $\gamma$-ray observation within 12 days (the average cadence between optical observations across all sources). The $\gamma$-ray activity level of each source was determined by computing the median flux over the RINGO3 observation timescale. Any group of two or more points above 3 times the median absolute deviation (including lower error limits) are considered to be in a flaring state. This is indicated on the light curves in Figures \ref{fig:IC310_lc} - \ref{fig:3C454.3_lc} as a dashed purple line.

We note that no host galaxy corrections have been performed on the data. In general, blazars outshine their host galaxies by several orders of magnitude so no correction is required. We acknowledge that some sources in our sample, within the HSP BL Lac class, do have resolvable host galaxies but these are not variable. This means that for a small amount of data, a change in position on magnitude/flux axes would occur following host correction, but the correlation analysis and variability observed would remain unchanged. We note that this approach might not be appropriate for multi-facility analysis, especially in the presence of significantly variable seeing.

\subsection{Optical -- gamma-ray flux}
\label{sec: flux correlations}

\begin{figure*}
\centering
\includegraphics[width=\linewidth]{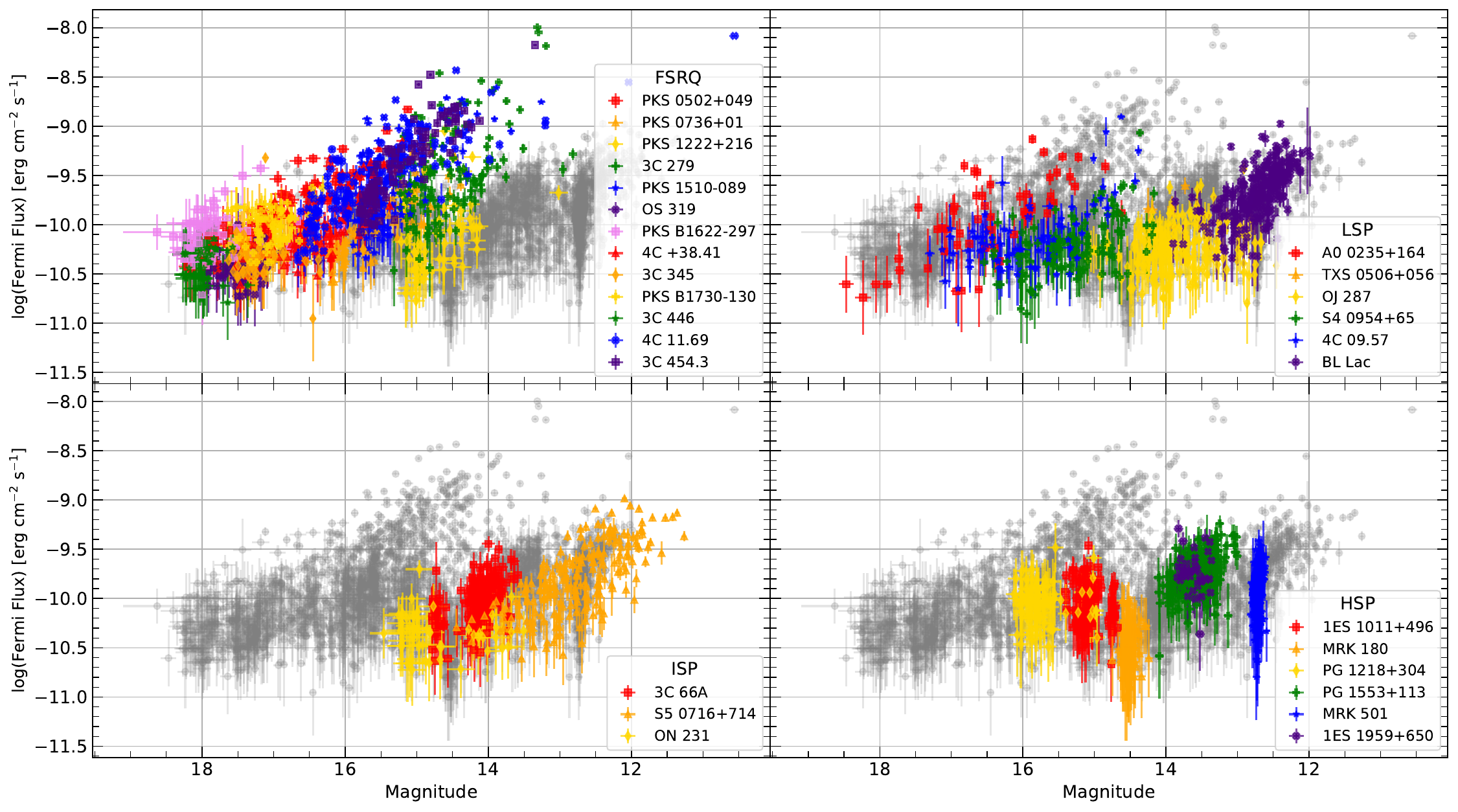}
\caption{Fermi $\gamma$-ray flux vs. optical \textit{r*} magnitude for the sample. Each panel highlights the data for the different blazar subclasses separately with the upper left, upper right, lower left, and lower right highlighting data for FSRQs, BL Lac LSPs, ISP BL Lacs, and HSP BL Lacs respectively. In each panel, the data for the other subclasses is shown as faint grey circles.}
\label{fig:flux-fermi_magnitude}
\end{figure*}

The $\gamma$-ray flux and \textit{r*}-band magnitude correlations are shown in Fig. \ref{fig:flux-fermi_magnitude}, with each panel showing sources of different classifications (i.e. high-, intermediate-, low- synchrotron peak BL Lac types and FSRQs) and each colour within the panel showing different sources. The data from other object classifications are shown as grey points. There is a clear distinction between the BL Lac and FSRQ sources, with some overlap between the FSRQs and BL Lac LSPs. The FSRQ sources generally have higher $\gamma$-ray fluxes; however, this may just be due to the biased nature of the sample selection (sources were added to the observing campaign if they showed high levels of $\gamma$-ray activity with Fermi). 

Using the redshifts, $z$, from Table \ref{sample_data}, the $\gamma$-ray fluxes and optical magnitudes can be calibrated for distance. To do this the luminosity distance, $d_\mathrm{L}$, was calculated for each object using the \textsc{WMAP9 cosmology} module in \textsc{Astropy}. This module assumes a flat universe, with a Hubble constant of $H_0=69.32\,\mathrm{km\,s^{-1}\,Mpc^{-1}}$ and the matter density parameter set at $\Omega_\mathrm{m}=0.2865$ \citep{hinshaw2013}. 

The $\gamma$-ray luminosity was calculated by 

\begin{equation}
    L = (\Gamma-1)\,4\pi F d_L^2 (1+z)^\Gamma
\end{equation}
from \cite{hovatta2014} where $\Gamma$ is the power-law index (taken from the Fermi LCR), F is the $\gamma$-ray flux given in $\mathrm{erg}\,cm^{-2}\,s^{-1}$ in the 1--100 GeV photon energy range, and $d_\mathrm{L}$ is the luminosity distance given in cm. The absolute magnitude was calculated by 
\begin{equation}
\label{abs_mag}
    M = m-5\log d_\mathrm{L} + 5
\end{equation}
where $m$ is the apparent magnitude, and $d_\mathrm{L}$ is the luminosity distance in pc. 

\begin{figure*}
\centering
\includegraphics[width=\linewidth]{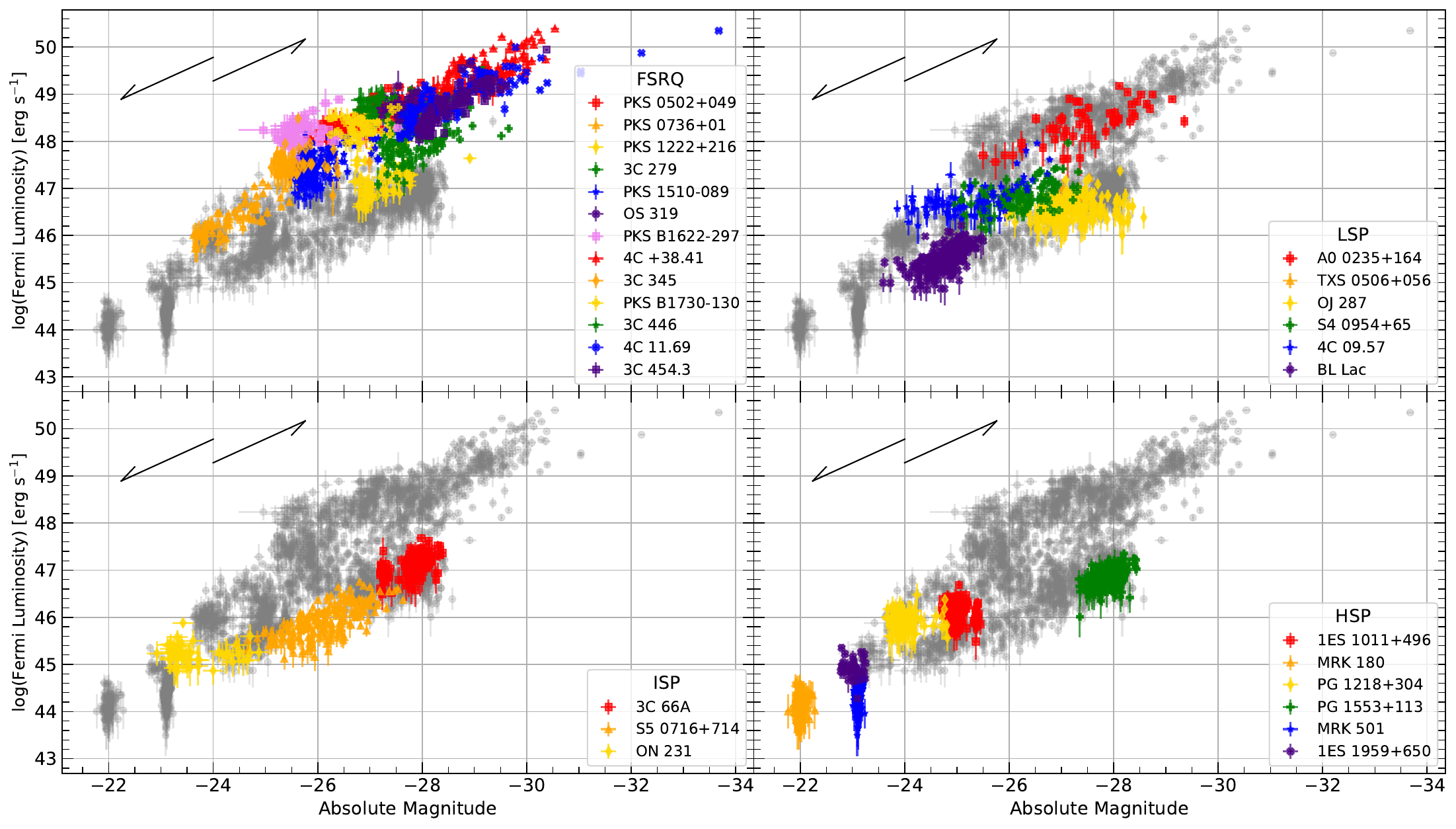}
\caption{As Fig. \ref{fig:flux-fermi_magnitude}, but with Fermi $\gamma$-ray flux and \textit{r*} magnitude calibrated for distance ($\gamma$-ray luminosity and \textit{r*} absolute magnitude respectively). Additionally, redshift vectors are included to show the direction and distance the data for a given source would move if the redshift value given in Table \ref{sample_data} was incorrect. The decreasing line indicates we've overestimated the redshift by double and the increasing underestimated by half.}
\label{fig:flux-fermi_abs_magnitude}
\end{figure*}

The resulting distance-calibrated data are shown in Fig. \ref{fig:flux-fermi_abs_magnitude}. As in Fig. \ref{fig:flux-fermi_magnitude} the data are displayed in each panel according to subclass (e.g. FSRQ, BL Lac, HSP, ISP, LSP), with grey points showing the other subclasses. Similarly to Fig. \ref{fig:flux-fermi_magnitude}, the distance-calibrated FSRQs are generally brighter at $\gamma$-ray frequencies but also dominate at the brightest optical absolute magnitudes.

We note that by definition, BL Lac objects have difficult-to-determine redshifts due to the comparably small (or absent) emission lines in their optical spectra. This means the calculation of absolute magnitude and $\gamma$-ray luminosity is subject to this uncertainty. To account for this, 1000 redshift values between $\frac{z}{2}<z<2z$ (where $z$ is the redshift value stated in Table \ref{sample_data}) were used to calculate the average absolute magnitude and $\gamma$-ray luminosity as functions of redshift. These were plotted and fitted showing the data followed a linear relationship of the form $y=Ax+B$ where $A=-0.504\pm0.012$ and $B=33.90\pm0.20$ across all sources. Furthermore, the vector distance the data shifted was the same for redshift values $\frac{z}{2}$ and $2z$ at $1.971\pm0.023$ units in Fig. \ref{fig:flux-fermi_abs_magnitude}. Therefore, redshift vectors in the upper left corner of each panel of Fig. \ref{fig:flux-fermi_abs_magnitude} are displayed. These vectors show the distance and direction the data for a given source would shift for a 50 per cent reduction ($\frac{z}{2}$) and a 100 per cent increase ($2z$) in the redshift value stated in Table \ref{sample_data}. If the true redshift was half the stated value, implying an overestimation, the object would be intrinsically fainter and so would shift towards the point (0,33.90) in Fig. \ref{fig:flux-fermi_abs_magnitude}. Conversely, if the true redshift was double the stated value, implying an underestimation, the object would be intrinsically brighter and would shift away from the point (0,33.90) in Fig. \ref{fig:flux-fermi_abs_magnitude}.

Table \ref{fermi_flux_stats} in the appendix shows the Spearman rank $p$ values and coefficients ($c$) for correlations between the RINGO3 wavebands and $\gamma$-ray flux for each source in the sample; excluding IC310, 1ES 1426+428, and MRK421 for the reasons discussed previously. Of these 28 sources, 21 showed significant correlations between $\gamma$-ray flux and each optical \textit{b*g*r*} magnitude. All significant correlations were positive. Breaking these correlations down by subclassification we find that 33 per cent of HSP BL Lac sources showed significant correlations, increasing to 66 per cent for ISP BL Lac sources. All BL Lac LSP sources showed significant correlations between the optical and $\gamma$-ray fluxes, along with 85 per cent of FSRQs.

\subsection{Optical spectral index -- flux}
\label{sec: colour}

\begin{figure*}
\centering
\includegraphics[width=0.9\linewidth]{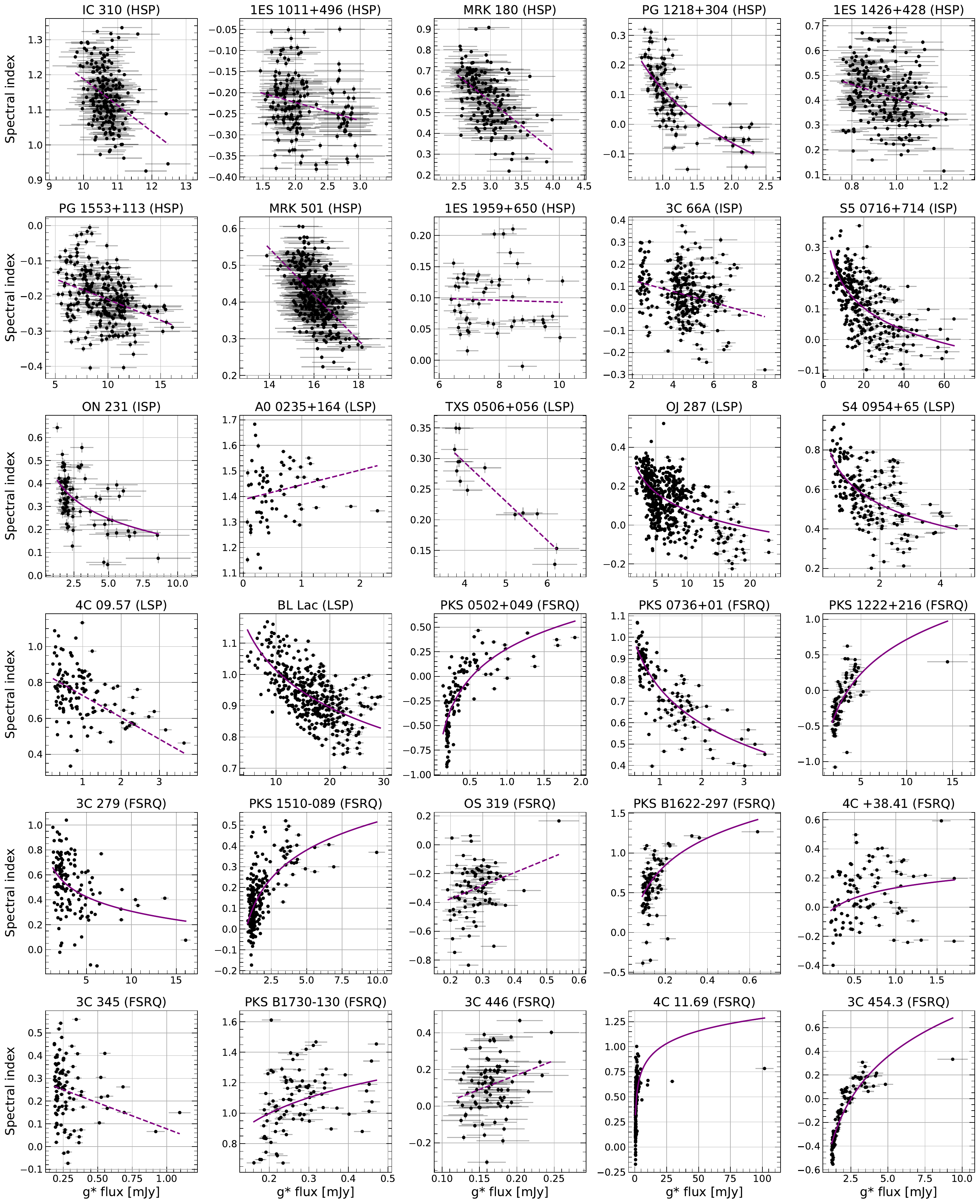}
\caption{Spectral index vs. optical \textit{g*}-band flux for each object in our sample. A best-fit line is fitted to each set of data after having linear and logarithmic fits qualitatively assessed using AIC and BIC coefficients. A preferential linear fit is shown by a dotted fit line, while a logarithmic fit is shown by a solid line. The Spearman rank correlation coefficients associated with each correlation (linearised in the case of a logarithmic fit) are given in Table \ref{colour_mag_stats_full}.}
\label{fig:all_colour}
\end{figure*}

Fig. \ref{fig:all_colour} shows the changes in the spectral index, $\alpha$, with \textit{g*}-band flux for all sources in our sample. The spectral index was calculated assuming a single power law as defined by the following equation
\begin{equation}
    F_\nu \propto \nu^{-\alpha}
\end{equation}
where $F_\nu$ is the flux at wavelength $\nu$, $\nu$ is the central wavelength of the RINGO3 bands, and $\alpha$ is the spectral index. A two-point spectral index was calculated by taking the gradient of log flux vs. log frequency, giving $-\alpha$. This was done using the \textit{r*} and \textit{b*} data, and by fitting a linear least-squares regression. It is important to not include the \textit{g*}-band data in this calculation to not induce false correlations in the subsequent analysis arising from correlated errors (see Appendix \ref{false_col_cor} for more details). The uncertainty of the spectral index at each epoch was calculated using Monte Carlo resampling where, at each epoch, 1000 pairs of randomly generated \textit{r*} and \textit{b*} flux values within the respective error limits were generated and the spectral index was calculated. The standard error on these 1000 values was taken as the error on the spectral index. 

From the data presented in Fig. \ref{fig:all_colour}, it is clear that in many cases, a linear fit is not well suited to describe the relationship between optical spectral index and flux. For this reason, the spectral index and \textit{g*}-band flux were fitted with a linear least-squares regression and logarithmic function of the form 
\begin{equation}
\label{eqn: log_fit}
    \alpha = A\,\mathrm{ln}\left(F_\mathrm{g*}\right) + B
\end{equation}
where $A$ and $B$ are free parameters. For both the linear and logarithmic fits, an Akaike Information Criterion \cite[AIC;][]{akaike1974} and Bayesian Information Criterion \cite[BIC;][]{schwarz1978}, were calculated to quantify the quality of the fits on the data. The number of free parameters for both fits was two, and lower AIC and BIC values indicated better fits. Furthermore, Spearman rank correlation coefficients were calculated for each dataset. For those data that were better fitted logarithmically, the alpha values were logged before the Spearman rank calculations to make the data linear.

Table \ref{colour_mag_stats_full} in the appendix shows the results of the above analysis, with the `Fit' column describing which model fits the data better according to the AIC and BIC values. We note that there were no cases with conflicting AIC and BIC values. The `Trend' column describes the colour relationship observed given the obtained preferred fit and sign on the Spearman rank strength coefficient, $c$. Negative strengths indicate bluer-when-brighter (BWB) behaviour, implying the spectral index flattens during periods of heightened flux. Conversely, positive strength coefficients indicate redder-when-brighter (RWB) behaviour implying the spectral index steepens during periods of heightened flux. In both cases, where a log fit is preferred over a linear one, the behaviour becomes more stable as brightness increases meaning the rate at which the colour changes decreases, or altogether flattens; bluer-stable-when-brighter and redder-stable-when-brighter (BSWB and RSWB, respectively). The preferred fit for each source is included in Fig. \ref{fig:all_colour}. A dotted line indicates the linear fit and a solid line indicates a logarithmic fit.

Of the 17 BL Lac types (8 HSPs, 3 ISPs and 6 LSPs), 15 show significant long-term colour--flux relationships; all with negative correlation coefficients. Two BL Lac-type sources did not display significantly correlated behaviour: 1ES 1959+650 (HSP) and A0 0235+164 (LSP). In most cases, these significant correlations are best fitted linearly, but six objects show the BSWB relationship, meaning their colour becomes more stable during high flux periods. All 13 FSRQ sources show significant long-term colour--flux behaviour. Two sources show a BWB trend, and one shows the BSWB trend. Eight show the RSWB, while only two FSRQs show linear RWB colour relationships.

\subsection{Optical spectral index -- gamma-ray}
\label{sec: colour gamma}

\begin{figure*}
\centering
\includegraphics[width=0.9\linewidth]{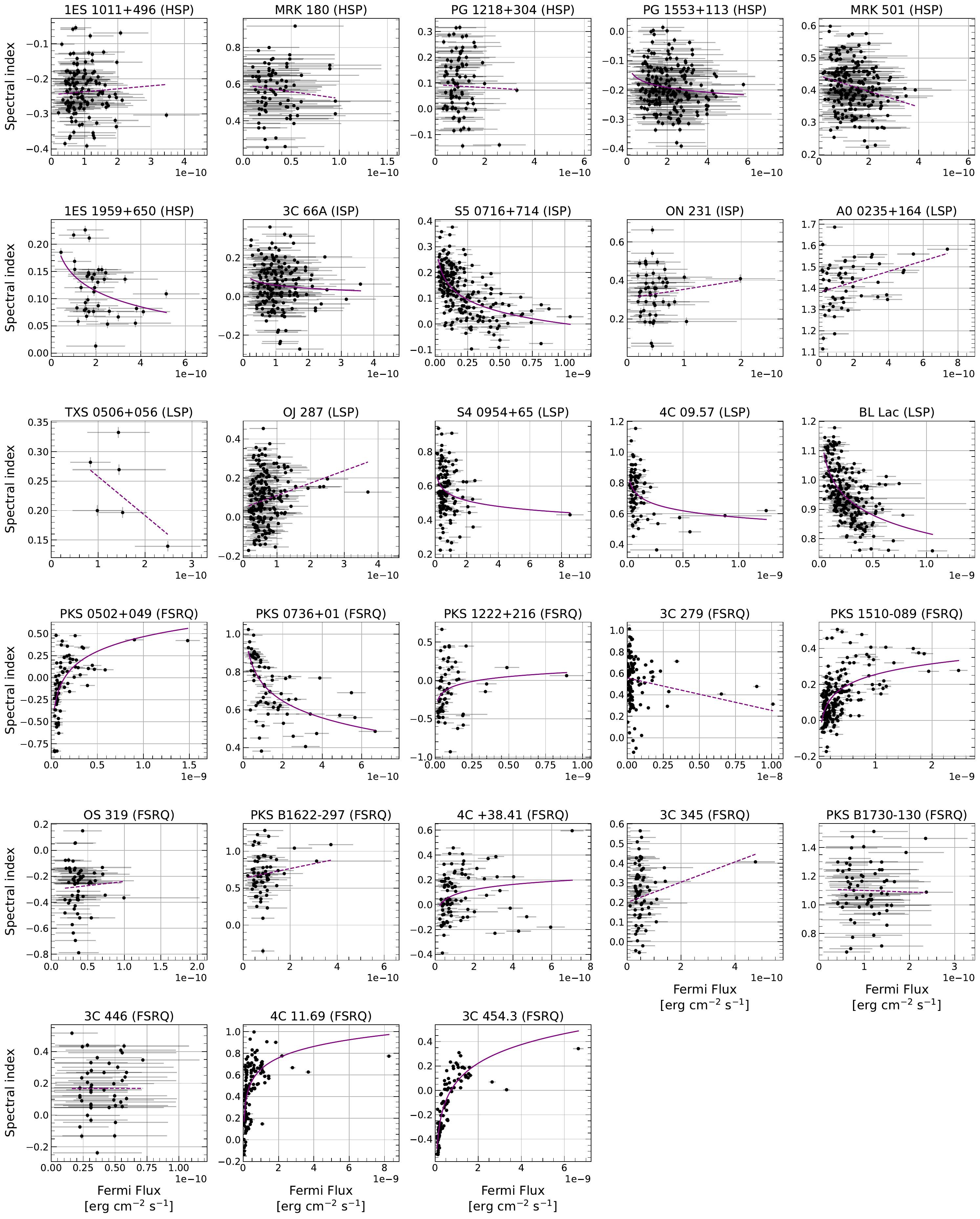}
\caption{As Fig. \ref{fig:all_colour} but for spectral index vs. $\gamma$-ray flux. The Spearman rank coefficients associated with each correlation are given in Table \ref{fermi_flux_colour_stats}.}
\label{fig:all_fermi_colour}
\end{figure*}

Fig. \ref{fig:all_fermi_colour} shows the optical spectral index vs. $\gamma$-ray flux correlations for each object in our sample, and Table \ref{fermi_flux_colour_stats} in the appendix shows their correlation strengths and significances. The $\gamma$-ray flux data are independent of the optical, meaning the optical spectral index could be calculated using all three optical flux colours (\textit{b*g*r*}), avoiding correlating errors (see Appendix \ref{false_col_cor} for more details). The error calculation for the spectral index per epoch, and the determination of linear/logarithmic fit preference was the same as detailed in Section \ref{sec: colour}. To reduce the chance of false correlations, we remove repeated instances where the same $\gamma$-ray flux value has been assigned to multiple optical epochs. 

Of the 28 sources, 15 show significant correlations between the optical spectral index and $\gamma$-ray flux. The strengths of these correlations range from very weak ($\sim\,$|0.164|) to very strong ($\sim\,$|0.897|) with both positive and negative correlation strengths. 

Of the 15 BL Lac type sources, eight showed significant behaviour (2 HSPs, 1 ISP, and all 5 LSPs). Additionally, all but two of the significantly correlated sources were negative in strength, implying an increase in $\gamma$-ray emission correlated with a decreasing optical spectral index (i.e. the sources became optically bluer when $\gamma$-ray brighter). Furthermore, five objects demonstrated the BSWB trend. Two sources, namely A0 0235+164 (LSP) and OJ287 (LSP), showed the opposite trend; these objects became optically redder when brighter in $\gamma$-rays, but followed the linear relationship. For the 13 FSRQs, seven showed correlated behaviour between the optical spectral index and $\gamma$-ray flux. When brighter in $\gamma$-rays one of these correlations showed the BSWB trend, while the remaining six showed the RSWB trend.

\subsection{Optical -- gamma-ray time lags}
\label{sec: lags}

The Discrete Correlation Function \citep[DCF;][]{edelson1988} was used to explore time lags between the optical and $\gamma$-ray data as described in \cite{mccall2024}. The DCF points are fitted using the \textsc{GaussianProcessRegressor} module and the Rational Quadratic kernel from \textsc{scikit-learn} in Python \citep{scikit-learn}. The maximum absolute value from this fit is chosen as the lag. 
The optical \textit{r*}-band flux is shifted with respect to the $\gamma$-ray flux, meaning a negative lag implies the $\gamma$-ray emission is leading the optical, and a positive lag implies the optical emission is leading the $\gamma$-ray. 

The results are shown in Table \ref{tab: lags} in the appendix. If a lag is inconsistent with zero at the $3\sigma$ level and has a strength coefficient of greater than 0.40 (see Table \ref{srank}) we label the lag as potentially significant (yes*). The $\Delta$Peak column shows the error on the peak value, given as the average cadence between successive optical observations. In all cases, this value was larger than the error associated with the calculation of the DCF. The sources that meet these criteria are 1ES 1959+650, ON231, PKS 1510-089, and PKS B1622-297. To determine whether the measured lags were real, they were applied to the optical data and the correlation was re-tested using the Spearman rank coefficient. We make the assumption that if the lags were real, the optical and $\gamma$-ray fluxes would become, or remain, significantly correlated with a larger strength value. These results are shown in Table \ref{tab: lag_adj_stats} where the Spearman rank correlation statistics before and after applying the detected lags are shown. We found that the correlations for PKS 1510-089 and PKS B1622-297 did not increase in strength after shifting, whereas for both 1ES 1959+650 and ON231, the optical--$\gamma$-ray correlations became significantly correlated. The coefficients suggested an inverse correlation for 1ES 1959+650, and direct for ON231. 

\begin{table}
\centering
\caption{Spearman rank correlation coefficients before and after application of the detected interband time lag between the optical and $\gamma$-ray fluxes.}
\label{tab: lag_adj_stats}
\resizebox{\columnwidth}{!}{
\begin{tabular}{ccccccc}
\hline
Source & Lag [days] & $\Delta$Lag [days] & \multicolumn{2}{c}{$p$} & \multicolumn{2}{c}{$c$} \\
       &     &             & Before & After & Before & After  \\
\hline
1ES 1959+650  & 29.75   & 9.34  & 0.197                & $4.87\times10^{-05}$ & -0.185 & -0.542 \\
ON 231        & 118.09  & 16.20 & 0.250                & $7.65\times10^{-03}$ & -0.128 & 0.293  \\
PKS 1510-089  & -55.41  & 9.44  & $2.34\times10^{-21}$ & $1.28\times10^{-11}$ & 0.565  & 0.421  \\
PKS B1622-297 & -129.82 & 17.29 & 0.012                & $7.46\times10^{-02}$ & 0.272  & 0.196  \\
\hline
\end{tabular}
}
\end{table}

We note the following caveats to the lag analysis results. The 1ES 1959+650 light curve shows that from MJD$\sim$57500--57560 (while in an optically fainter state) the data were obtained at a much higher cadence (one observation every two days) than the rest of the observations (one observation every two weeks). After correcting for the potential lag, this period then aligns with an active state in the $\gamma$-ray light curve. Outside this period there is no correlated behaviour. Given this, we conclude that the lag is likely a false detection caused by irregular sampling dominating the correlation statistics. In the case of ON231, we find the variability in the $\gamma$-ray light curve is consistent with noise, so the correlation results for this source are dubious. We therefore determine no significant long-term time lags between the optical and Fermi $\gamma$-ray fluxes in our data.

\section{Discussion}
\label{sec: discussion}
Correlations were explored using the Spearman rank test and were considered significant at the 95 per cent confidence interval, where $p\leq$0.05. We omitted those sources which did not have data accessible in the Fermi LAT LCR (IC 310, 1ES 1426+428) or did not have calibratable optical magnitude data (MRK 421). Table \ref{tab:all stats} shows the results of all four correlation tests performed in this work. The table gives a `Yes' or `No' describing whether or not a significant correlation was detected, or the observed colour trend for those significantly correlated.

\begin{table*}
\centering
\caption{Results of the different correlations explored in this work. The source name and subclass are given, along with `Yes' or `No' to describe whether or not the given correlation was statistically significant. The asterisk in the $3\sigma$ lag column indicates the potential significance of the detected lag, warranting further analysis.}
\label{tab:all stats}
\resizebox{\textwidth}{!}{
\begin{tabular}{cccccc}
\hline
Source        & Type       & opt flux -- $\gamma$-ray flux & $\alpha$ -- opt flux & $\alpha$ -- $\gamma$-ray flux & 3$\sigma$ Lag \\
\hline
IC 310        & HSP        & -                 & BWB      & -         & -           \\
1ES 1011+496  & HSP        & No                & BWB      & No        & No          \\
MRK 421       & HSP        & -                 & -        & -         & -           \\
MRK 180       & HSP        & No                & BWB      & No        & No          \\
PG 1218+304   & HSP        & No                & BSWB     & No        & No          \\
1ES 1426+428  & HSP        & -                 & BWB      & -         & -           \\
PG 1553+113   & HSP        & Yes               & BWB      & No        & No          \\
MRK 501       & HSP        & Yes               & BWB      & BWB       & No         \\
1ES 1959+650  & HSP        & No                & No       & BSWB      & Yes*        \\
3C 66A        & ISP        & Yes               & BWB      & No        & No          \\
S5 0716+714   & ISP        & Yes               & BSWB     & BSWB      & No          \\
ON 231        & ISP        & No                & BSWB     & No        & Yes*        \\
A0 0235+164   & LSP        & Yes               & No       & RWB       & No          \\
TXS 0506+056  & LSP        & Yes               & BWB      & No        & No          \\
OJ 287        & LSP        & Yes               & BSWB     & RWB       & No         \\
S4 0954+65    & LSP        & Yes               & BSWB     & BSWB      & No         \\
4C 09.57      & LSP        & Yes               & BWB      & BSWB      & No          \\
BL Lac        & LSP        & Yes               & BSWB     & BSWB      & No          \\
PKS 0502+049  & LSP (FSRQ) & Yes               & RSWB     & RSWB      & No          \\
PKS 0736+01   & LSP (FSRQ) & Yes               & BSWB     & BSWB      & No          \\
PKS 1222+216  & LSP (FSRQ) & Yes               & RSWB     & RSWB      & No          \\
3C 279        & LSP (FSRQ) & Yes               & BSWB     & No        & No          \\
PKS 1510-089  & LSP (FSRQ) & Yes               & RSWB     & RSWB      & Yes*        \\
OS 319        & LSP (FSRQ) & Yes               & RWB      & No        & No          \\
PKS B1622-297 & LSP (FSRQ) & Yes               & RSWB     & No        & Yes*        \\
4C +38.41     & LSP (FSRQ) & Yes               & RSWB     & RSWB      & No          \\
3C 345        & LSP (FSRQ) & No                & BWB      & No        & No          \\
PKS B1730-130 & LSP (FSRQ) & No                & RSWB     & No        & No          \\
3C 446        & LSP (FSRQ) & Yes               & RWB      & No        & No          \\
4C 11.69      & LSP (FSRQ) & Yes               & RSWB     & RSWB      & No          \\
3C 454.3      & LSP (FSRQ) & Yes               & RSWB     & RSWB      & No          \\
\hline
\end{tabular}
}
\end{table*}

\subsection{Optical--gamma-ray analysis}
Figures \ref{fig:flux-fermi_magnitude} and \ref{fig:flux-fermi_abs_magnitude} show the optical and $\gamma$-ray correlations of the RINGO3 sample (both for the whole sample and separated by blazar subclasses) as functions of both apparent and absolute magnitude. We find 21 out of 28 (i.e. 75 percent) of sources showed significant positive correlations ranging in strength from $0.217\leq\textit{c}\leq0.891$. Exploring the correlated behaviour between optical and $\gamma$-ray flux allows the exploration of emission processes within the jets.
In the leptonic scenario, higher-energy $\gamma$-ray emission is a result of inverse-Compton upscattering of lower energy seed photons via relativistic particles in the jet \citep{maraschi1992,bloom1996,bottcher2013b}. If the seed photons are from the same population of photons responsible for lower-energy synchrotron emission, one would expect changes at optical and $\gamma$-ray wavelengths to be positively correlated over long timescales. Conversely, in hadronic models, the high-energy emission is produced through proton synchrotron emission or proton-proton interactions. In this case, long-term correlations between optical and $\gamma$-ray emission would be less likely \citep{mannheim1992,aharonian2000,bottcher2013b}. In this work we found significant positive correlations between optical and $\gamma$-ray flux for the majority of sources: We therefore conclude that the dominant emission mechanism within our blazar jet sample is likely leptonic. 

Sources with high synchrotron peaks showed fewer significant correlations (33 per cent for HSPs). The number of $\gamma$-ray--optical correlations increases to 67 per cent of ISP sources and 89 per cent of LSP objects (which includes all BL Lac objects and 85 per cent of FSRQs). \cite{ringo2paper} explored optical-$\gamma$-ray correlations for a sample of 15 blazars monitored with the RINGO2 and DIPOL polarimeters and determined that significant positive correlations were found in 68 per cent of cases. When considering object sub-classes, they found 43 per cent, 50 per cent, 88 per cent, and 60 per cent of HBL, IBL, LBL, and FSRQ sources, respectively, showed positively correlated behaviour. The overall results are therefore similar between the RINGO2 and RINGO3 analyses. 
The decrease in the strength of optical--$\gamma$-ray emission correlations on increasing synchrotron peak frequency (blazar subclass) can be understood as host-galaxy contamination in the case of HSPs. The host-galaxy emission would dilute the optical behaviour of the jet and as such result in weaker correlations with the $\gamma$-ray flux \citep{gaur2014,chang2019,otero-santos2022}.
However, one cannot rule out the possibility that the intrinsic properties of HSPs differ from those of ISP and LSP objects, leading to weaker correlations.

Based on the optical and $\gamma$-ray flux properties, most sources in this sample occupied one of two regions in figures \ref{fig:flux-fermi_magnitude} and \ref{fig:flux-fermi_abs_magnitude}. These regions are attributed to the FSRQ/BL~Lac classification of blazars; with FSRQs typically displaying brighter $\gamma$-ray fluxes than BL Lacs. There are three sources, however, that appear to occupy the region in between these two blazar subclasses: A0 0235+164, S4 0954+64 and 4C 09.57. A0 0235+164 was originally classified as a BL Lac LSP source by \cite{spinrad1975} due to its featureless spectrum, however, \cite{ruan2014} indicated that A0 0235+164 may belong to the FSRQ transitional class of blazars. This supports the conclusions of \cite{ackermann2012} who model the A0 0235+164 SED during a flaring episode in 2008-2009 and find that the source's isotropic luminosity is more indicative of FSRQs than BL Lacs, and the dominant mechanism for the high energy emission is more likely to be external Compton processes; a signature of FSRQs, rather than synchrotron self-Compton. Similar to A0 0235+164, S4 0954+65 was also classified as a BL Lac object \citep{stickel1991}. \cite{ghisellini2011} classify this source as a low-frequency peaked blazar (LBL) due to the absence of prominent emission lines and the appearance of its SED. However, \cite{hervet2016} classify the sources as an FSRQ by analysing the kinematic features of its radio jet. Furthermore, \cite{MAGIC2018} model the multiwavelength emission of S4 0954+65 and compare it to other sources, concluding that it shows many behavioural similarities to FSRQs and other suspected transitional/masquerading BL Lac objects. \cite{ghisellini2011} classify 4C 09.57 as a BL Lac LSP object based on the shape of its SED, however, they also observe emission lines with equivalent widths up to 12.5\AA. 4C 09.57 may also be a transitional object between the two subclasses \citep{uemura2017}.

The correlations presented in Figs. \ref{fig:flux-fermi_magnitude} and \ref{fig:flux-fermi_abs_magnitude} also agree with the results of \cite{hovatta2014} in that the flux-flux correlations appear tighter for HSP and ISP sources compared to that of LSP objects (BL Lacs and FSRQs). In the SSC case, the low- and high-energy emission originates from the same region of the blazar jet so the emission is subject to the same level of Doppler boosting. This would mean those sources which have $\gamma$-ray production dominated by SSC emission should show tight correlations between the optical and $\gamma$-ray fluxes. Any EC emission should be more strongly boosted, obscuring any linear dependence between the low- and high-energy emission \citep{dermer1995}. In this EC case, the flux-flux correlations would appear more scattered. Our results support HSP and ISP sources having SSC as the dominant $\gamma$-ray emission mechanism, whereas LSP sources would show significant EC emission.

\subsection{Spectral analysis (vs. optical and gamma-ray)}

Fig. \ref{fig:all_colour} shows the spectral index vs. optical flux diagrams for all sources in the sample. Significant correlations were found for 28 of the 30 sources (15 BL Lacs and all 13 FSRQs). For the BL Lac objects, all 15 showed BWB behaviour with six (40 per cent) displaying a logarithmic trend, indicating the stabilisation of the colour at higher fluxes (bluer-stable-when-brighter; BSWB). Of the FSRQs, two showed linear RWB trends and one showed a linear BWB relationship. The rest (85 per cent) showed stable trends. One of the stable trends was BSWB, while the remaining eight were RSWB. The Spearman rank strengths of all significant correlations ranged from $\sim\,|0.195|$--$\sim\,|0.953|$ (weak--very strong).  

The exact mechanisms behind the optical colour behaviour of blazars are still debated and can be explained by both one and two-component synchrotron models. In the one-component model energy injection into the emitting regions increases the number of high-energy electrons, shifting the synchrotron SED peak to higher energies. This shift makes the object appear bluer \citep{ikejiri2011}. The most accepted reasoning for this energy injection would be internal shocks travelling through the jet \citep{mastichiadis2002,guetta2004}. In the two-component model, the total emission comprises radiation from different regions of the blazar, notably an underlying thermal contribution originating from the accretion disc and a more variable, non-thermal contribution from the jet \citep{fiorucci2004}. If the flare component has a higher synchrotron peak frequency than that of the thermal emission, BWB trends would be observed. An observed feature in the SED of FSRQs is the UV bump \citep{gu2006}, an excess of thermal emission which flattens the thermal/non-thermal composite spectrum at optical wavelengths. The increase in the thermal emission likely originates from the accretion disk and BLR. It follows that when the source brightens and the non-thermal emission increases, the spectrum steepens resulting in RWB trends \citep{ramirez2004,gu2006}.

Our work shows more complex behaviour than a linear relationship between colour and flux, where colour changes stall, or become altogether absent, during heightened optical activity. \cite{zhang2022} suggest a unified model to explain observed blazar colour behaviour based on a two-component scenario made up of a less-variable thermal emission component from the accretion disk, and a highly variable non-thermal component from the jet synchrotron emission. They model the observed changes to the spectral index as a logarithmic relation given by 
\begin{equation}
    \alpha_\mathrm{obs} = 2.67\,\mathrm{ln}\left[a+\frac{b}{F_\mathrm{obs,R}}\right] 
\end{equation}
where $a$ and $b$ are free parameters, and $F_\mathrm{obs,R}$ is the observed flux in $R$ band. For a given source the spectral index, $\alpha_\mathrm{obs}$, depends only on $F_\mathrm{obs,R}$. This is based on the assumption that the two spectral index components (thermal and less variable, non-thermal and highly variable) are constant.

Our results agree with the work by \cite{zhang2022} and \cite{zhang2023}: non-linear fits can better describe the relationship between the spectral index and flux in some sources, and the spectral index flattens during high states in all blazar classes. Where this is the case, the data can be fitted well by a single logarithmic curve with two free parameters. For those sources where no SWB features are observed, a lack of observations during high or low states may explain the seemingly linear trend.

This same analysis was used to look for non-linear relationships between the optical spectral index and $\gamma$-ray flux, the results of which are shown in Fig. \ref{fig:all_fermi_colour} with best fits and correlation statistics shown in Table \ref{fermi_flux_colour_stats}. We found 15 sources to have significant correlations (54 per cent), eight BL Lacs (two HSP, one ISP, and five LSP) and seven FRSQs. Only three of the BL Lac objects displayed linear relationships (one BWB and two RWB), while the remaining five BL Lacs and all seven FSRQs showed stable features (all BSWB for the BL Lacs types and one FSRQ, and six RSWB for the remaining FSRQs).  

Taking the leptonic scenario as the dominant source of high energy emission, the $\gamma$-ray emission from blazars originates from inverse-Compton processes occurring within the jet. While the optical emission can seed the $\gamma$-ray, causing the relationships observed in Section \ref{sec: flux correlations}, it can also be composed of accretion disc variability or host galaxy emission. It follows that if the $\gamma$-ray emission is correlated with changes in the optical spectral index, then both emissions are more likely to be a result of jet activity, rather than coinciding disc and jet processes. 



\subsection{Time lag analysis}
Leptonic modelling of blazar jet emission requires the high-energy $\gamma$-ray emission to be a result of inverse-Compton scattering of photons from the lower-energy synchrotron electrons. The seed photons may come from two locations; either the synchrotron photons from within the jet \cite[synchrotron-self Compton;][]{maraschi1992} or photons from outside the jet \cite[external Compton;][]{dermer1993}. It follows that temporal lags between the optical and $\gamma$-ray emission may allow the distinction between the two high-energy emission processes given the difference in separation between the low- and high-energy emitting regions \citep{cohen2014}.

Time lags were tested for all sources with optical and $\gamma$-ray data. As summarized in Table \ref{tab:all stats}, we found little evidence of significantly lagged behaviour. Although four sources showed a potential lag based on the discrete correlation function, further analysis showed that all were likely false correlations. This included re-correlating the optical and $\gamma$-ray fluxes after applying the detected lag to the optical data, and inspecting the light curves for irregularities in cadence which could dominate the time lag correlation statistics. In summary, no evidence of long-term time lags with high confidence was found in our sample.

Our analysis differs from that of related work in that we look for long-term lag behaviour between the optical and $\gamma$-ray bands, rather than individual flares. However, our analysis is in agreement with recent work \citep{cohen2014,liodakis2019,dejaeger2023} in that we find little evidence for temporal lags between the optical and $\gamma$-ray bands that are not consistent with zero days at the $3\sigma$ level. This means our data are suggestive of SSC processes dominating the jet flux, but further analysis into the characteristics of individual optical flares, and any gamma-ray correlations, could reveal more detail.


\subsection{Correlations summary}
Table \ref{tab:all stats} shows the results of all statistical tests performed in this work for each object in our sample. We notice that the majority of sources show the same relationships across tests, which could be used to infer the dominant radiation mechanisms over long timescales. 

13 sources show significant correlations between the spectral index and optical flux but not between the spectral index and $\gamma$-ray flux. Of these 13, six (three HSPs, one ISP, and two FSRQs) do not show optical and $\gamma$-ray flux correlations, possibly due to relatively inactive or indiscernible $\gamma$-ray behaviour during our observations. The remaining seven sources (one HSP, one ISP, one LSP, and four FSRQs) do show significant optical and $\gamma$-ray flux correlations with no detected time lag. This implies their $\gamma$-ray brightening episodes are optically colourless; a possible indication of $\gamma$-ray emission originating from seed photon fields outside the jet and a signature of EC processes.

12 sources (one HSP, one ISP, three LSPs, and seven FSRQs) show significant correlations between the optical and $\gamma$-ray fluxes with no detectable time lag, along with significant spectral index correlations for both optical and $\gamma$-ray flux, with matching spectral trends (ie. both B(S)WB or both R(S)WB). This implies a strong connection between the jet's higher and lower energy behaviour in these sources, likely an indication of SSC-dominated emission.

Two objects (1ES 1959+650; HSP, and A0 0235+164; LSP) demonstrate significant spectral index vs. $\gamma$-ray flux correlations, but not spectral index vs. optical flux correlations. Furthermore, 1ES 1959+650 shows no significant correlation between the optical and $\gamma$-ray fluxes whereas A0 0235+164 does. In the case of 1ES 1959+650, the correlations indicate that optical colour variability from jet emission may be present, but would not be temporally consistent with any optical flux, but rather consistent with preceding or delayed $\gamma$-ray activity. Our time lag result for this source hinted at such behaviour but was ultimately disregarded due to unevenly sampled optical data. More observations of this object with regular sampling need to be obtained, ideally during a range of activity states, in order to make more definitive conclusions. For A0 0235+164, the majority of our observations occurred during a prolonged optical and $\gamma$-ray heightened state. It is possible that our optical observations only detected the flat region of a logarithmic RSWB colour trend and therefore appeared colourless, as not enough data were obtained during quiescent and/or low states. Furthermore, it has been previously reported that this object has a much brighter accretion disc than other LSP BL Lacs, and is more comparable to that of FSRQs \citep{ghisellini2010,zhang2022}. This would explain why RWB spectral index vs. $\gamma$-ray flux behaviour was observed, in contrast to the generally accepted, almost exclusively BWB behaviour of BL Lac type blazars.

Finally, one object (OJ287; LSP) showed significantly correlated optical and $\gamma$-ray fluxes, but conflicting trends in the spectral index vs. optical flux (BSWB) and spectral index vs. $\gamma$-ray flux (RWB). We note that spectral index vs. $\gamma$-ray flux is very weakly correlated, and may be slightly skewed, attributed to an expected accretion disk impact flare by a hypothesised secondary SMBH companion \citep{lehto1996}. This will be explored in more detail in a future publication utilising the full RINGO3 photopolarimetric data set.

\section{Conclusions}

A sample of 31 blazars made up of 18 BL Lac and 13 FSRQ types were observed over a period of seven years using the multicolour, simultaneous polarimeter RINGO3 on the Liverpool Telescope. We combined our optical photometric data with Fermi $\gamma$-ray data and summarise our findings as follows:

\begin{itemize}
    \item 75 per cent of sources show significant optical--$\gamma$-ray flux correlations, which consisted of 67 per cent of BL Lac types and 85 per cent of FSRQs. The greater scatter in the correlations for LSP objects compared to ISP or HSP sources indicates the possible presence of a more significant external Compton contribution in our sample.
    
    \item Significant spectral behaviour was found in 93 per cent of sources: 88 per cent of BL Lacs and 100 per cent of FSRQs. We find evidence to suggest that in the majority of cases, the behaviour is best fit logarithmically rather than linearly, implying a transition between BWB/RWB to SWB spectral behaviour in higher activity states. We also conclude that poor sampling or lack of high-activity states during our observation periods might result in poor identification of the stable spectral behaviour responsible for logarithmic relationships. 
    
    \item Significant correlations between the optical spectral index and $\gamma$-ray flux were found in 54 per cent of sources. This consisted of 53 per cent of BL Lacs and 54 per cent of FSRQs. We detect here both BWB and RWB behaviour and in the majority of cases a SWB tendency is present at the highest activity states, resulting in the best fit for the data being logarithmic.
    
    \item We find no indication of significant interband time lags (not consistent with zero days) at the $3\sigma$ level between the optical and $\gamma$-ray fluxes, which is indicative of synchrotron-self Compton processes dominating the observed flux.
\end{itemize}

In this work, we have distinguished between radiation mechanisms and particle populations in blazars jets using a large photometric dataset of blazars. RINGO3 is also a polarimeter and polarimetric observations of blazars can be instrumental in disentangling the jet's synchrotron emission from any thermal contributions from other parts of the AGN. These polarimetric properties will be explored in the next publication. The results presented in this work encourage further high-cadence photopolarimetric observations of all blazar subclasses to ensure adequate monitoring of all activity states.

\section*{Acknowledgements}
We thank the anonymous reviewer for their constructive comments. The Liverpool Telescope is operated on the island of La Palma by Liverpool John Moores University in the Spanish Observatorio del Roque de los Muchachos of the Instituto de Astrofisica de Canarias with financial support from the UKRI Science and Technology Facilities Council (STFC) (ST/T00147X/1). 

I.A. acknowledges financial support from the grant CEX2021-001131-S funded by MCIN/AEI/10.13039/501100011033 to the Instituto de Astrof\'isica de Andaluc\'ia-CSIC and from MICIN grants PID2019-107847RB-C44 and PID2022-139117NB-C44.

U.B.A. acknowledges the receipt of a FAPERJ State Scientist Fellowship nr. E-26/200.532/2023 and a CNPq Productivity Fellowship nr. 309053/2022-6

T.H. was supported by the Research Council of Finland projects 317383, 320085, 322535, and 345899.

G.P.L. is supported by Royal Society Dorothy Hodgkin Fellowship (grant Nos. DHF-R1-221175 and DHFERE-221005)

\section*{Data Availability}
All data presented here are available through the Liverpool Telescope data archive at https://telescope.astro.ljmu.ac.uk. The data underlying this article will be shared upon reasonable request to the corresponding author.

\bibliographystyle{mnras}
\bibliography{Paper}

\appendix
\section{False correlations in colour-magnitude analysis}
\label{false_col_cor}

When correlation analysis is performed on datasets where the same data are used on two axes, they contain a common uncertainty which may produce a false correlation due to correlated errors. To demonstrate this we use the example of a simple colour magnitude correlation. $b$ vs. $b-r$ and $r$ vs. $b-r$ plots were made with RINGO3 data to compare with the analysis made by \cite{gupta2016}. The correlations suggested that sources were both bluer- and redder-when-brighter, depending on the chosen x-axis (\textit{b*} magnitude and \textit{r*} magnitude respectively). This highlights the necessity to use two independent data sets when performing correlation statistics (i.e. both data sets must have independent uncertainties)

\begin{figure*}
\centering
\includegraphics[width=\linewidth]{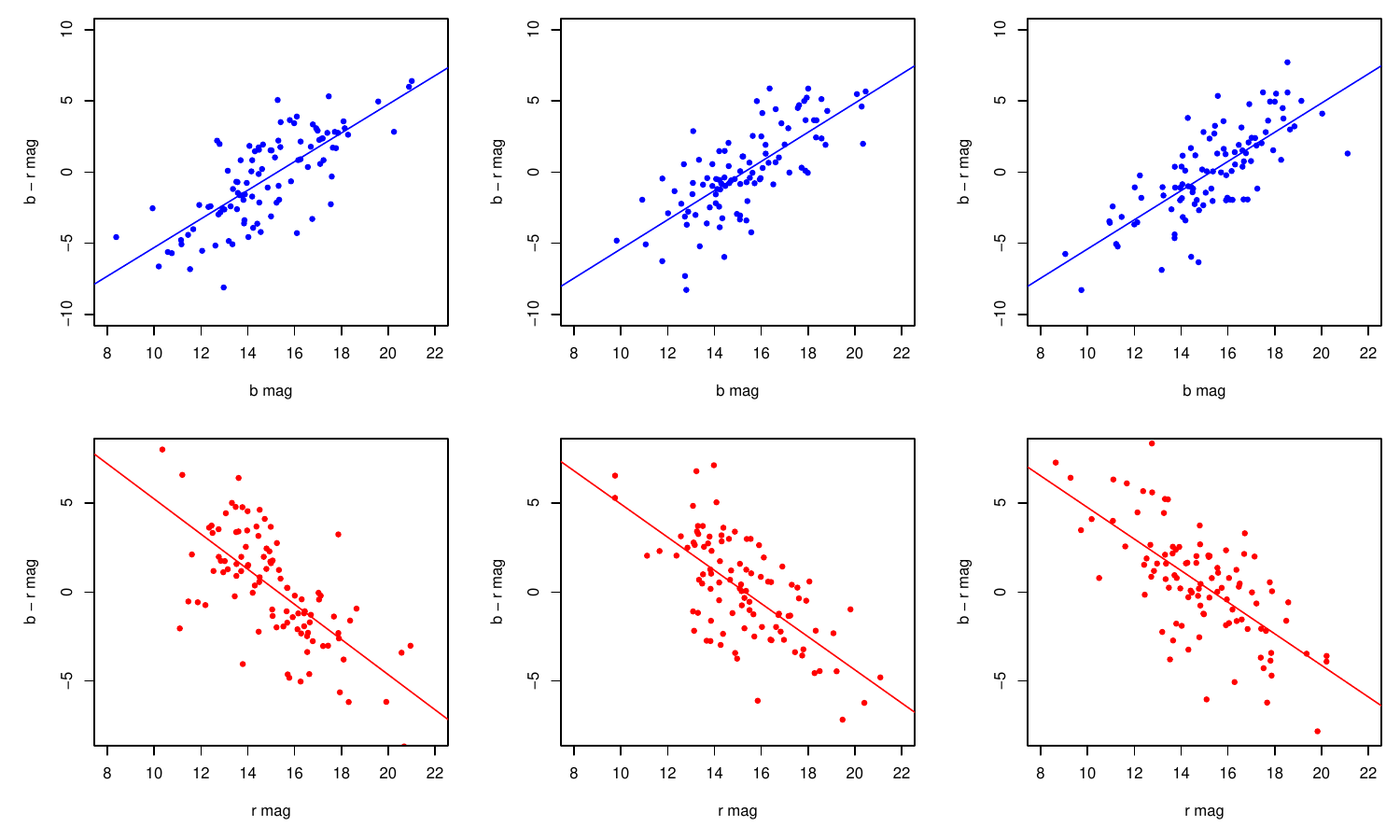}
\caption{Plots of randomly generated $b$ and $r$ values in the form x = $b$ y = $b$-$r$ (blue) and x = $r$ y = $b$-$r$ (red). This shows three iterations of the code and the tendency of the fit to be positive (blue) or negative (red).}
\label{fig:col_prob}
\end{figure*}

To explore this property independently from RINGO3 data, random magnitude values ($b$ and $r$) were generated using Monte Carlo methods, along with randomly generated error values, and plotted against each other in the form $b$ vs. $b$-$r$ and $r$ vs. $b$-$r$. A linear regression was fitted to each set of data points and this showed the preference for x-axis-dependent correlations. Fig. \ref{fig:col_prob} shows an example of 3 repetitions of this process with one hundred values centred on 15, and error values centred on 2. These trends highlight the necessity to have truly independent data for correlation analysis. For this reason the optical spectral index (a similar measurement to colour) calculated in this work uses the \textit{r*} and \textit{b*} flux only, so as to not include a common uncertainty on both the x and y axes when correlated against the \textit{g*}-band flux.

\section{Figures and Tables}

\begin{table*}
\centering
\caption{Spearman correlation coefficients for optical flux vs. Fermi $\gamma$-ray flux, where $c$ is the correlation coefficient and $p$ is the corresponding significance coefficient}. The number of optical data points (per camera) used in these correlation calculations is also shown. Note there are no Fermi data available for IC 310 or 1ES 1426+428.
\label{fermi_flux_stats}
\resizebox{\textwidth}{!}{%
\begin{tabular}{ccccccccc}
\hline
Source & Type & \textit{r*} \textit{c} & \textit{r*} \textit{p} & \textit{g*} \textit{c} & \textit{g*} \textit{p} & \textit{b*} \textit{c} & \textit{b*} \textit{p} & Points\\
\hline
IC 310        & HSP        & -      & -                     & -      & -                     & -      & -                     & 206    \\
1ES 1011+496  & HSP        & -0.113 & 0.144                 & -0.088 & 0.252                 & -0.107 & 0.165                 & 170    \\
MRK 421       & HSP        & -  & -                 & -  & -                 & -  & -                 & 281    \\
MRK 180       & HSP        & 0.117  & 0.126                 & 0.110  & 0.152                 & 0.137  & 0.074                 & 171    \\
PG 1218+304   & HSP        & 0.043  & 0.653                 & 0.002  & 0.984                 & 0.027  & 0.777                 & 110    \\
1ES 1426+428  & HSP        & -      & -                     & -      & -                     & -      & -                     & 219    \\
PG 1553+113   & HSP        & 0.482  & 2.04$\times 10^{-16}$ & 0.480  & 3.00$\times 10^{-16}$ & 0.471  & 1.11$\times 10^{-15}$ & 258    \\
MRK 501       & HSP        & 0.366  & 1.21$\times 10^{-11}$ & 0.391  & 3.14$\times 10^{-13}$ & 0.365  & 1.42$\times 10^{-11}$ & 322    \\
1ES 1959+650  & HSP        & -0.186 & 0.196                 & -0.145 & 0.315                 & -0.140 & 0.332                 & 50     \\
3C 66A        & ISP        & 0.438  & 6.02$\times 10^{-15}$ & 0.419  & 1.21$\times 10^{-13}$ & 0.439  & 5.22$\times 10^{-15}$ & 288    \\
S5 0716+714   & ISP        & 0.705  & 2.99$\times 10^{-44}$ & 0.710  & 3.92$\times 10^{-45}$ & 0.720  & 6.29$\times 10^{-47}$ & 286    \\
ON 231        & ISP        & -0.129 & 0.247                 & -0.179 & 0.108                 & -0.128 & 0.251                 & 82     \\
A0 0235+164   & LSP        & 0.674  & 2.94$\times 10^{-10}$ & 0.674  & 3.08$\times 10^{-10}$ & 0.660  & 9.20$\times 10^{-10}$ & 68     \\
TXS 0506+056  & LSP        & 0.569  & 0.034                 & 0.670  & 8.78$\times 10^{-3}$  & 0.670  & 8.78$\times 10^{-3}$  & 14     \\
OJ 287        & LSP        & 0.266  & 4.43$\times 10^{-10}$ & 0.254  & 2.62$\times 10^{-9}$  & 0.229  & 8.76$\times 10^{-8}$  & 534    \\
S4 0954+65    & LSP        & 0.472  & 2.60$\times 10^{-12}$ & 0.476  & 1.50$\times 10^{-12}$ & 0.465  & 6.06$\times 10^{-12}$ & 197    \\
4C 09.57      & LSP        & 0.458  & 9.20$\times 10^{-8}$  & 0.452  & 1.41$\times 10^{-7}$  & 0.454  & 1.20$\times 10^{-7}$  & 124    \\
BL Lac        & LSP        & 0.661  & 8.46$\times 10^{-50}$ & 0.674  & 2.44$\times 10^{-52}$ & 0.681  & 8.68$\times 10^{-54}$ & 385    \\
PKS 0502+049  & LSP (FSRQ) & 0.824  & 7.75$\times 10^{-30}$ & 0.824  & 6.12$\times 10^{-30}$ & 0.829  & 1.36$\times 10^{-30}$ & 116    \\
PKS 0736+01   & LSP (FSRQ) & 0.844  & 2.05$\times 10^{-35}$ & 0.837  & 3.37$\times 10^{-34}$ & 0.852  & 1.31$\times 10^{-36}$ & 126    \\
PKS 1222+216  & LSP (FSRQ) & 0.438  & 4.74$\times 10^{-5}$  & 0.405  & 1.98$\times 10^{-4}$  & 0.398  & 2.60$\times 10^{-4}$  & 80     \\
3C 279        & LSP (FSRQ) & 0.465  & 9.01$\times 10^{-9}$  & 0.478  & 2.94$\times 10^{-9}$  & 0.464  & 9.74$\times 10^{-9}$  & 138    \\
PKS 1510-089  & LSP (FSRQ) & 0.564  & 2.49$\times 10^{-21}$ & 0.559  & 7.33$\times 10^{-21}$ & 0.521  & 6.35$\times 10^{-18}$ & 237    \\
OS 319        & LSP (FSRQ) & 0.217  & 0.049                 & 0.353  & 1.06$\times 10^{-3}$  & 0.264  & 0.016                 & 83     \\
PKS B1622-297 & LSP (FSRQ) & 0.272  & 0.012                 & 0.222  & 0.042                 & 0.231  & 0.035                 & 84     \\
4C +38.41     & LSP (FSRQ) & 0.668  & 6.13$\times 10^{-13}$ & 0.639  & 1.21$\times 10^{-11}$ & 0.644  & 7.50$\times 10^{-12}$ & 90     \\
3C 345        & LSP (FSRQ) & 0.089  & 0.381                 & 0.035  & 0.729                 & 0.024  & 0.815                 & 99     \\
PKS B1730-130 & LSP (FSRQ) & 0.026  & 0.805                 & 0.057  & 0.593                 & 0.071  & 0.506                 & 90     \\
3C 446        & LSP (FSRQ) & 0.318  & 4.52$\times 10^{-3}$  & 0.332  & 2.99$\times 10^{-3}$  & 0.261  & 0.021                 & 78     \\
4C 11.69      & LSP (FSRQ) & 0.761  & 2.36$\times 10^{-42}$ & 0.761  & 3.24$\times 10^{-42}$ & 0.760  & 4.01$\times 10^{-42}$ & 217    \\
3C 454.3      & LSP (FSRQ) & 0.891  & 4.72$\times 10^{-37}$ & 0.887  & 2.50$\times 10^{-36}$ & 0.869  & 3.63$\times 10^{-33}$ & 105    \\
\hline
\end{tabular}%
}
\end{table*}

\begin{table*}
\centering
\caption{Correlation strengths for the spectral index vs. \textit{g*}-band flux correlations. The source name and subclass are given in columns one and two. The best fit functional form} determined by AIC and BIC values is shown in column three. Columns four and five give the Spearman rank strength and significance correlation coefficients after having linearised the dataset if better fitted with a logarithmic curve. Column six gives the colour trend of the object given the fit and Spearman rank coefficients. Column seven gives the average spectral index, and column 8 gives the number of data points.
\label{colour_mag_stats_full}
\resizebox{\textwidth}{!}{%
\begin{tabular}{cccccccc}
\hline
Source        & Type       & Fit    & $c$      & $p$                     & Trend & $\alpha_\mathrm{av}$ & Points \\
\hline
IC 310        & HSP        & linear & -0.288 & 2.72$\times 10^{-5}$  & BWB   & 1.134                & 206    \\
1ES 1011+496  & HSP        & linear & -0.251 & 9.51$\times 10^{-4}$  & BWB   & -0.233               & 170    \\
MRK 421       & HSP        & -      & -      & -                     & -     & -                    & 281    \\
MRK 180       & HSP        & linear & -0.527 & 1.29$\times 10^{-13}$ & BWB   & 0.578                & 171    \\
PG 1218+304   & HSP        & log    & -0.735 & 6.11$\times 10^{-20}$ & BSWB  & 0.091                & 110    \\
1ES 1426+428  & HSP        & linear & -0.266 & 6.84$\times 10^{-5}$  & BWB   & 0.426                & 219    \\
PG 1553+113   & HSP        & linear & -0.378 & 3.51$\times 10^{-10}$ & BWB   & -0.211               & 258    \\
MRK 501       & HSP        & linear & -0.527 & 2.32$\times 10^{-24}$ & BWB   & 0.416                & 322    \\
1ES 1959+650  & HSP        & linear & -0.088 & 0.543                 & -     & 0.093                & 50     \\
3C 66A        & ISP        & linear & -0.195 & 9.01$\times 10^{-4}$  & BWB   & 0.065                & 288    \\
S5 0716+714   & ISP        & log    & -0.639 & 3.38$\times 10^{-34}$ & BSWB  & 0.123                & 286    \\
ON 231        & ISP        & log    & -0.487 & 3.49$\times 10^{-6}$  & BSWB  & 0.336                & 82     \\
A0 0235+164   & LSP        & linear & 0.164  & 0.182                 & -     & 1.431                & 68     \\
TXS 0506+056  & LSP        & linear & -0.881 & 3.11$\times 10^{-5}$  & BWB   & 0.271                & 14     \\
OJ 287        & LSP        & log    & -0.494 & 2.92$\times 10^{-34}$ & BSWB  & 0.149                & 534    \\
S4 0954+65    & LSP        & log    & -0.598 & 1.74$\times 10^{-20}$ & BSWB  & 0.576                & 197    \\
4C 09.57      & LSP        & linear & -0.421 & 1.11$\times 10^{-6}$  & BWB   & 0.750                & 124    \\
BL Lac        & LSP        & log    & -0.637 & 3.51$\times 10^{-45}$ & BSWB  & 0.931                & 385    \\
PKS 0502+049  & LSP (FSRQ) & log    & 0.825  & 5.64$\times 10^{-30}$ & RSWB  & -0.172               & 116    \\
PKS 0736+01   & LSP (FSRQ) & log    & -0.816 & 2.35$\times 10^{-31}$ & BSWB  & 0.774                & 126    \\
PKS 1222+216  & LSP (FSRQ) & log    & 0.769  & 7.98$\times 10^{-17}$ & RSWB  & -0.154               & 80     \\
3C 279        & LSP (FSRQ) & log    & -0.374 & 6.18$\times 10^{-6}$  & BSWB  & 0.524                & 138    \\
PKS 1510-089  & LSP (FSRQ) & log    & 0.566  & 1.76$\times 10^{-21}$ & RSWB  & 0.136                & 237    \\
OS 319        & LSP (FSRQ) & linear & 0.227  & 0.039                 & RWB   & -0.290               & 83     \\
PKS B1622-297 & LSP (FSRQ) & log    & 0.476  & 4.85$\times 10^{-6}$  & RSWB  & 0.599                & 84     \\
4C +38.41     & LSP (FSRQ) & log    & 0.261  & 0.013                 & RSWB  & 0.048                & 90     \\
3C 345        & LSP (FSRQ) & linear & -0.210 & 0.037                 & BWB   & 0.243                & 99     \\
PKS B1730-130 & LSP (FSRQ) & log    & 0.426  & 2.78$\times 10^{-5}$  & RSWB  & 1.077                & 90     \\
3C 446        & LSP (FSRQ) & linear & 0.243  & 0.032                 & RWB   & 0.120                & 78     \\
4C 11.69      & LSP (FSRQ) & log    & 0.723  & 2.16$\times 10^{-36}$ & RSWB  & 0.469                & 217    \\
3C 454.3      & LSP (FSRQ) & log    & 0.953  & 2.87$\times 10^{-55}$ & RSWB  & -0.167               & 105    \\
\hline
\end{tabular}
}
\end{table*}

\begin{table*}
\centering
\caption{As Table \ref{colour_mag_stats_full} but for spectral index vs. $\gamma$-ray flux.}
\label{fermi_flux_colour_stats}
\resizebox{\textwidth}{!}{%
\begin{tabular}{cccccccc}
\hline
Source        & Type       & Fit    & $c$      & $p$                     & Trend & $\alpha_\mathrm{av}$ & Points \\
\hline
IC 310        & HSP        & -      & -      & -                     & -     & -              & 206    \\
1ES 1011+496  & HSP        & linear & 0.035  & 0.698                 & -     & -0.242         & 127    \\
MRK 421       & HSP        & -      & -      & -                     & -     &  -             & 281     \\
MRK 180       & HSP        & linear & 0.001  & 0.994                 & -     & 0.570          & 82     \\
PG 1218+304   & HSP        & linear & 0.004  & 0.974                 & -     & 0.098          & 83     \\
1ES 1426+428  & HSP        & -      & -      & -                     & -     & -              & 219    \\
PG 1553+113   & HSP        & log    & -0.138 & 0.054                 & -     & -0.200         & 196    \\
MRK 501       & HSP        & linear & -0.165 & 0.017                 & BWB   & 0.405          & 209    \\
1ES 1959+650  & HSP        & log    & -0.342 & 0.033                 & BSWB  & 0.120          & 39     \\
3C 66A        & ISP        & log    & -0.061 & 0.435                 & -     & 0.056          & 169    \\
S5 0716+714   & ISP        & log    & -0.615 & 5.75$\times 10^{-22}$ & BSWB  & 0.125          & 198    \\
ON 231        & ISP        & linear & 0.047  & 0.745                 & -     & 0.351          & 50     \\
A0 0235+164   & LSP        & linear & 0.310  & 0.023                 & RWB   & 1.438          & 54     \\
TXS 0506+056  & LSP        & linear & -0.714 & 0.111                 & -     & 0.235          & 6      \\
OJ 287        & LSP        & linear & 0.164  & 0.024                 & RWB   & 0.093          & 188    \\
S4 0954+65    & LSP        & log    & -0.190 & 0.041                 & BSWB  & 0.564          & 116    \\
4C 09.57      & LSP        & log    & -0.261 & 0.022                 & BSWB  & 0.752          & 77     \\
BL Lac        & LSP        & log    & -0.598 & 4.11$\times 10^{-15}$ & BSWB  & 0.949          & 142    \\
PKS 0502+049  & LSP (FSRQ) & log    & 0.677  & 1.65$\times 10^{-10}$ & RSWB  & -0.056         & 69     \\
PKS 0736+01   & LSP (FSRQ) & log    & -0.673 & 4.27$\times 10^{-10}$ & BSWB  & 0.733          & 67     \\
PKS 1222+216  & LSP (FSRQ) & log    & 0.300  & 0.020                 & RSWB  & -0.140         & 60     \\
3C 279        & LSP (FSRQ) & linear & -0.158 & 0.091                 & -     & 0.541          & 116    \\
PKS 1510-089  & LSP (FSRQ) & log    & 0.527  & 9.67$\times 10^{-14}$ & RSWB  & 0.116          & 173    \\
OS 319        & LSP (FSRQ) & linear & -0.004 & 0.976                 & -     & -0.248         & 59     \\
PKS B1622-297 & LSP (FSRQ) & linear & 0.152  & 0.251                 & -     & 0.682          & 59     \\
4C +38.41     & LSP (FSRQ) & log    & 0.294  & 6.31$\times 10^{-3}$  & RSWB  & 0.045          & 85     \\
3C 345        & LSP (FSRQ) & linear & 0.094  & 0.440                 & -     & 0.255          & 70     \\
PKS B1730-130 & LSP (FSRQ) & linear & -0.035 & 0.782                 & -     & 1.089          & 65     \\
3C 446        & LSP (FSRQ) & linear & 0.132  & 0.377                 & -     & 0.158          & 47     \\
4C 11.69      & LSP (FSRQ) & log    & 0.633  & 7.86$\times 10^{-16}$ & RSWB  & 0.516          & 129    \\
3C 454.3      & LSP (FSRQ) & log    & 0.897  & 6.64$\times 10^{-37}$ & RSWB  & -0.155         & 101   \\
\hline
\end{tabular}
}
\end{table*}

\begin{table*}
\centering
\caption{DCF peak lag values with correlation strengths after computing on the optical and $\gamma$-ray fluxes. A negative lag implies the $\gamma$-ray emission is leading the optical. An asterisk in the $3\sigma$ lag column suggests the potential significance of the detected lag, warranting further analysis.}
\label{tab: lags}
\resizebox{\textwidth}{!}{
\begin{tabular}{cccccc}
\hline
Source        & Type       & Peak [days] & $\Delta$Peak [days]  & $c$    & sig? \\ 
\hline
IC 310        & HSP        & -               &  -                   & -      & -    \\
1ES 1011+496  & HSP        & 166.29          & 12.94                & -0.271 & no   \\
MRK 421       & HSP        & -               &  -                   & -      & -    \\
MRK 180       & HSP        & -68.19          & 12.86                & 0.202  & no   \\
PG 1218+304   & HSP        & 220.1           & 20.46                & 0.165  & no   \\
1ES 1426+428  & HSP        & -               &  -                   & -      & -    \\
PG 1553+113   & HSP        & -20.91          & 8.54                 & 0.493  & no   \\
MRK 501       & HSP        & -216.7          & 6.89                 & 0.373  & no   \\
1ES 1959+650  & HSP        & 29.75           & 9.34                 & -0.537 & yes* \\
3C 66A        & ISP        & -90.4           & 7.55                 & 0.367  & no   \\
S5 0716+714   & ISP        & -1.93           & 7.68                 & 0.57   & no   \\
ON 231        & ISP        & 118.09          & 16.2                 & 0.52   & yes* \\
A0 0235+164   & LSP        & -7.37           & 19.74                & 0.581  & no   \\
TXS 0506+056  & LSP        & 2.91            & 1.5                  & 0.161  & no   \\
OJ 287        & LSP        & -40.01          & 4.58                 & -0.285 & no   \\
S4 0954+65    & LSP        & -1.29           & 7.53                 & 0.315  & no   \\
4C 09.57      & LSP        & -8.27           & 11.69                & 0.692  & no   \\
BL Lac        & LSP        & -5.42           & 5.33                 & 0.57   & no   \\
PKS 0502+049  & LSP (FSRQ) & 1.92            & 11.47                & 0.576  & no   \\
PKS 0736+01   & LSP (FSRQ) & -0.08           & 6.94                 & 0.506  & no   \\
PKS 1222+216  & LSP (FSRQ) & 187.68          & 24.08                & 0.729  & no   \\
3C 279        & LSP (FSRQ) & -48.85          & 16.08                & 0.348  & no   \\
PKS 1510-089  & LSP (FSRQ) & -55.41          & 9.44                 & 0.659  & yes* \\
OS 319        & LSP (FSRQ) & 79.0            & 17.25                & 0.249  & no   \\
PKS B1622-297 & LSP (FSRQ) & -129.82         & 17.29                & 0.559  & yes* \\
4C +38.41     & LSP (FSRQ) & -5.84           & 15.62                & 0.639  & no   \\
3C 345        & LSP (FSRQ) & 89.44           & 14.72                & -0.254 & no   \\
PKS B1730-130 & LSP (FSRQ) & -0.42           & 15.93                & 0.485  & no   \\
3C 446        & LSP (FSRQ) & -118.37         & 15.82                & 0.262  & no   \\
4C 11.69      & LSP (FSRQ) & 6.23            & 6.09                 & 0.708  & no   \\
3C 454.3      & LSP (FSRQ) & -0.12           & 12.05                & 0.652  & no   \\
\hline
\end{tabular}
}
\end{table*}

\begin{figure*}
\centering
\includegraphics[width=\linewidth]{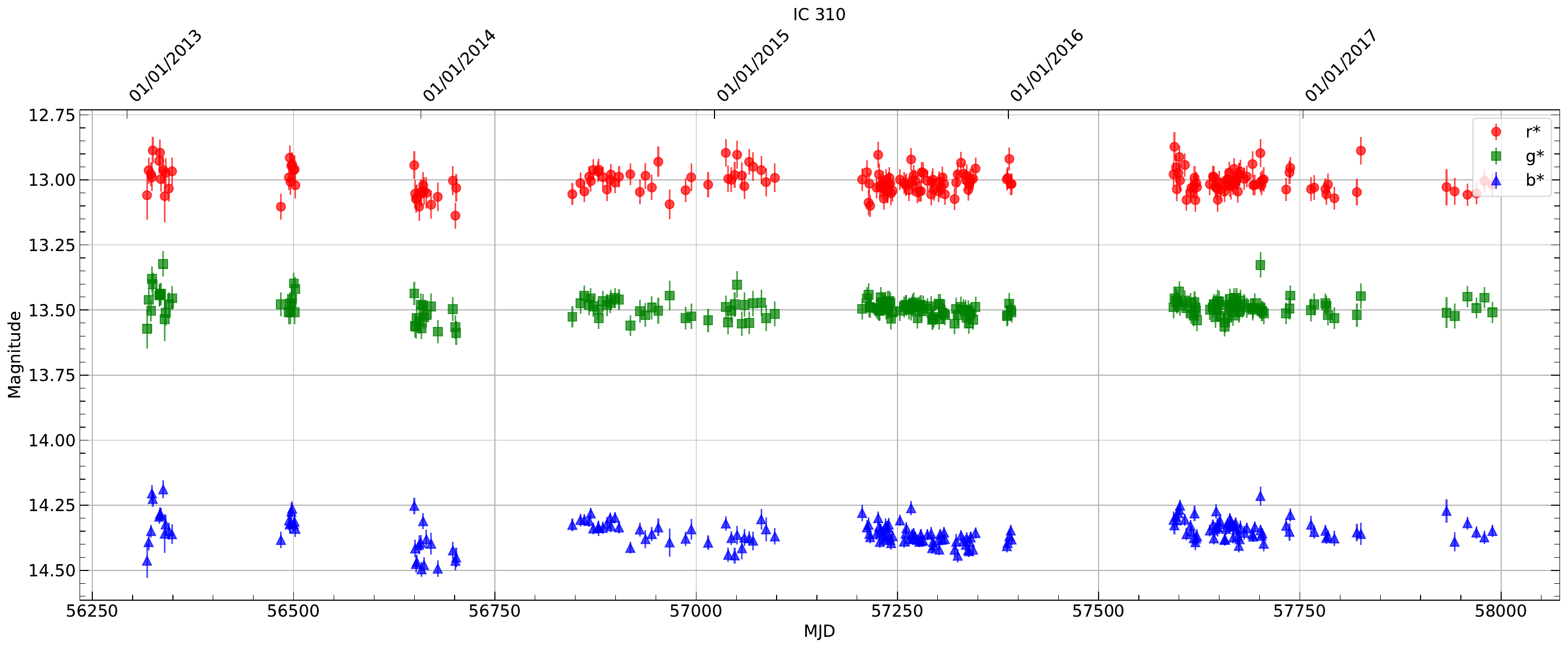}
\caption{Optical light curve for the source IC 310. The \textit{r*}, \textit{g*}, and \textit{b*} data were taken simultaneously with the RINGO3 polarimeter on the Liverpool Telescope.}
\label{fig:IC310_lc}
\end{figure*}

\begin{figure*}
\centering
\includegraphics[width=\linewidth]{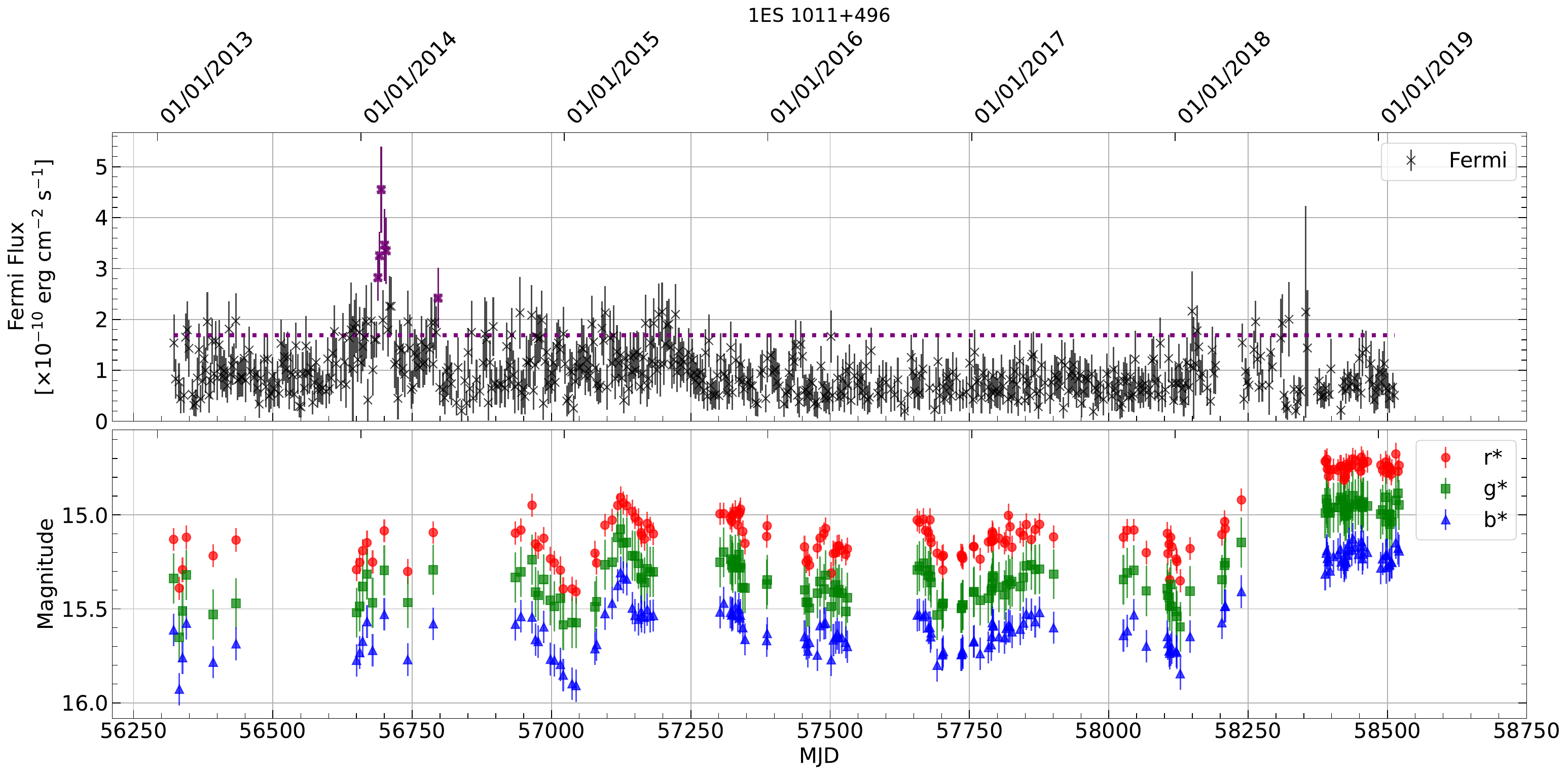}
\caption{Fermi $\gamma$-ray and optical  light curves for the source 1ES 1011+496. The top panel shows the $\gamma$-ray flux light curve, and the second shows the optical magnitude light curve. The optical light curve consists of simultaneous \textit{r*}, \textit{g*}, and \textit{b*} data observed with the RINGO3 polarimeter on the Liverpool Telescope. The Fermi light curve shows a dotted line indicating the threshold above which activity has been deemed part of a flaring state (median flux level over all time plus 3 times the median absolute deviation).}
\label{fig:1ES1011+496_lc}
\end{figure*}

\begin{figure*}
\centering
\includegraphics[width=\linewidth]{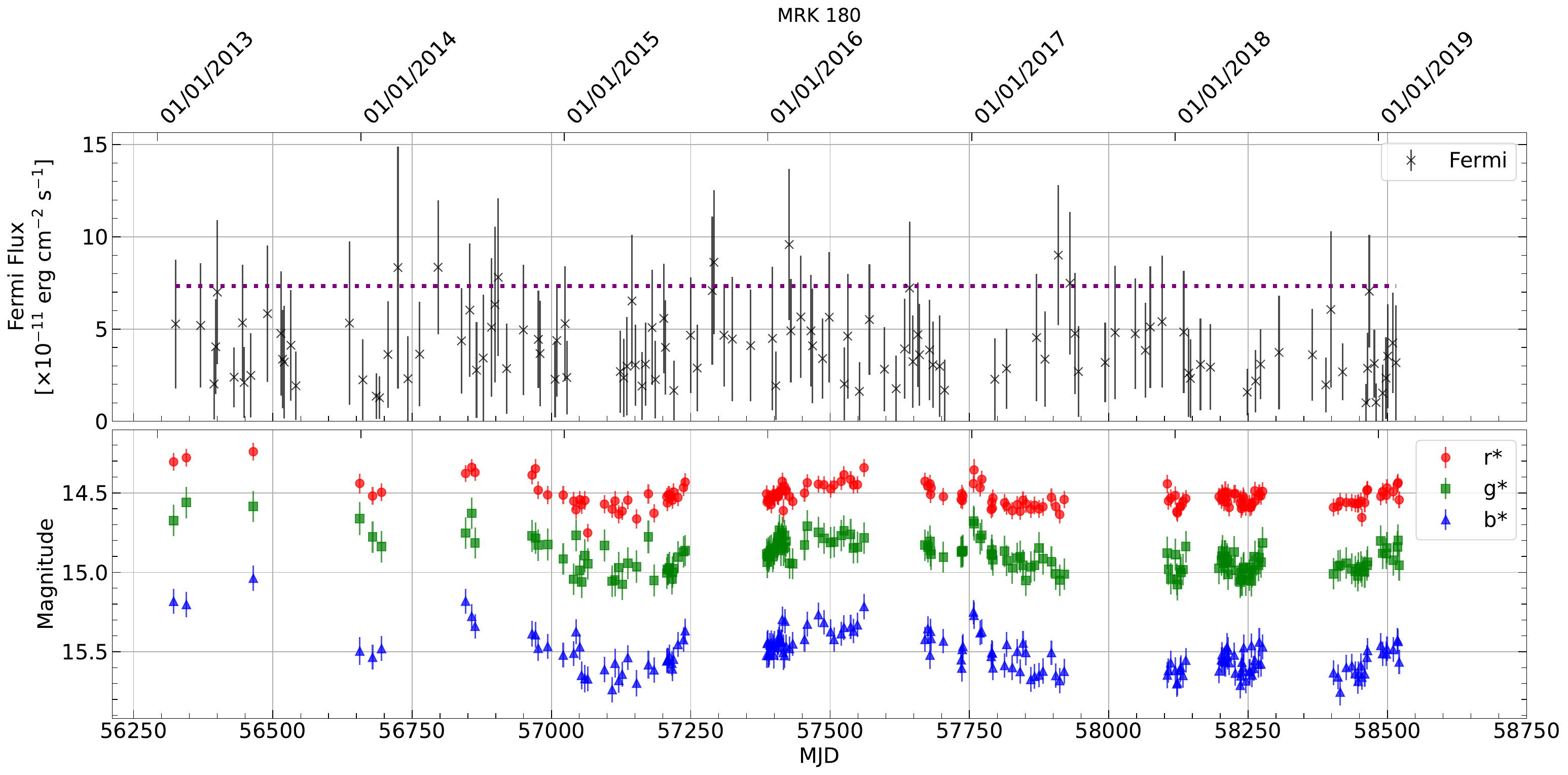}
\caption{As Fig. \ref{fig:1ES1011+496_lc} but for MRK 180.}
\label{fig:MRK180_lc}
\end{figure*}

\begin{figure*}
\centering
\includegraphics[width=\linewidth]{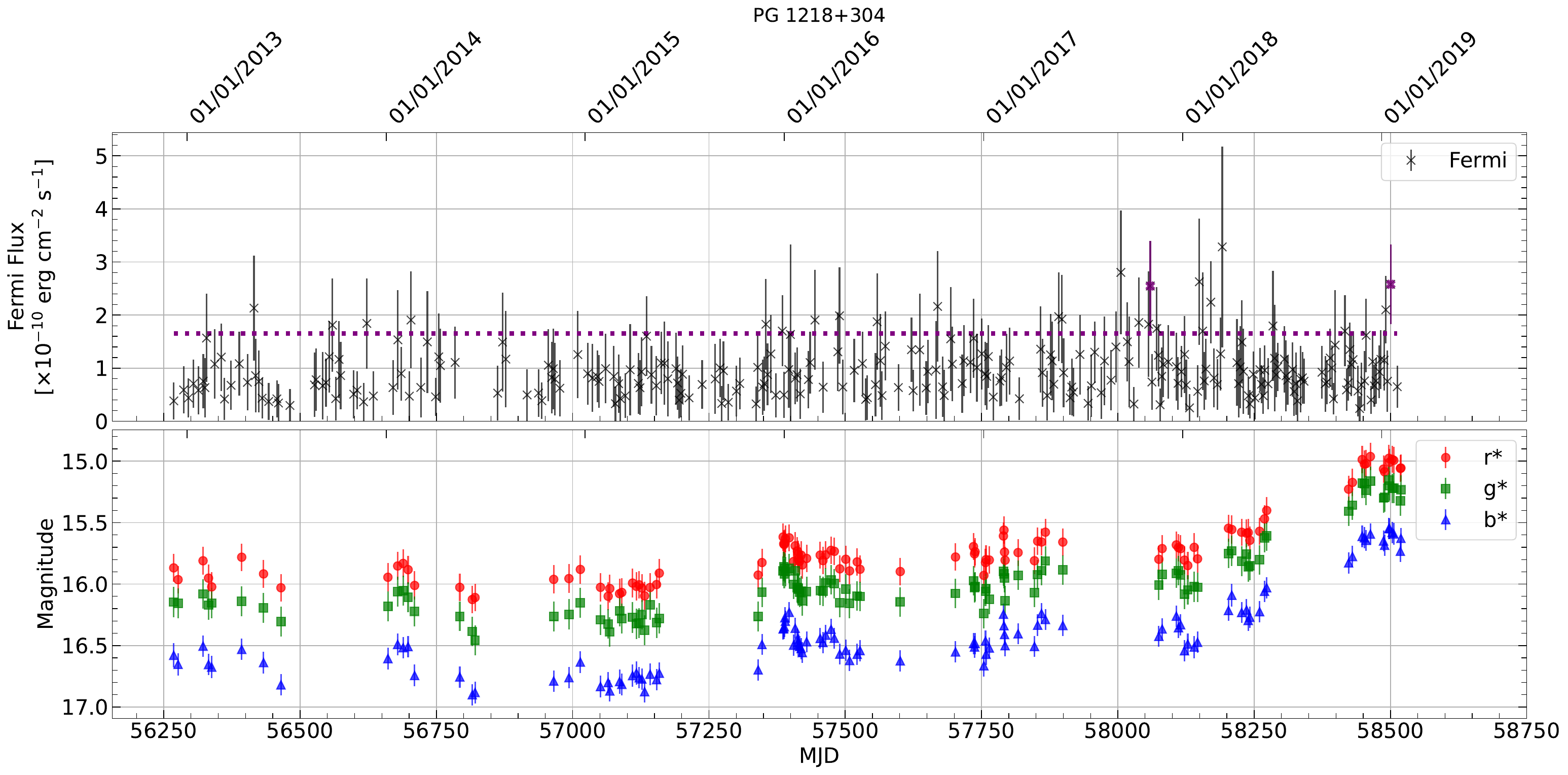}
\caption{As Fig. \ref{fig:1ES1011+496_lc} but for PG 1218+304.}
\label{fig:PG1218+304_lc}
\end{figure*}

\begin{figure*}
\centering
\includegraphics[width=\linewidth]{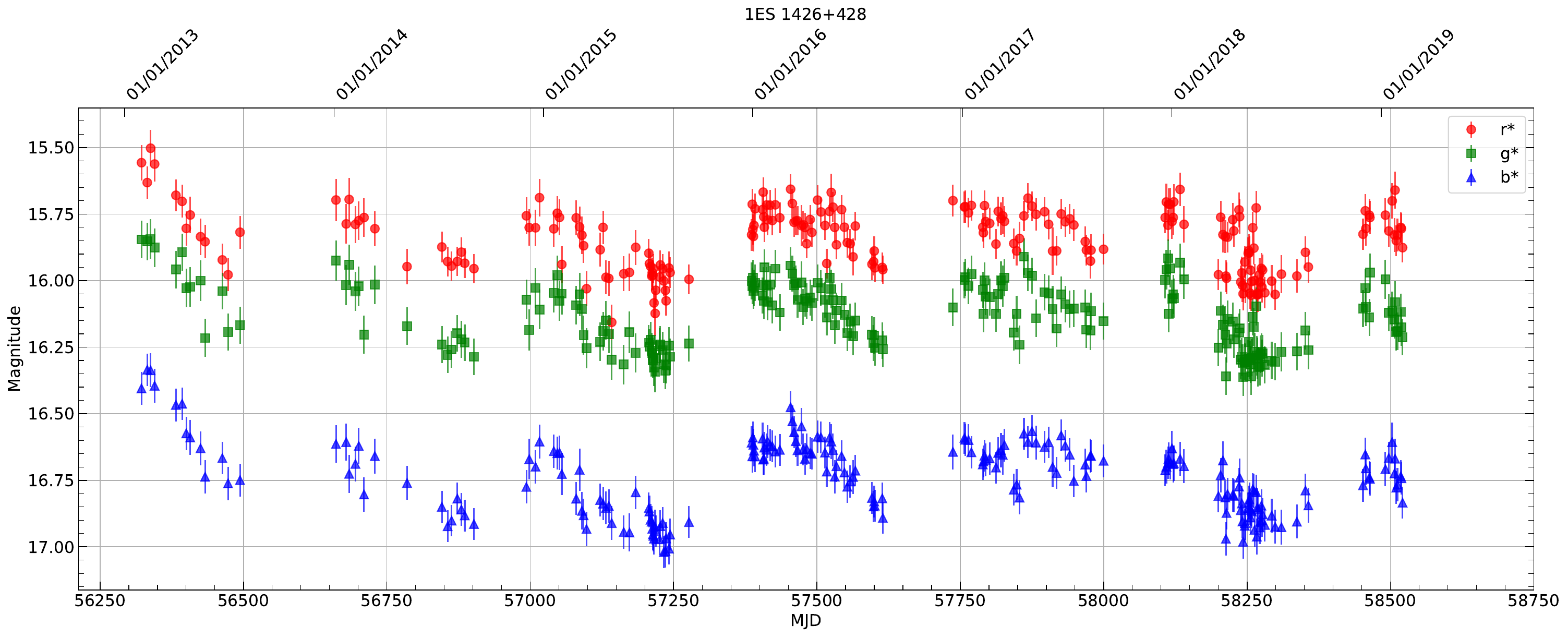}
\caption{As Fig. \ref{fig:IC310_lc} but for 1ES 1426+428.}
\label{fig:1ES1426+428_lc}
\end{figure*}

\begin{figure*}
\centering
\includegraphics[width=\linewidth]{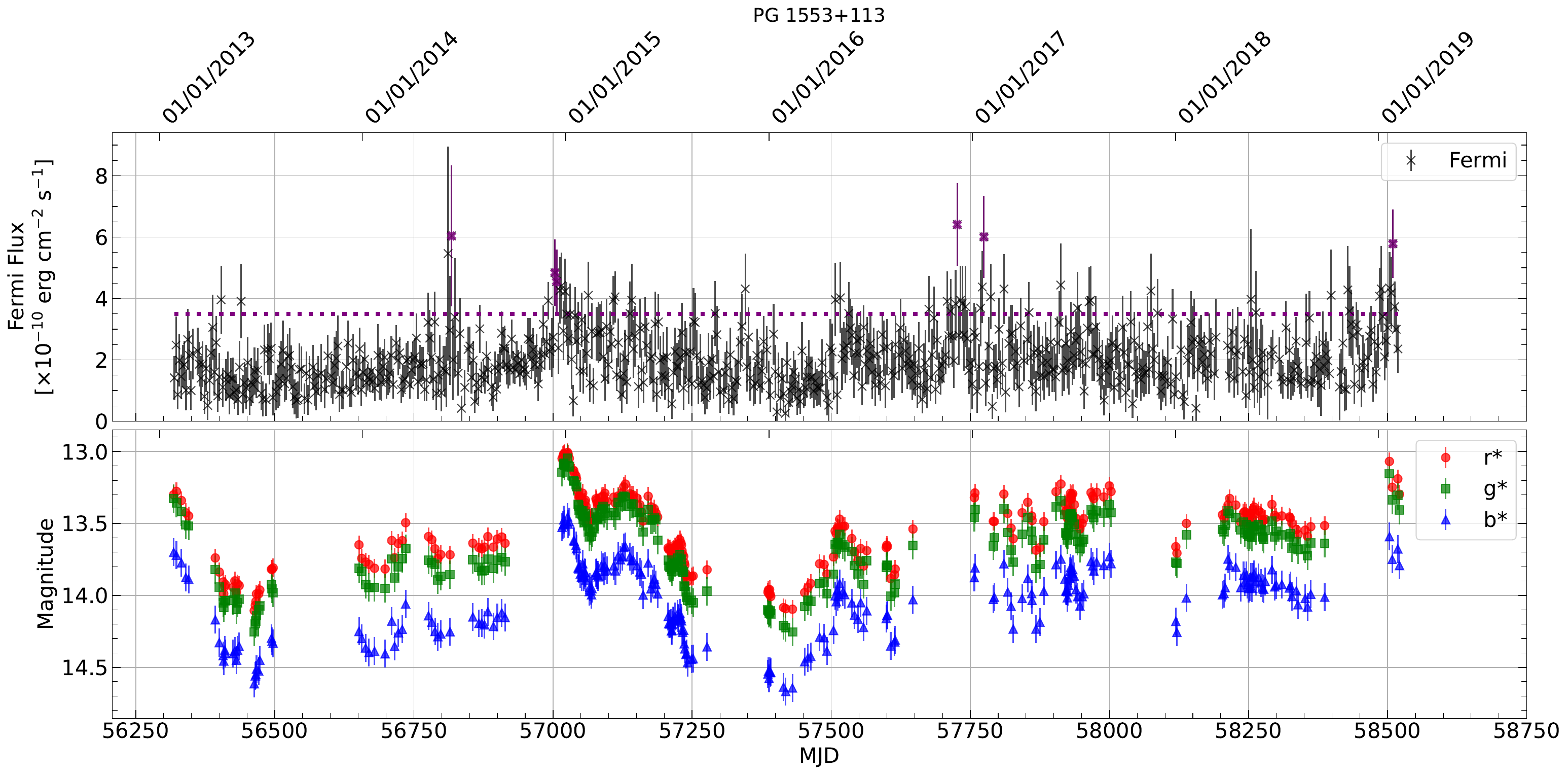}
\caption{As Fig. \ref{fig:1ES1011+496_lc} but for PG1553+113.}
\label{fig:PG1553+113_lc}
\end{figure*}

\begin{figure*}
\centering
\includegraphics[width=\linewidth]{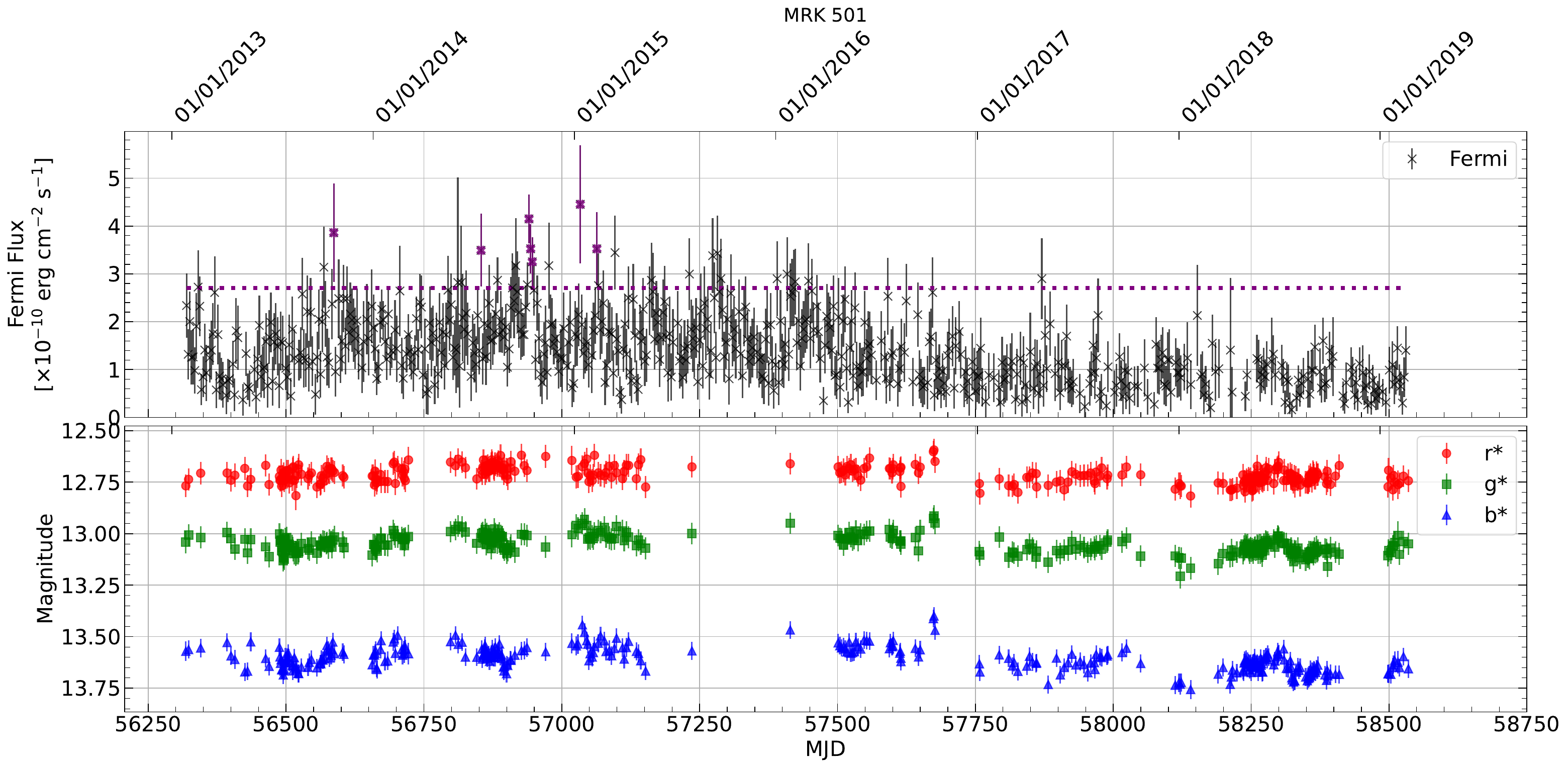}
\caption{As Fig. \ref{fig:1ES1011+496_lc} but for MRK 501.}
\label{fig:MRK501_lc}
\end{figure*}

\begin{figure*}
\centering
\includegraphics[width=\linewidth]{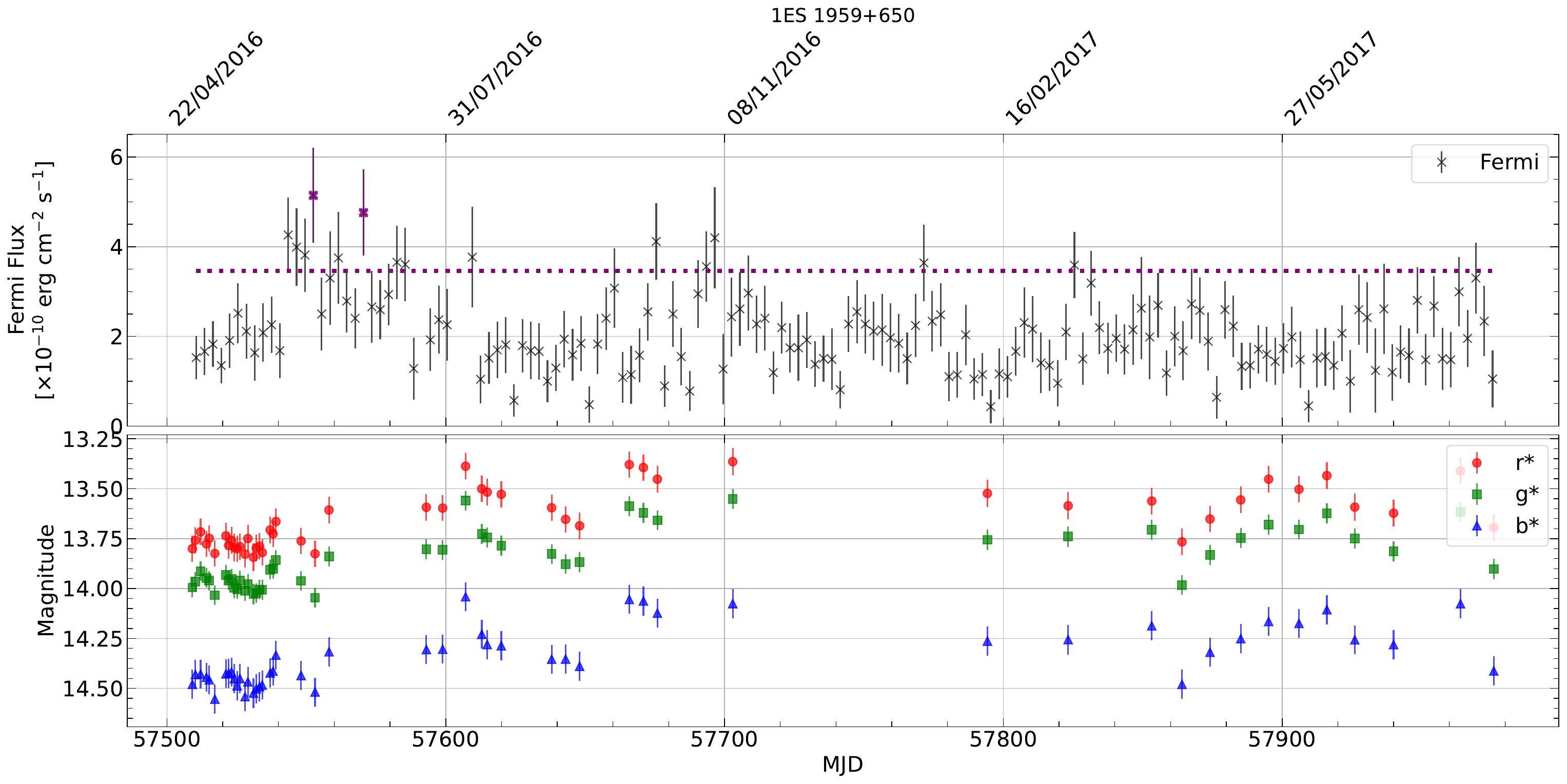}
\caption{As Fig. \ref{fig:1ES1011+496_lc} but for 1ES 1959+650.}
\label{fig:1ES1959+650_lc}
\end{figure*}

\begin{figure*}
\centering
\includegraphics[width=\linewidth]{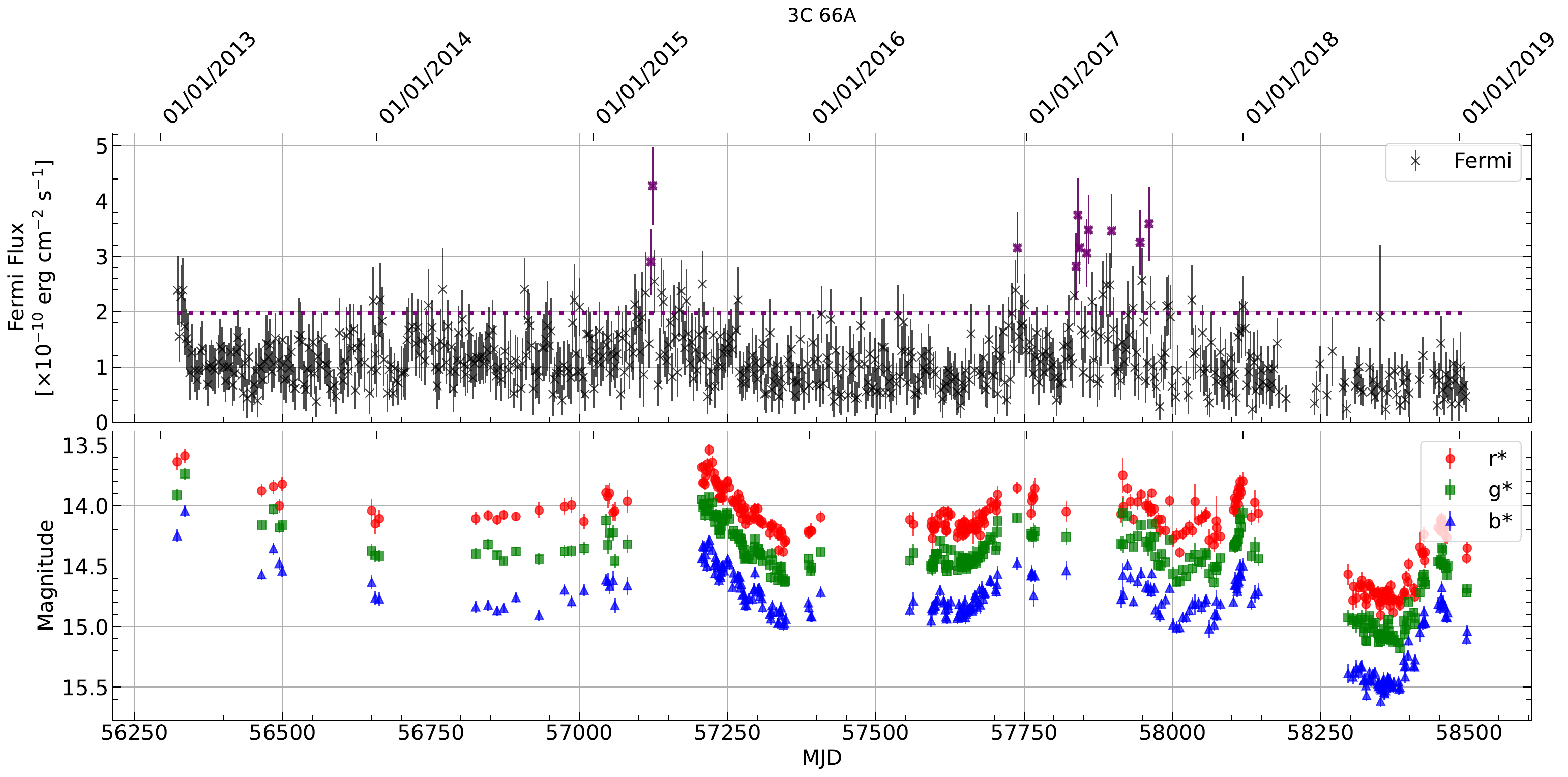}
\caption{As Fig. \ref{fig:1ES1011+496_lc} but for 3C 66A.}
\label{fig:3C66A_lc}
\end{figure*}

\begin{figure*}
\centering
\includegraphics[width=\linewidth]{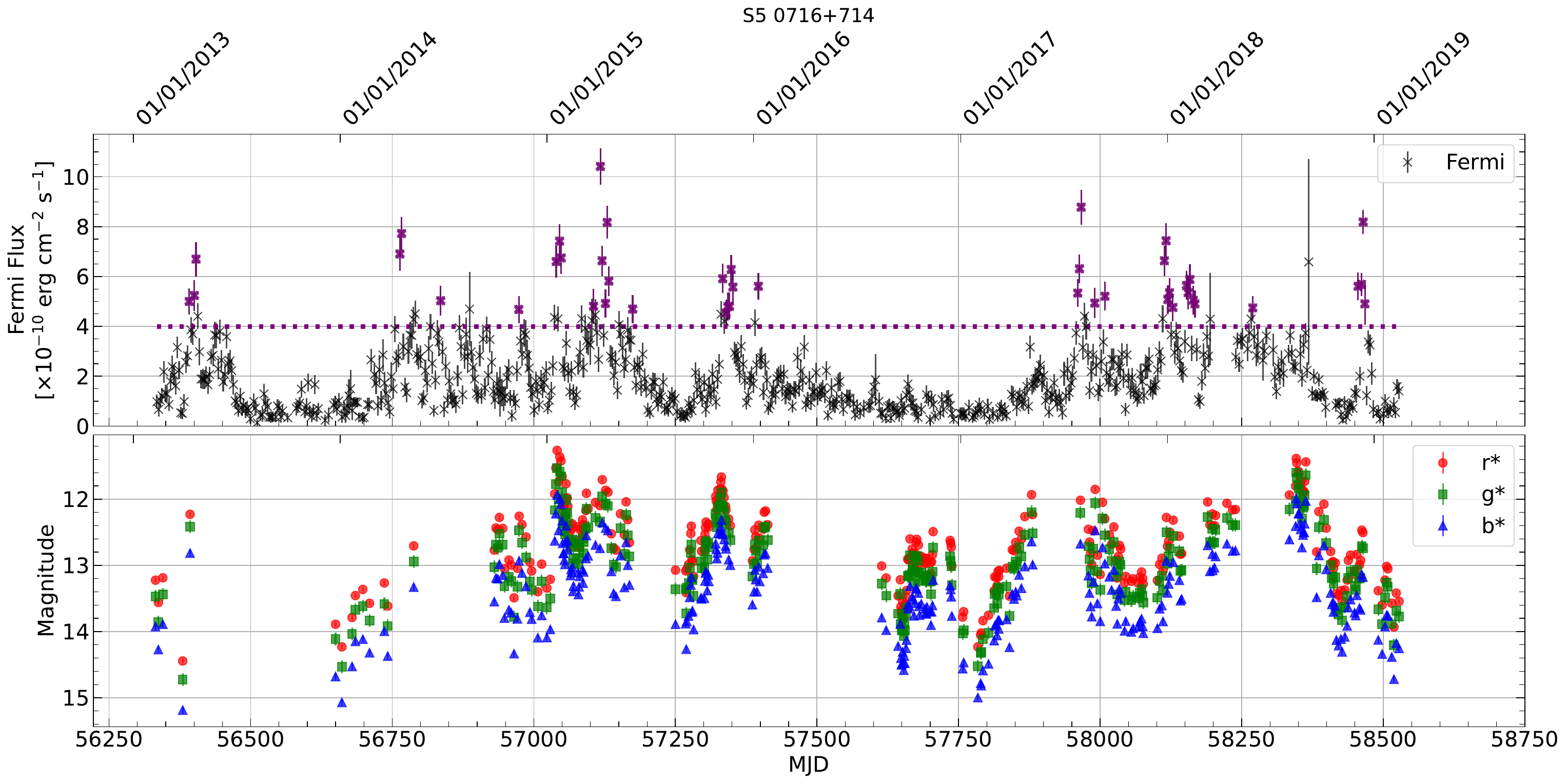}
\caption{As Fig. \ref{fig:1ES1011+496_lc} but for S5 0716+714.}
\label{fig:S50716+714_lc}
\end{figure*}

\begin{figure*}
\centering
\includegraphics[width=\linewidth]{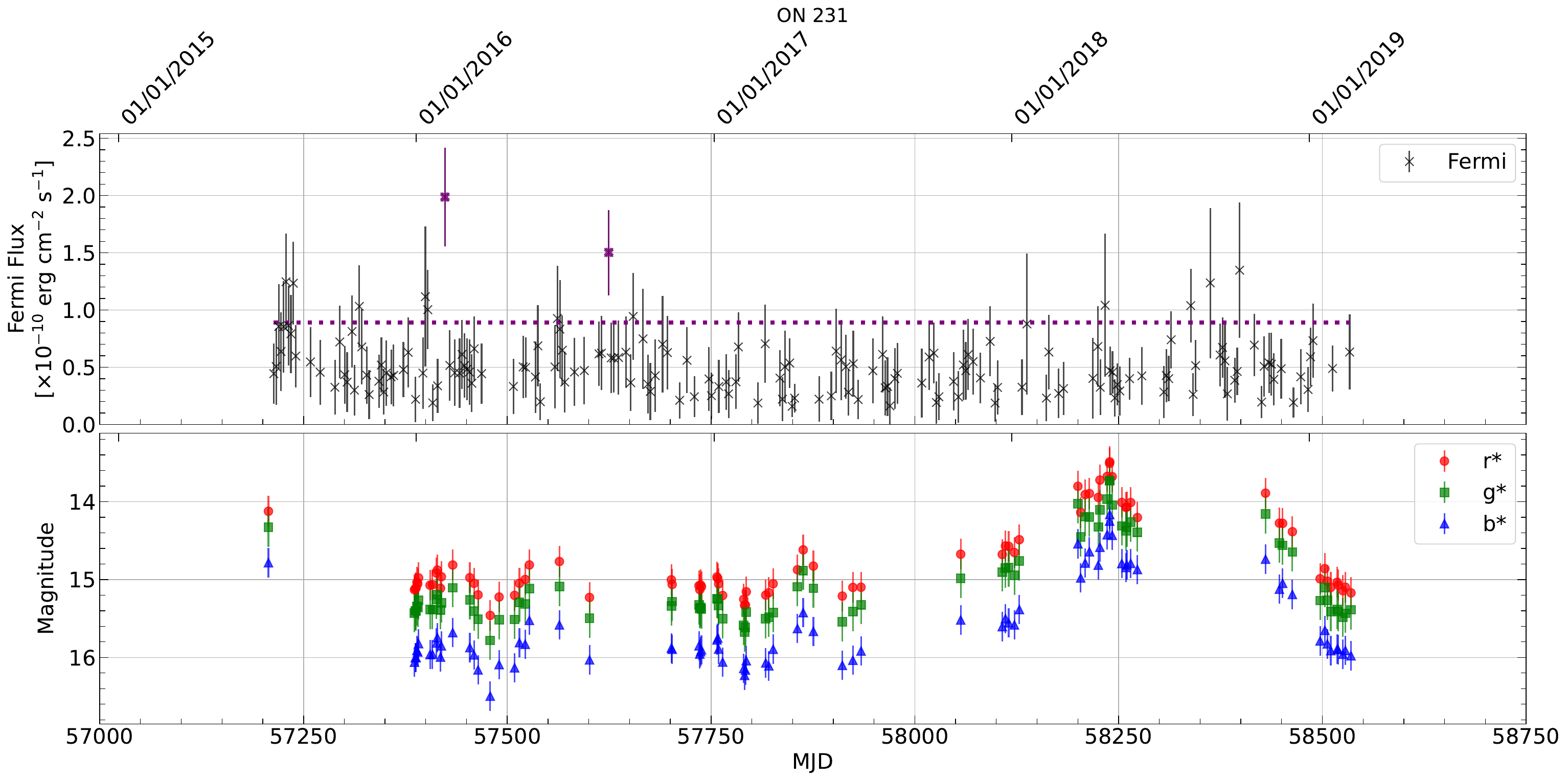}
\caption{As Fig. \ref{fig:1ES1011+496_lc} but for ON 231.}
\label{fig:ON231_lc}
\end{figure*}

\begin{figure*}
\centering
\includegraphics[width=\linewidth]{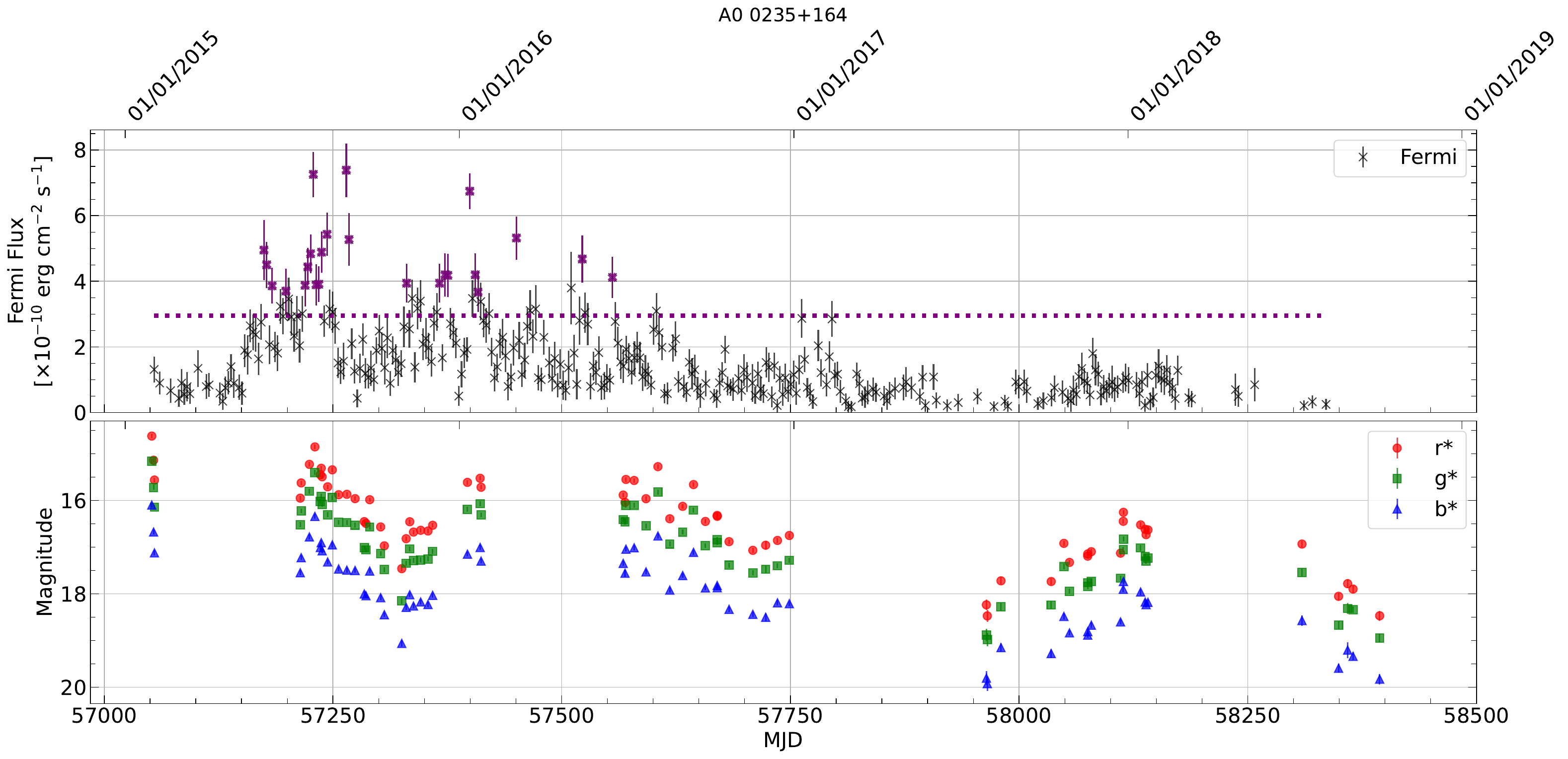}
\caption{As Fig. \ref{fig:1ES1011+496_lc} but for A0 0235+164.}
\label{fig:A0 0235+164_lc}
\end{figure*}

\begin{figure*}
\centering
\includegraphics[width=\linewidth]{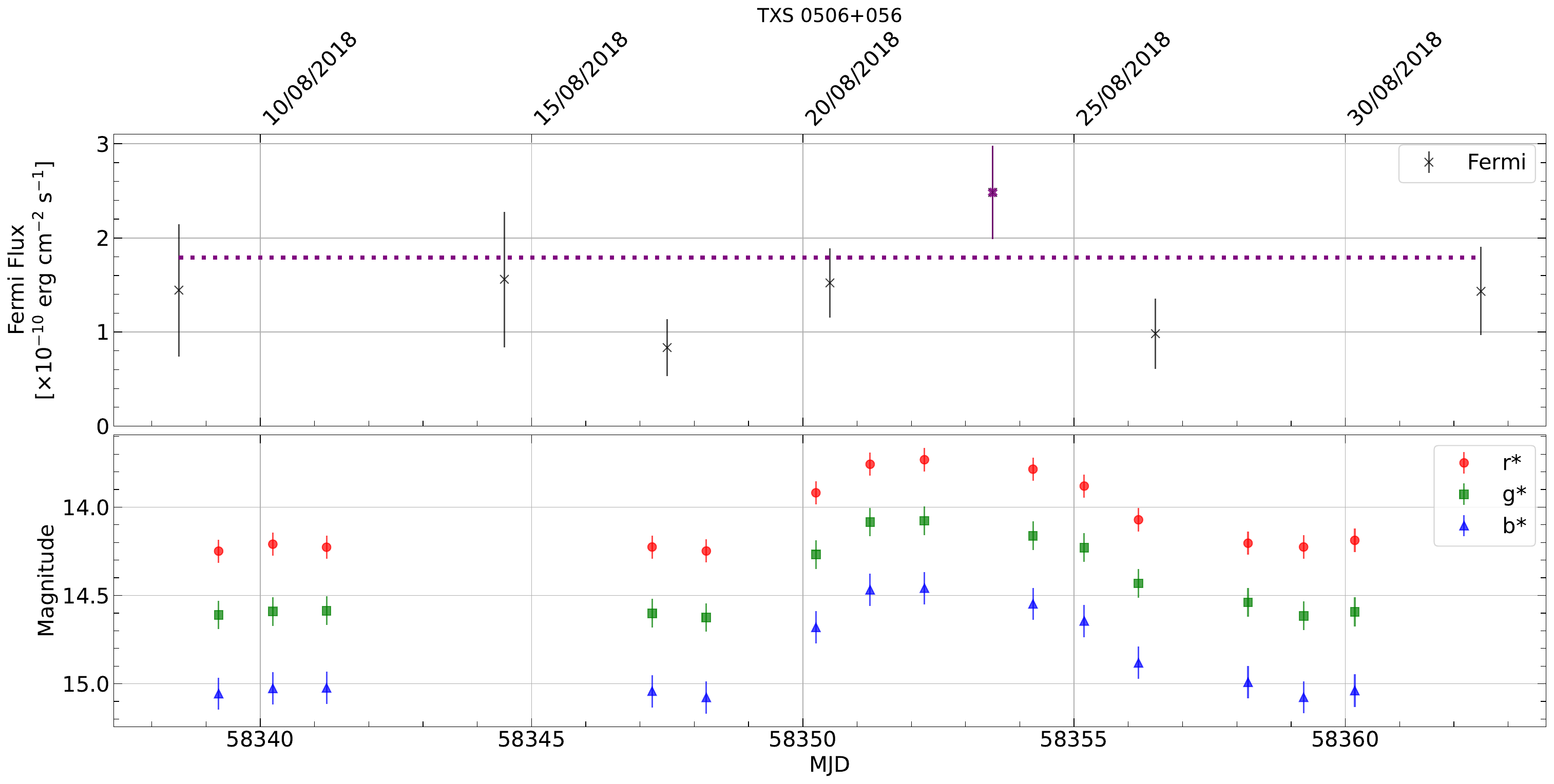}
\caption{As Fig. \ref{fig:1ES1011+496_lc} but for TXS 0506+056.}
\label{fig:TXS0506+056_lc}
\end{figure*}

\begin{figure*}
\centering
\includegraphics[width=\linewidth]{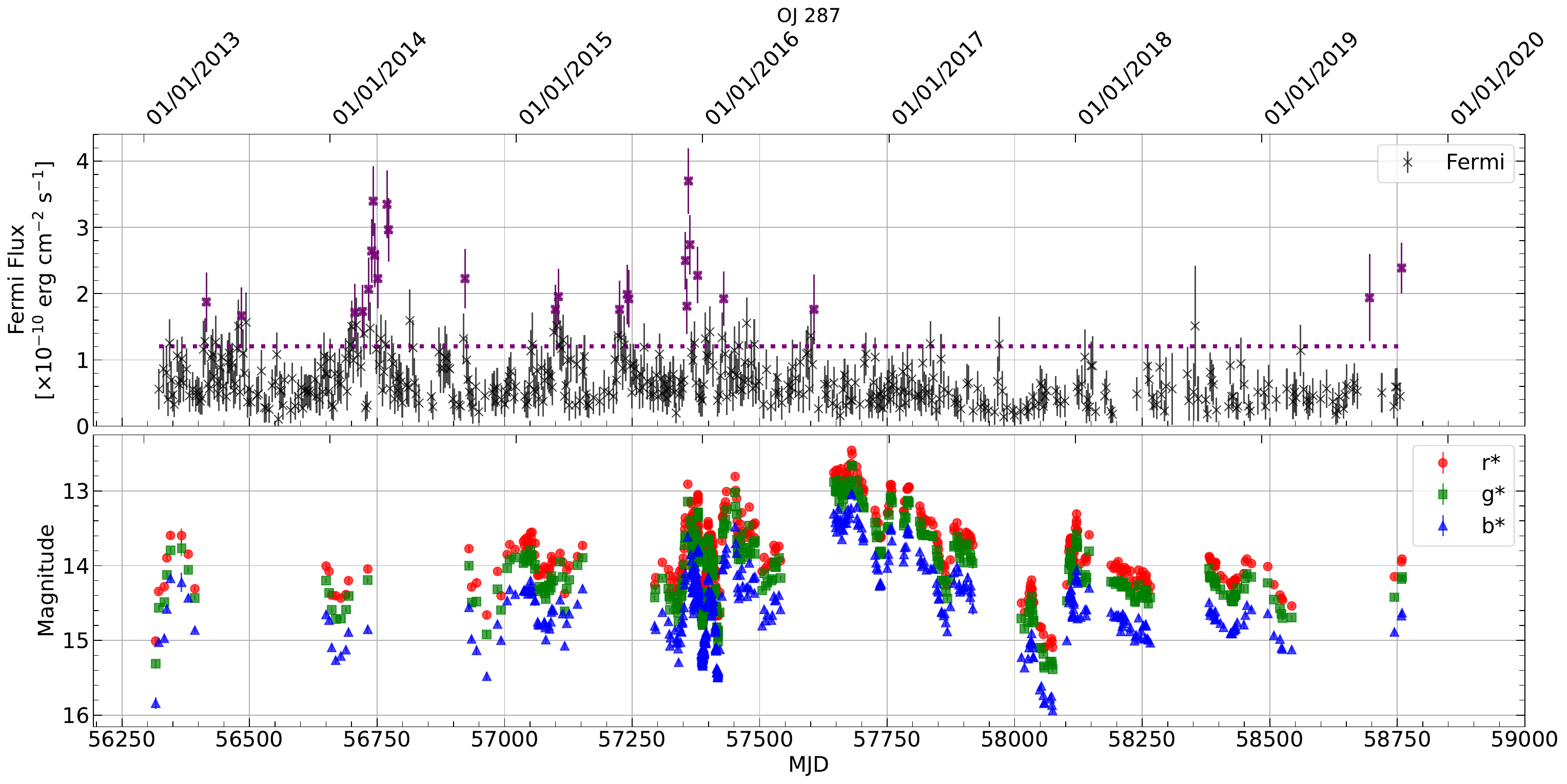}
\caption{As Fig. \ref{fig:1ES1011+496_lc} but for OJ 287.}
\label{fig:OJ287_lc}
\end{figure*}

\begin{figure*}
\centering
\includegraphics[width=\linewidth]{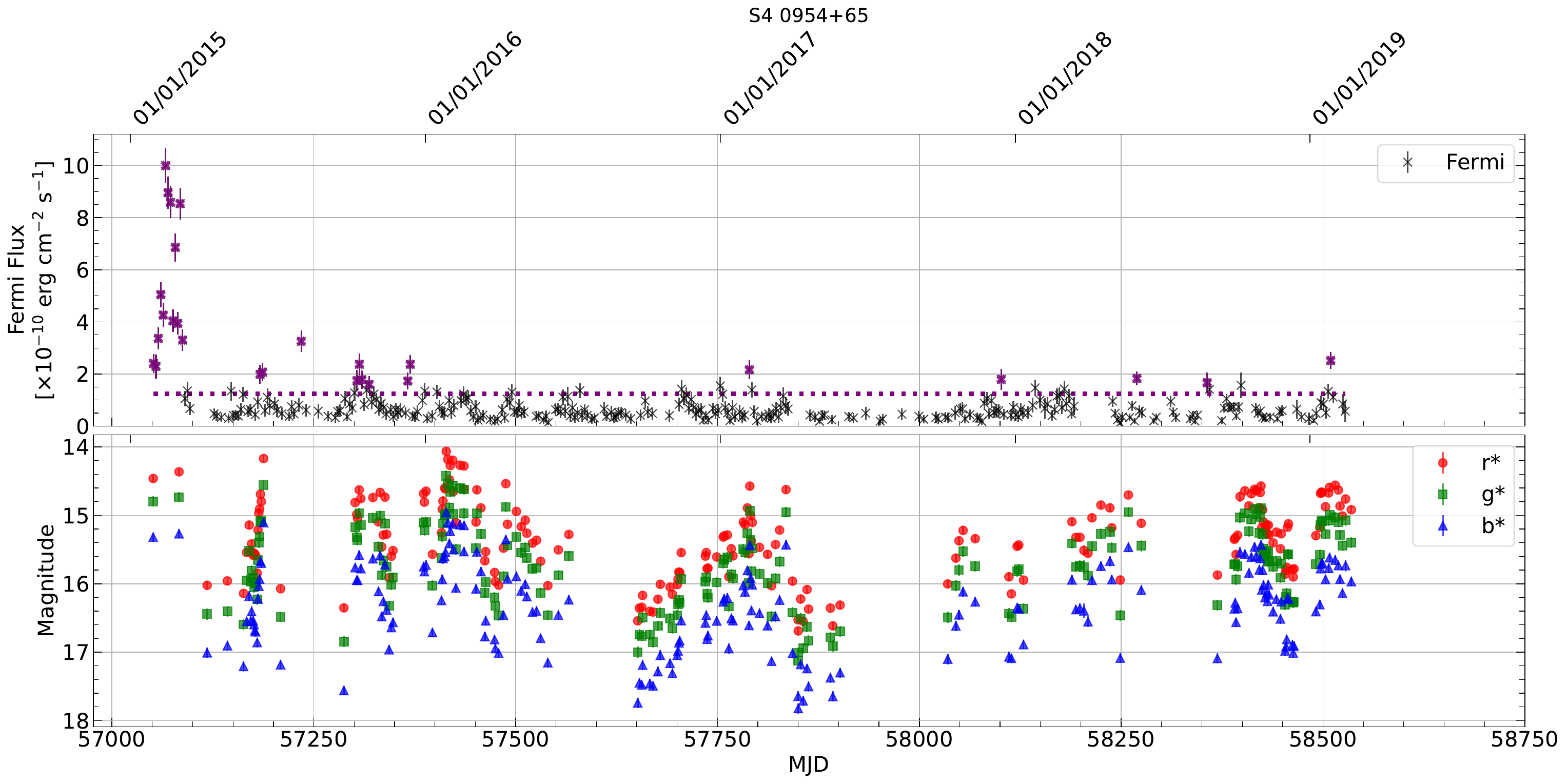}
\caption{As Fig. \ref{fig:1ES1011+496_lc} but for S4 0954+65.}
\label{fig:S40954+65_lc}
\end{figure*}

\begin{figure*}
\centering
\includegraphics[width=\linewidth]{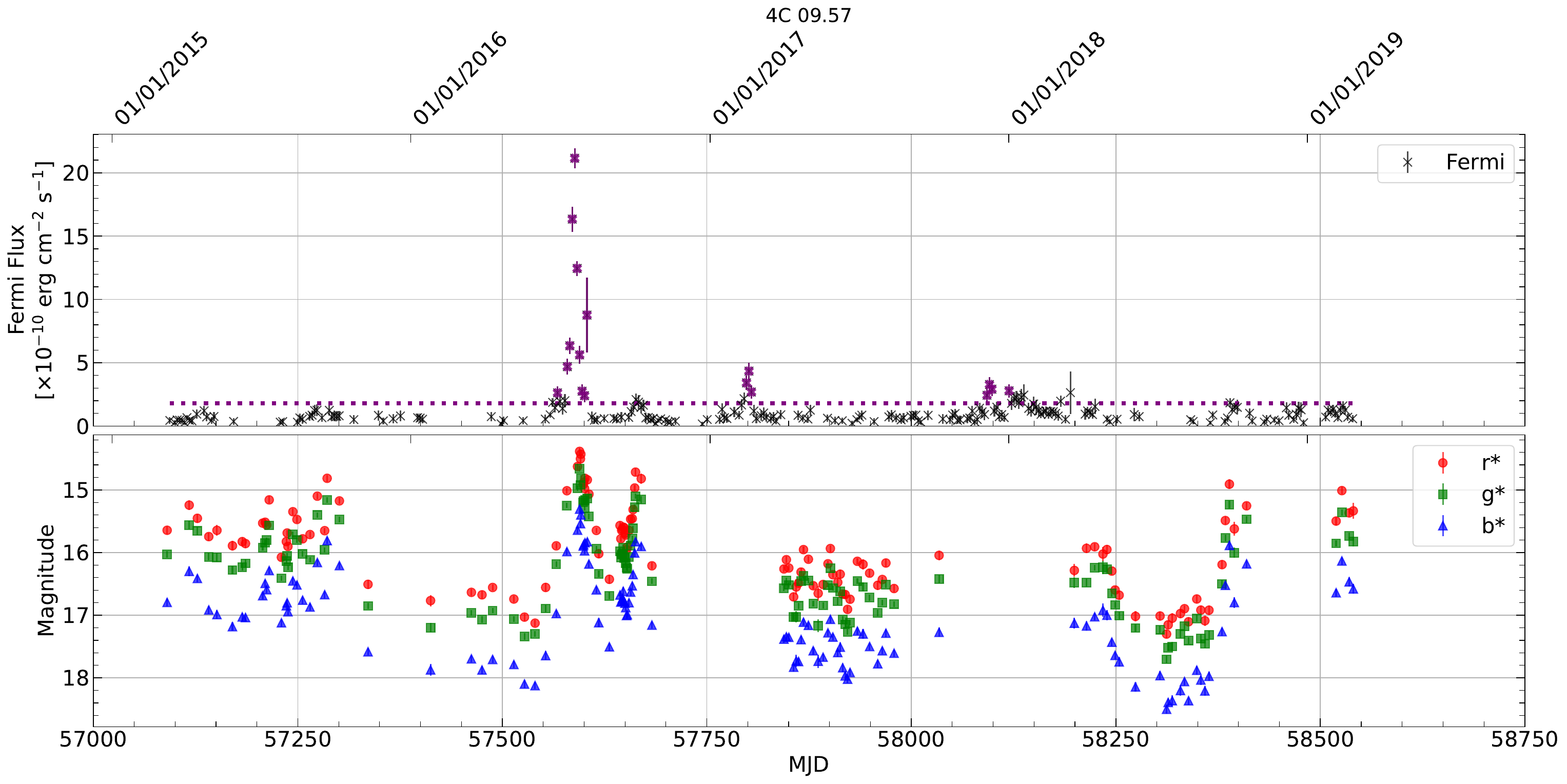}
\caption{As Fig. \ref{fig:1ES1011+496_lc} but for 4C 09.57.}
\label{fig:4C09.57_lc}
\end{figure*}

\begin{figure*}
\centering
\includegraphics[width=\linewidth]{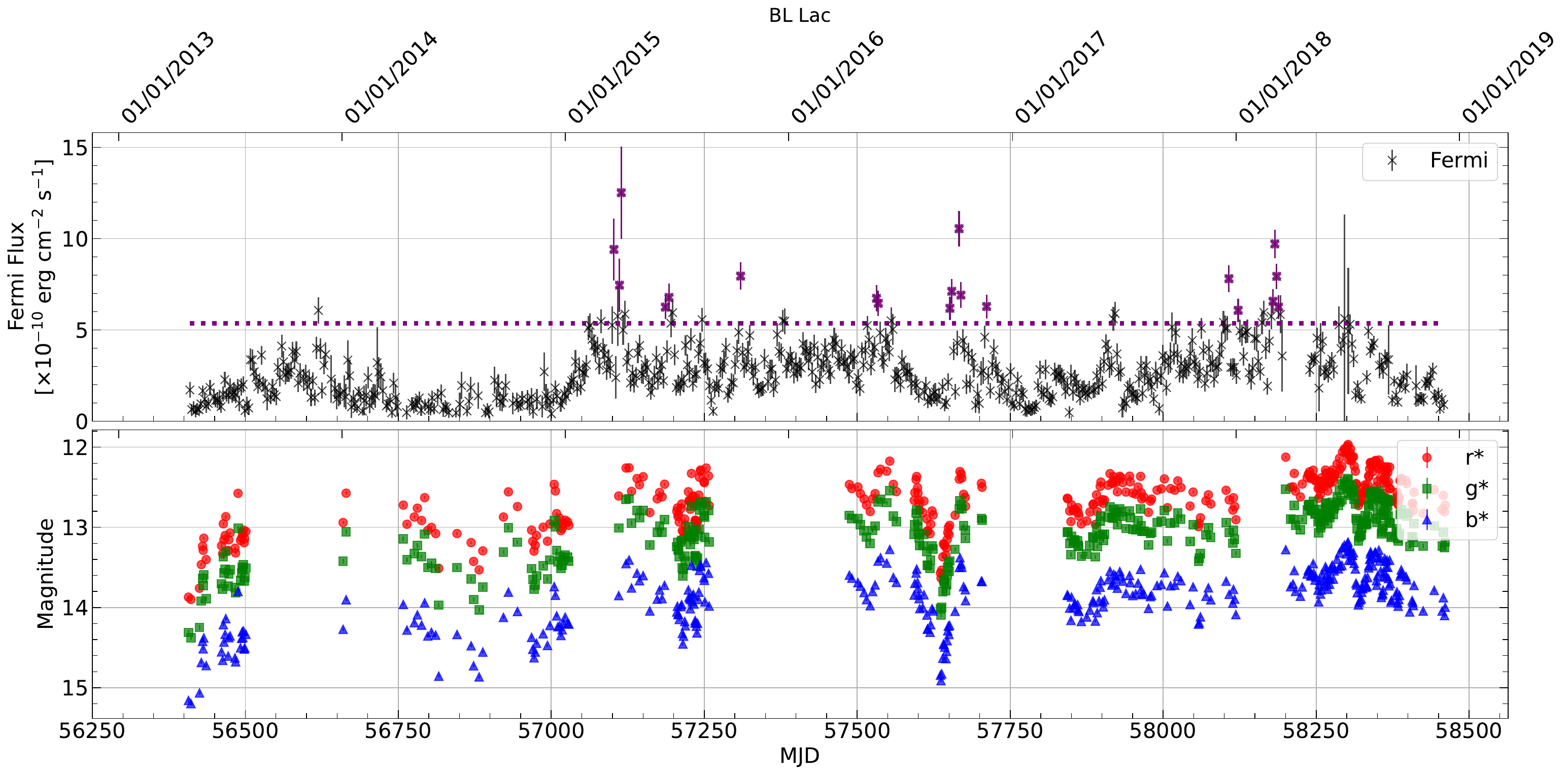}
\caption{As Fig. \ref{fig:1ES1011+496_lc} but for BL Lac.}
\label{fig:BLLac_lc}
\end{figure*}

\begin{figure*}
\centering
\includegraphics[width=\linewidth]{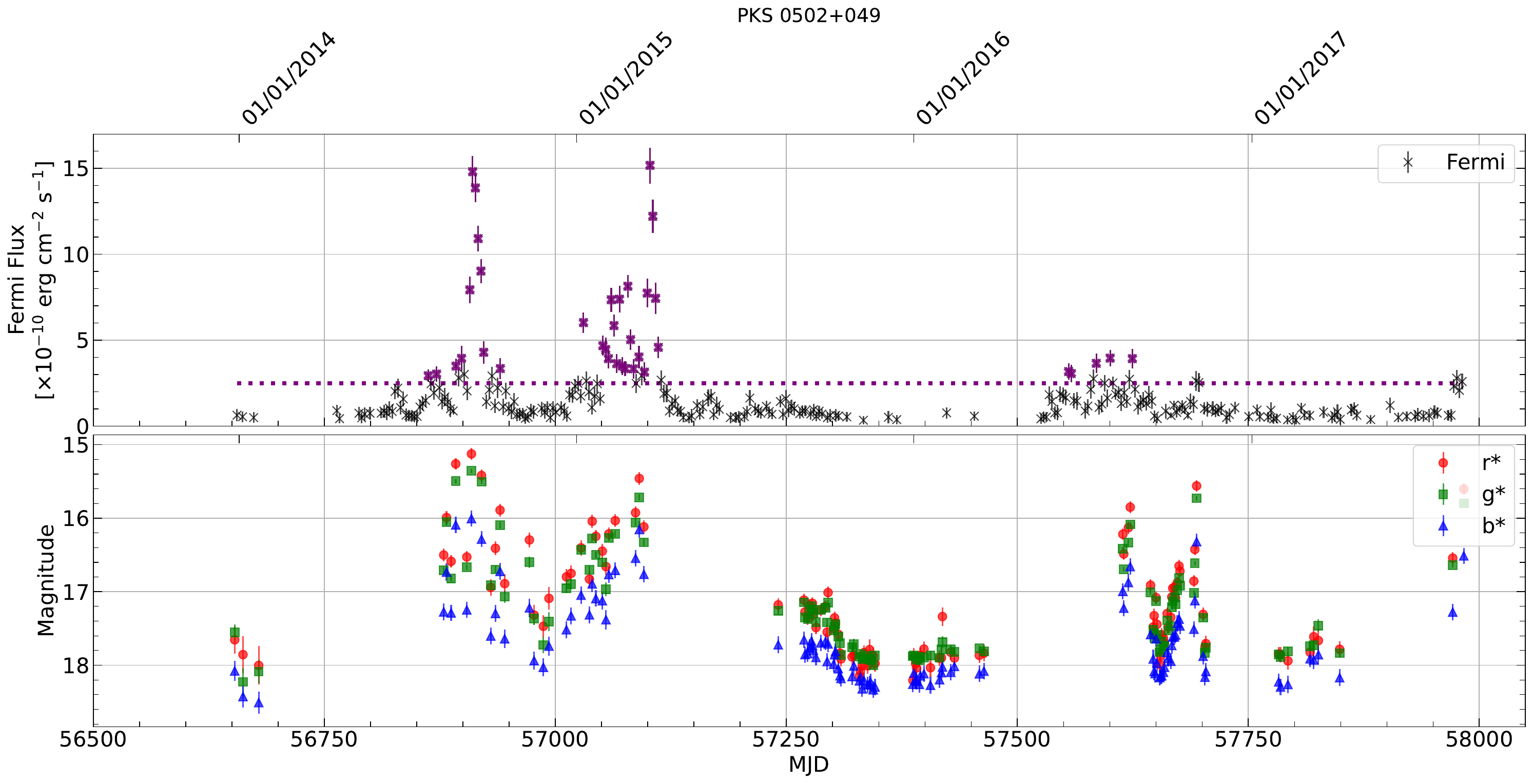}
\caption{As Fig. \ref{fig:1ES1011+496_lc} but for PKS0502+049.}
\label{fig:PKS0502+049_lc}
\end{figure*}

\begin{figure*}
\centering
\includegraphics[width=\linewidth]{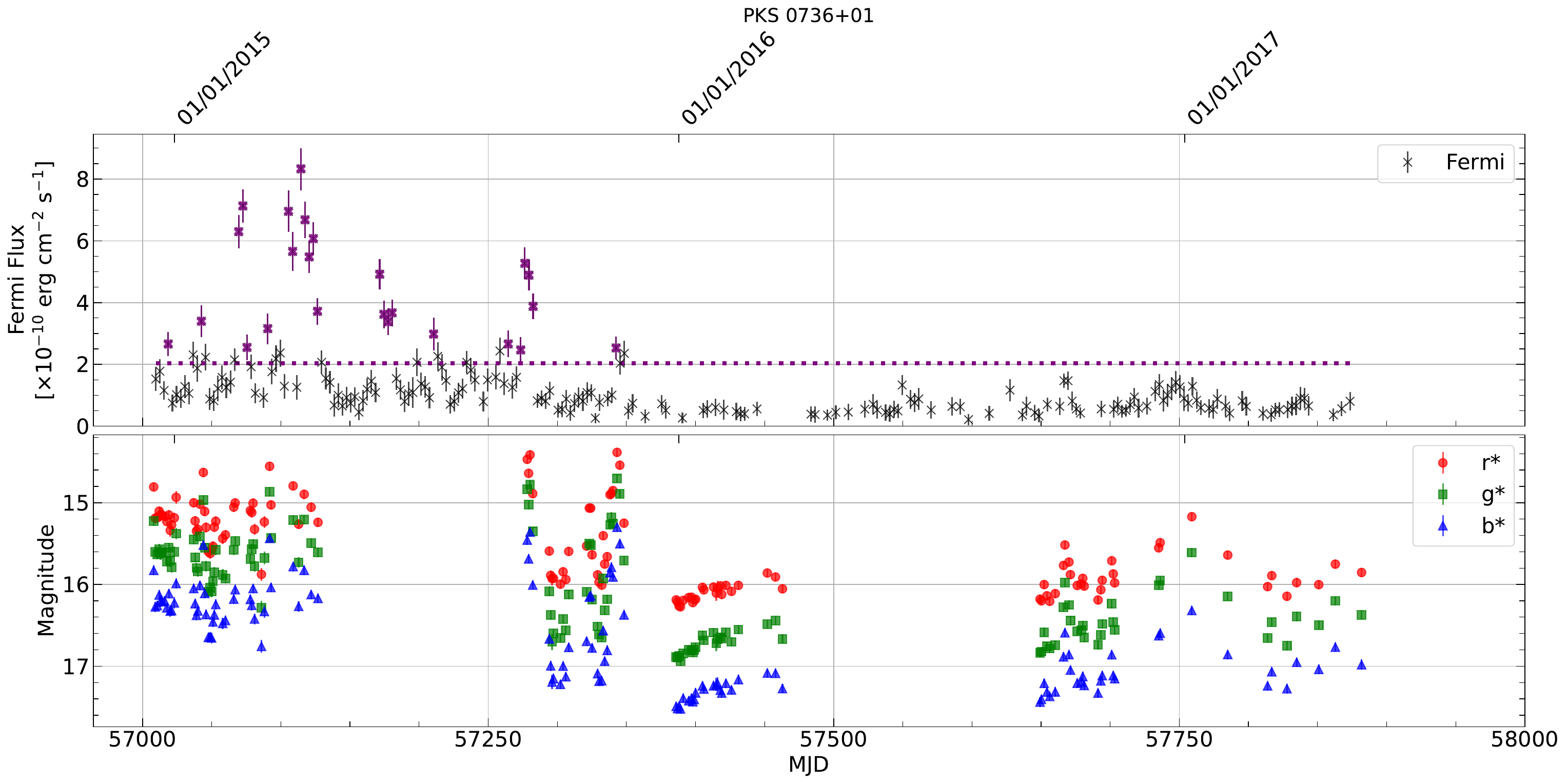}
\caption{As Fig. \ref{fig:1ES1011+496_lc} but for PKS 0736+01.}
\label{fig:PKS0736+01_lc}
\end{figure*}

\begin{figure*}
\centering
\includegraphics[width=\linewidth]{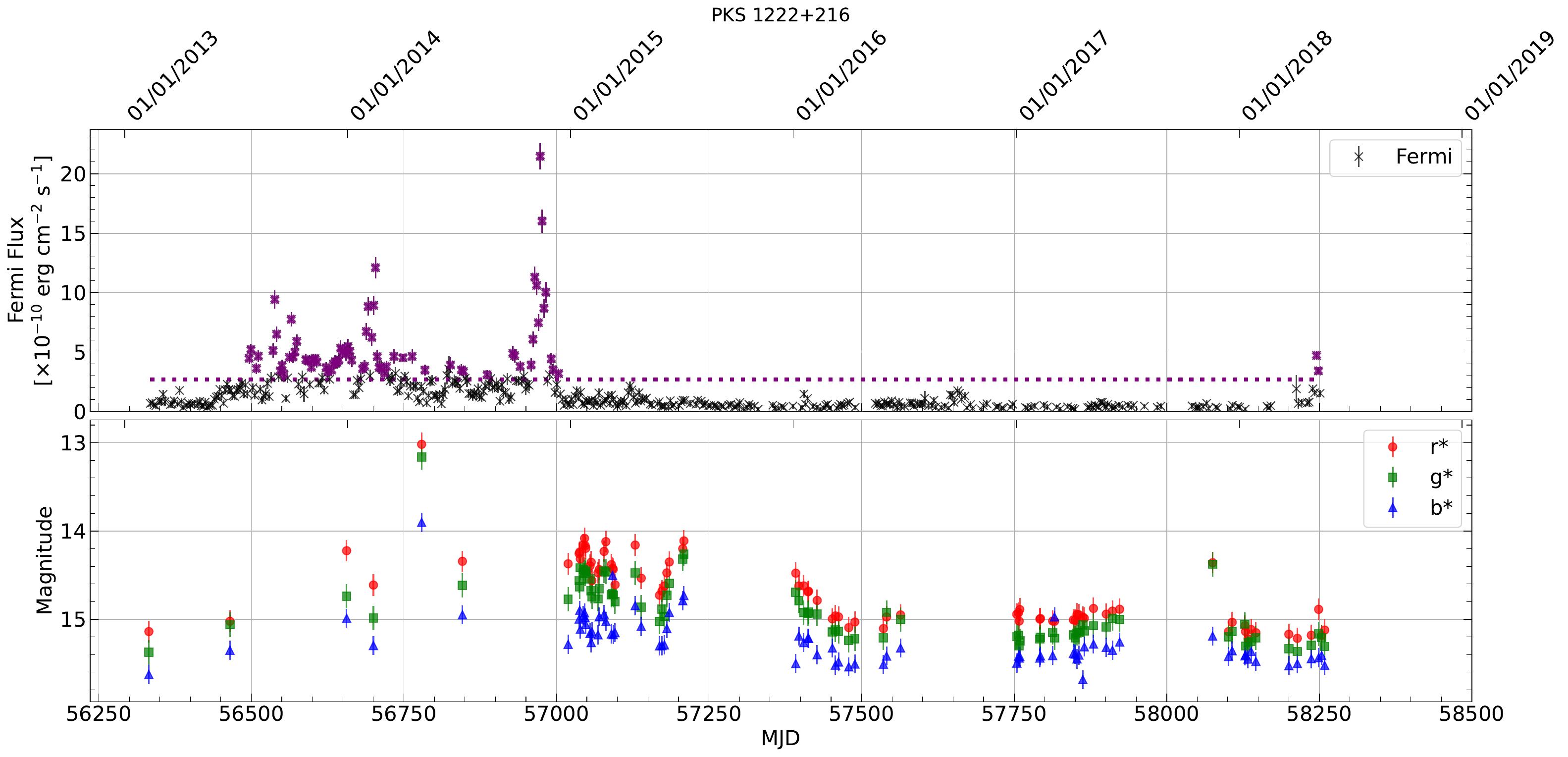}
\caption{As Fig. \ref{fig:1ES1011+496_lc} but for PKS 1222+216.}
\label{fig:PKS1222+216_lc}
\end{figure*}

\begin{figure*}
\centering
\includegraphics[width=\linewidth]{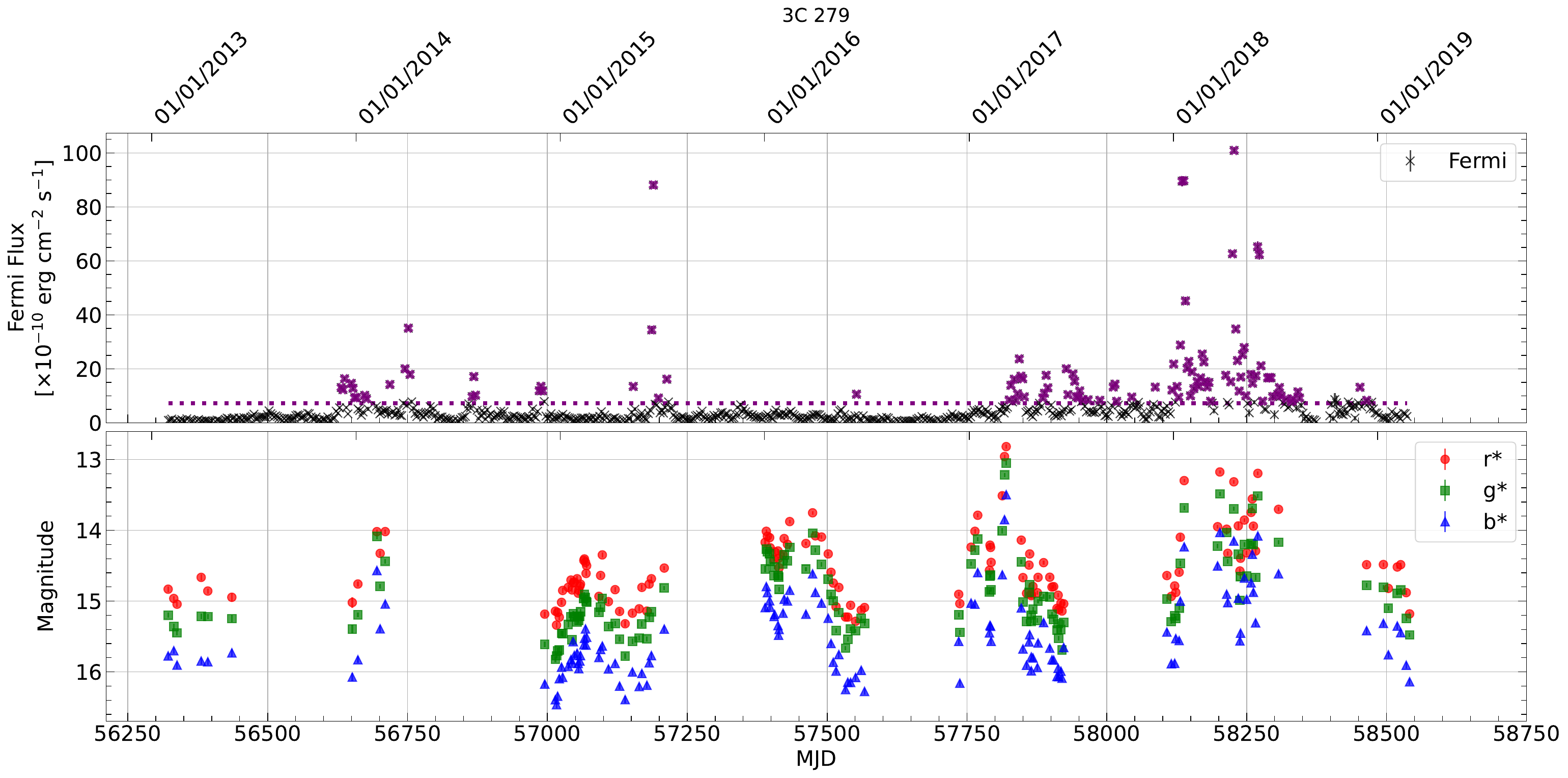}
\caption{As Fig. \ref{fig:1ES1011+496_lc} but for 3C 279.}
\label{fig:3C279_lc}
\end{figure*}

\begin{figure*}
\centering
\includegraphics[width=\linewidth]{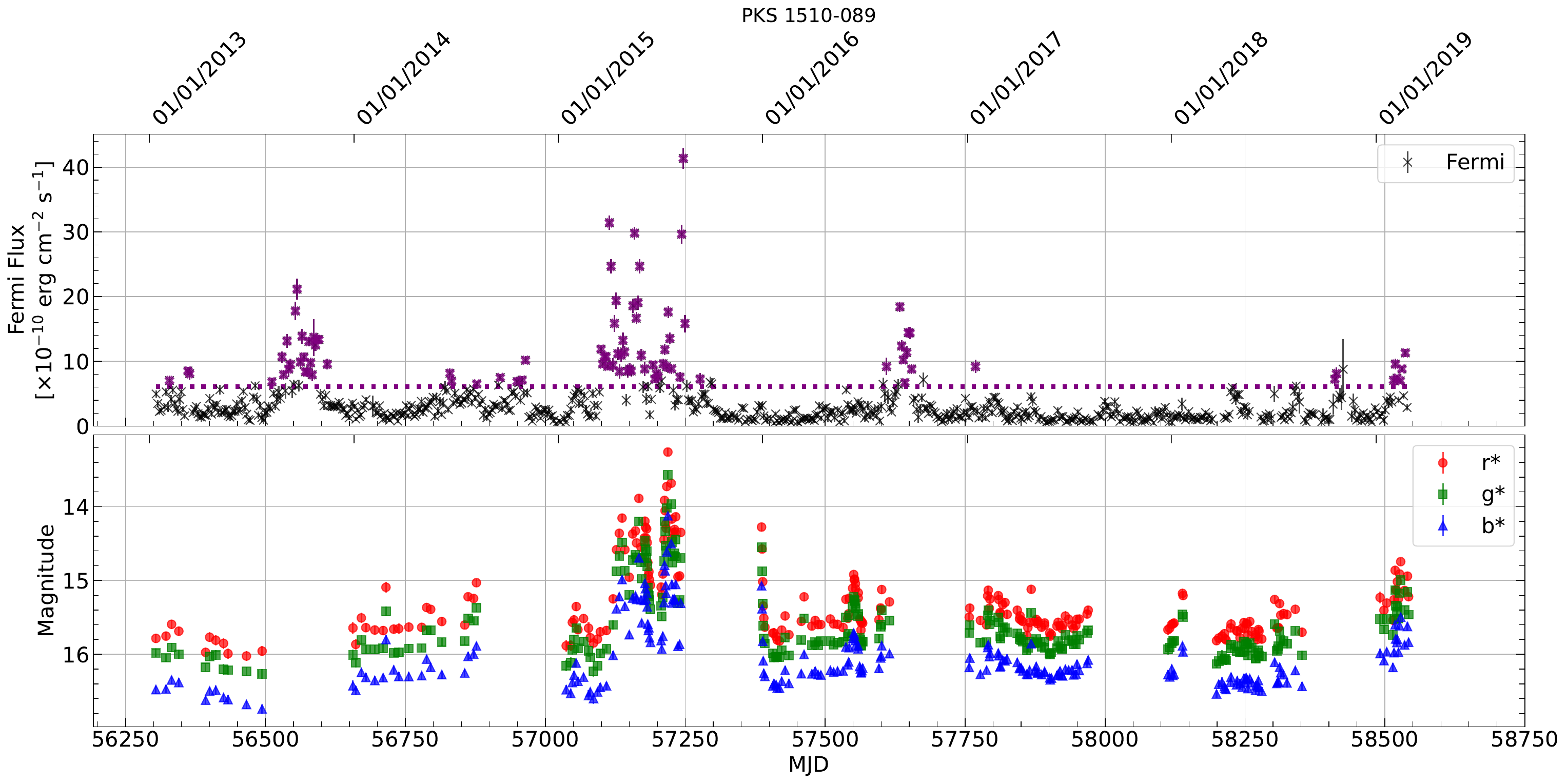}
\caption{As Fig. \ref{fig:1ES1011+496_lc} but for PKS 1510-089.}
\label{fig:PKS1510-089_lc}
\end{figure*}

\begin{figure*}
\centering
\includegraphics[width=\linewidth]{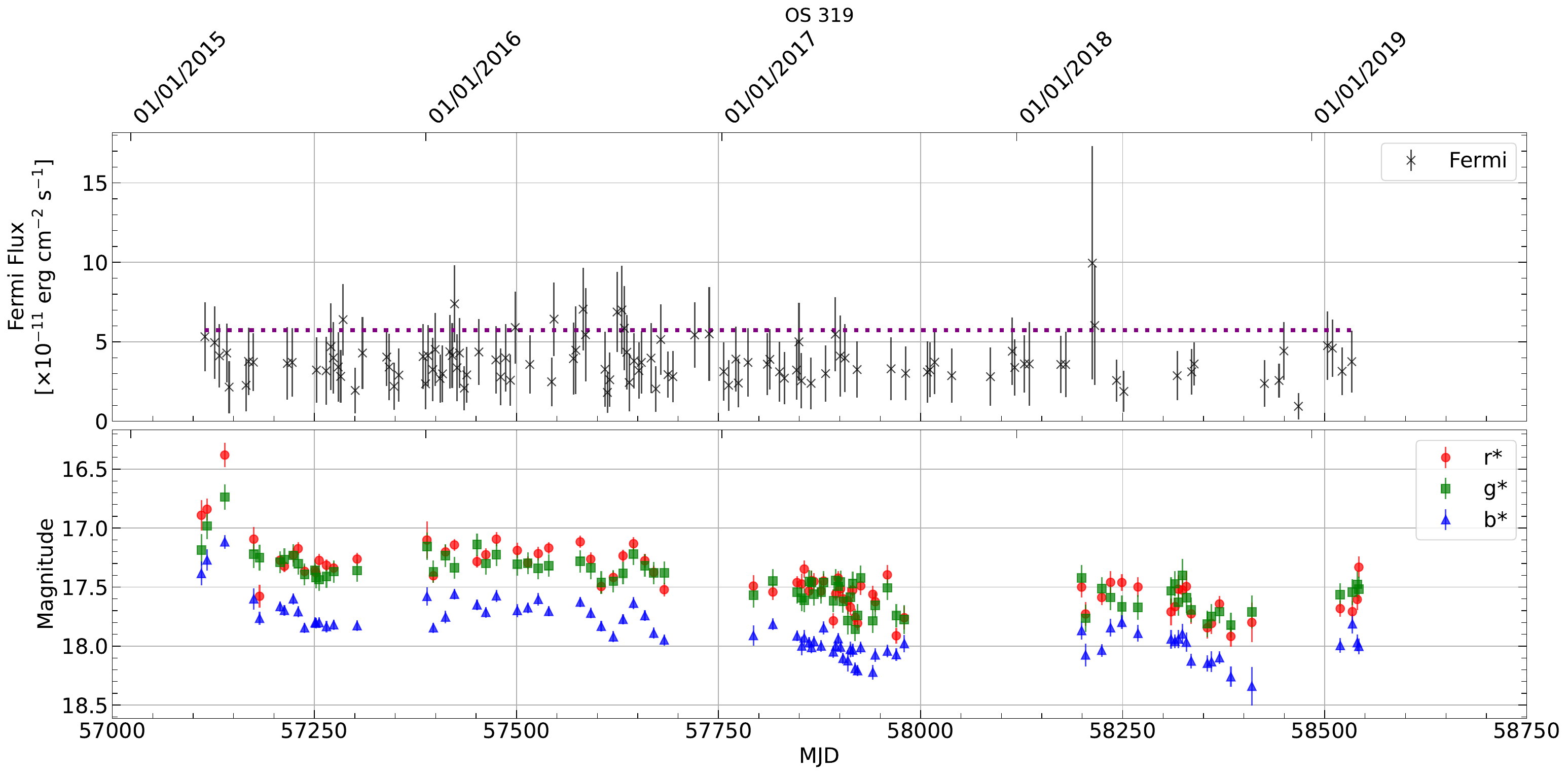}
\caption{As Fig. \ref{fig:1ES1011+496_lc} but for OS 319.}
\label{fig:1611+343_lc}
\end{figure*}

\begin{figure*}
\centering
\includegraphics[width=\linewidth]{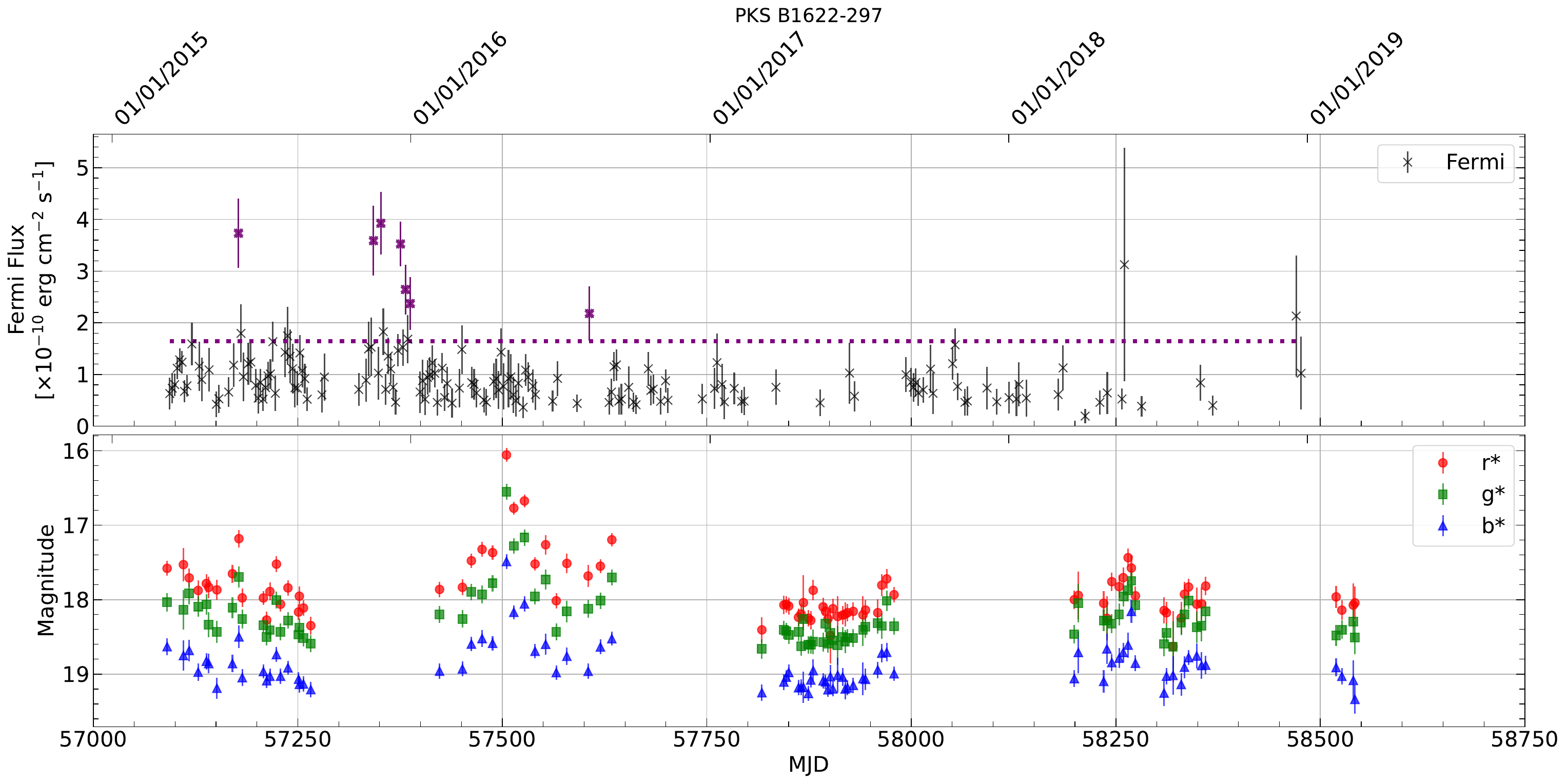}
\caption{As Fig. \ref{fig:1ES1011+496_lc} but for PKS B1622-297.}
\label{fig:1622-297_lc}
\end{figure*}

\begin{figure*}
\centering
\includegraphics[width=\linewidth]{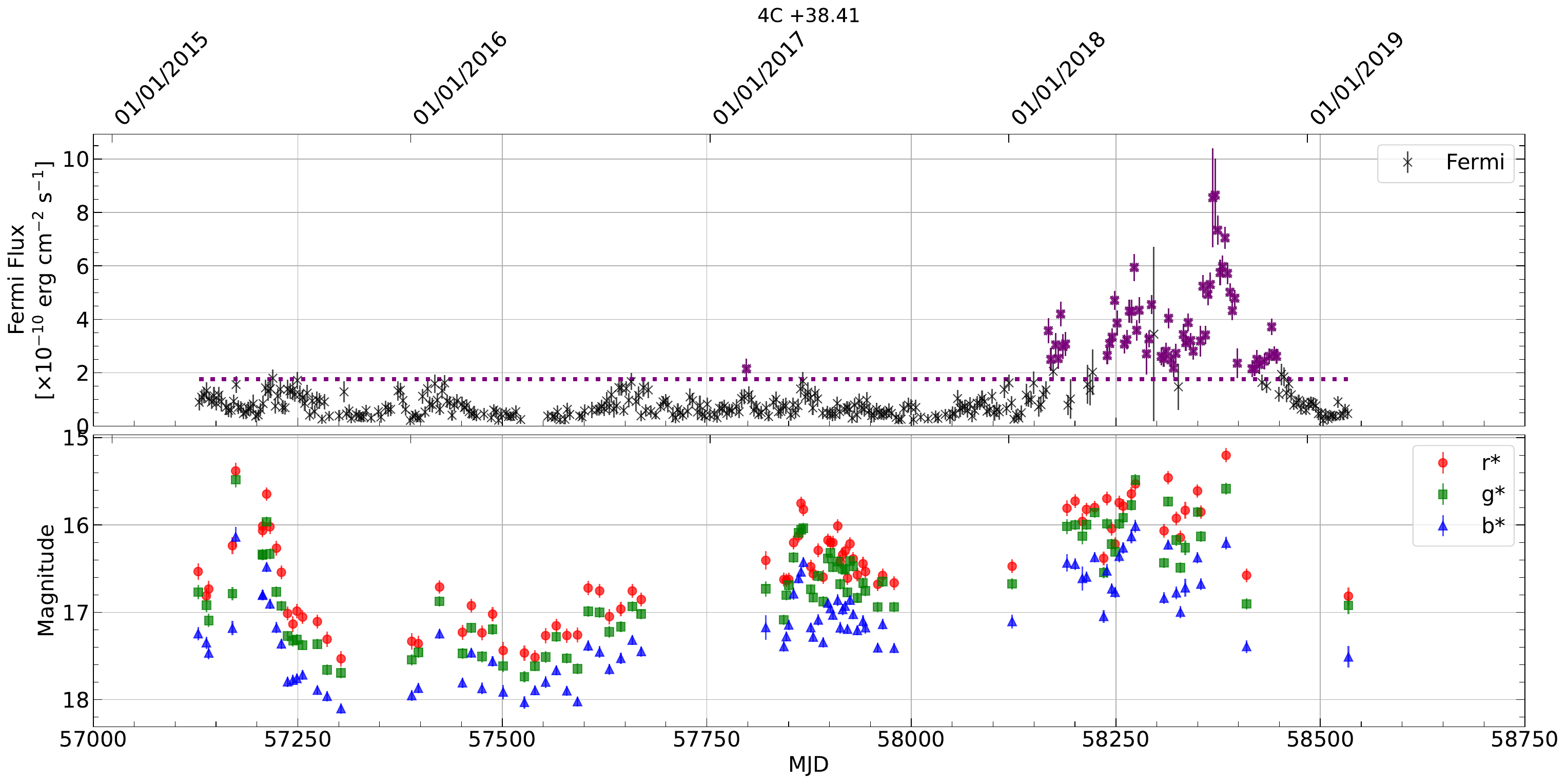}
\caption{As Fig. \ref{fig:1ES1011+496_lc} but for 4C +38.41.}
\label{fig:1633+382_lc}
\end{figure*}

\begin{figure*}
\centering
\includegraphics[width=\linewidth]{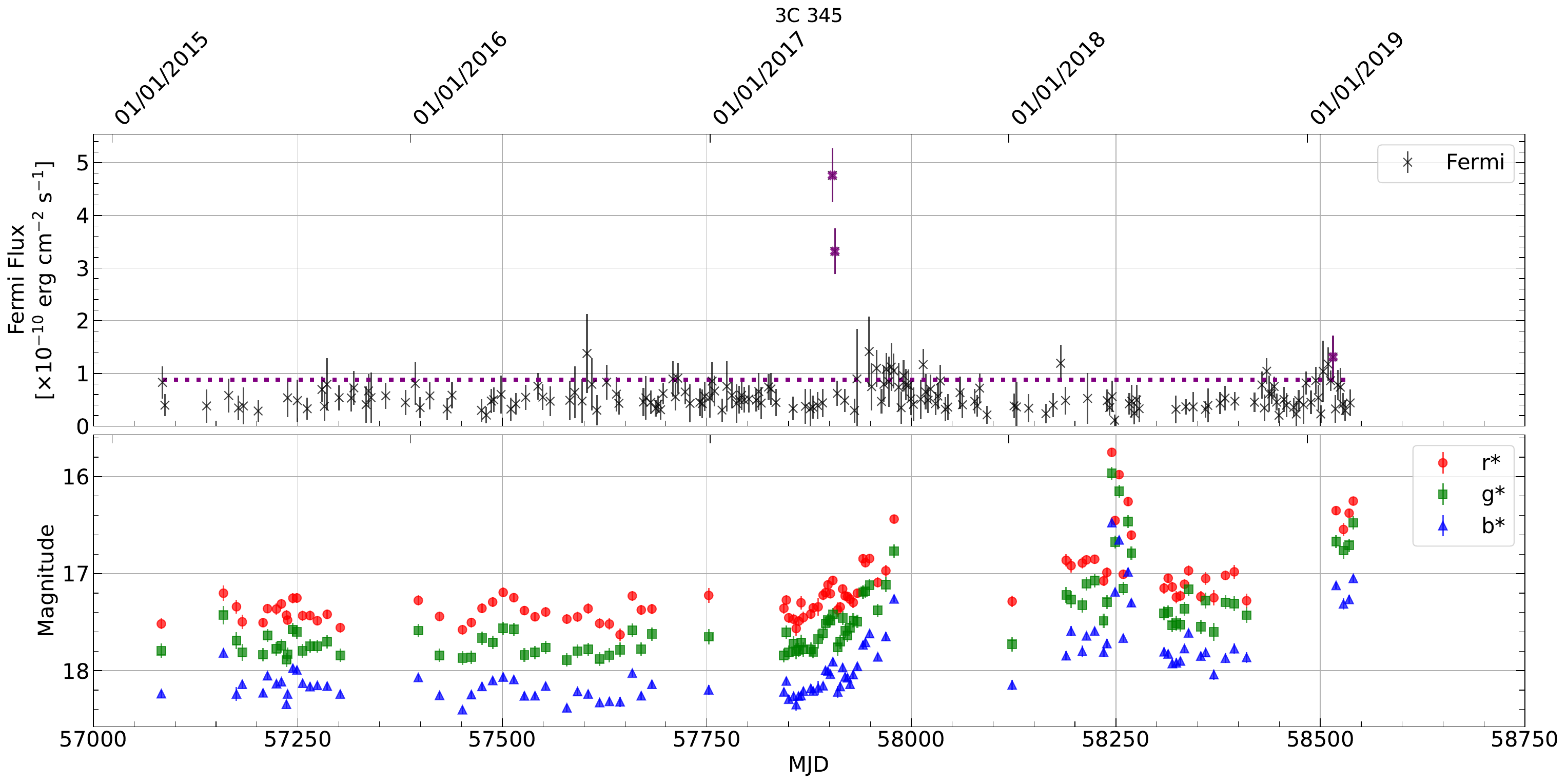}
\caption{As Fig. \ref{fig:1ES1011+496_lc} but for 3C 345.}
\label{fig:1641+339_lc}
\end{figure*}

\begin{figure*}
\centering
\includegraphics[width=\linewidth]{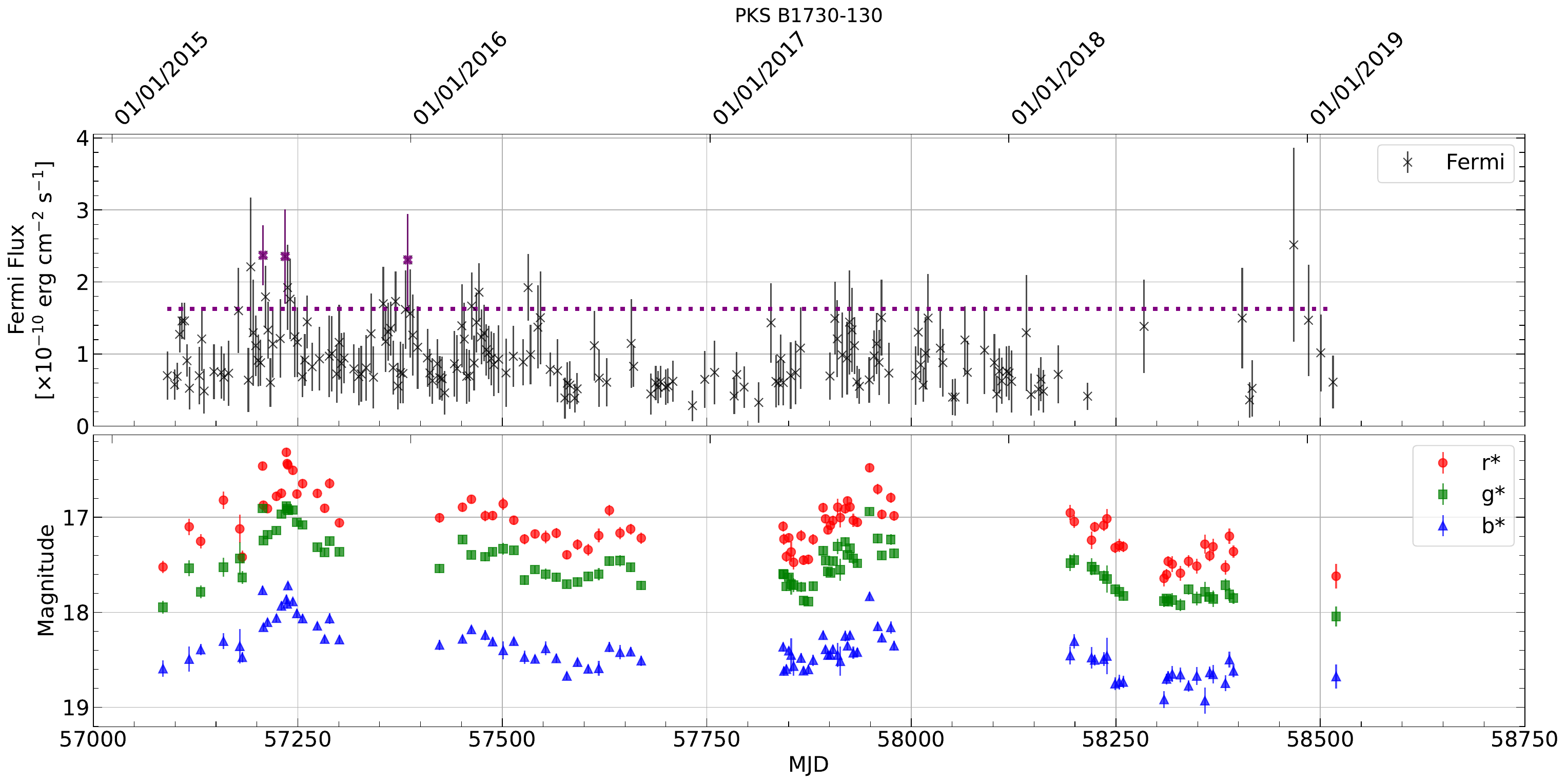}
\caption{As Fig. \ref{fig:1ES1011+496_lc} but for PKS B1730-130.}
\label{fig:1730-130_lc}
\end{figure*}

\begin{figure*}
\centering
\includegraphics[width=\linewidth]{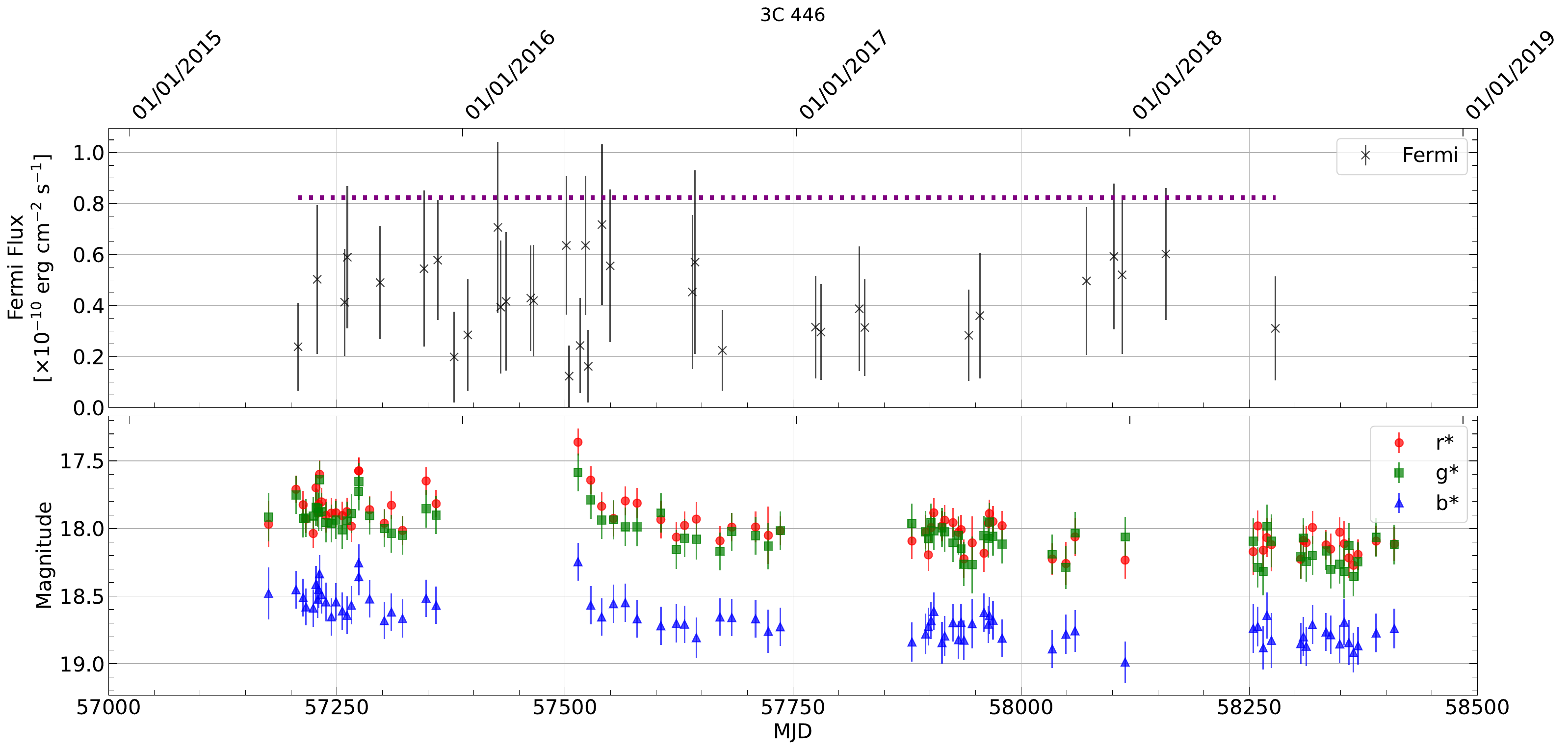}
\caption{As Fig. \ref{fig:1ES1011+496_lc} but for 3C 446.}
\label{fig:2223-052_lc}
\end{figure*}

\begin{figure*}
\centering
\includegraphics[width=\linewidth]{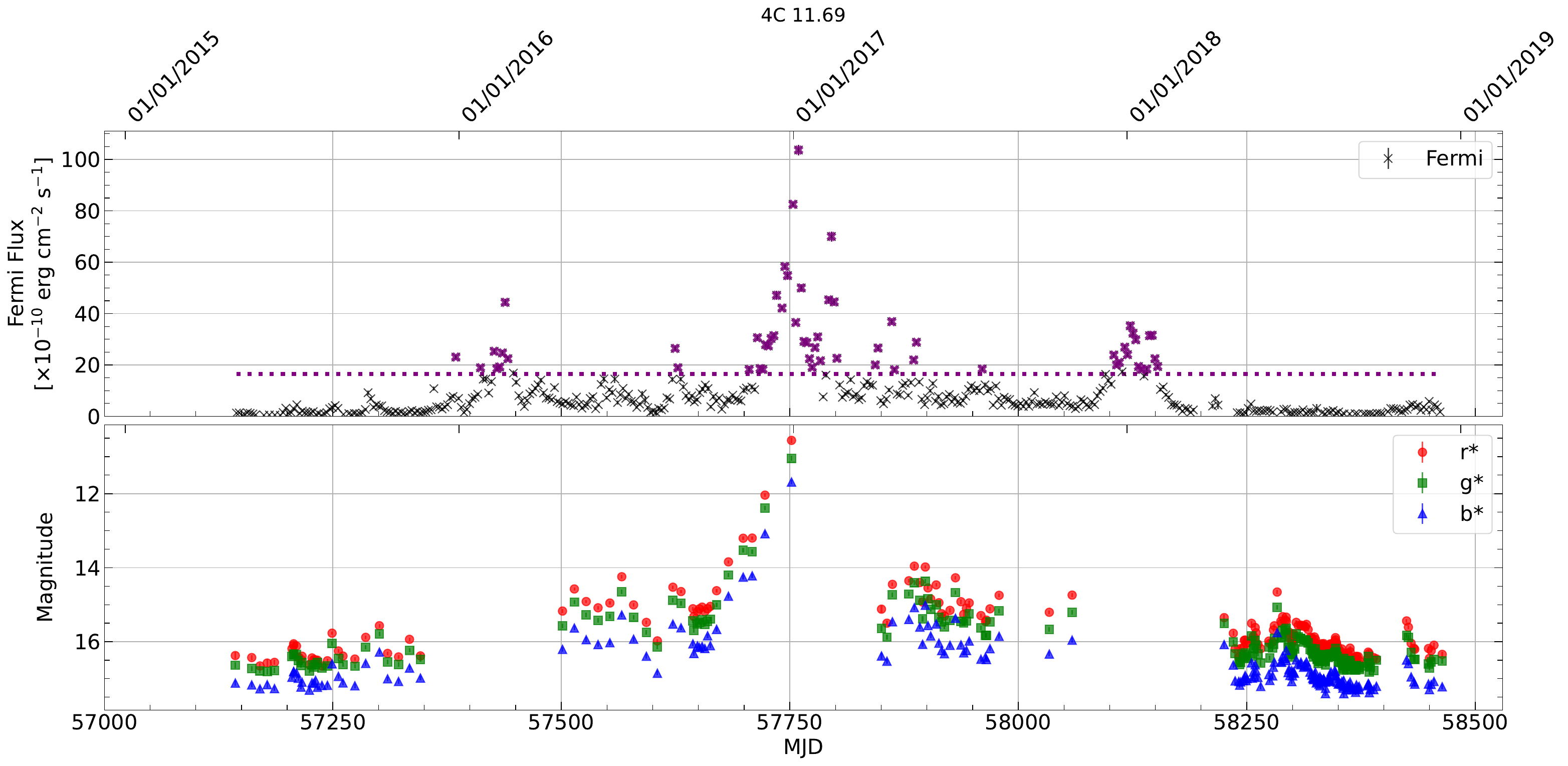}
\caption{As Fig. \ref{fig:1ES1011+496_lc} but for 4C 11.69.}
\label{fig:4C11.69_lc}
\end{figure*}

\begin{figure*}
\centering
\includegraphics[width=\linewidth]{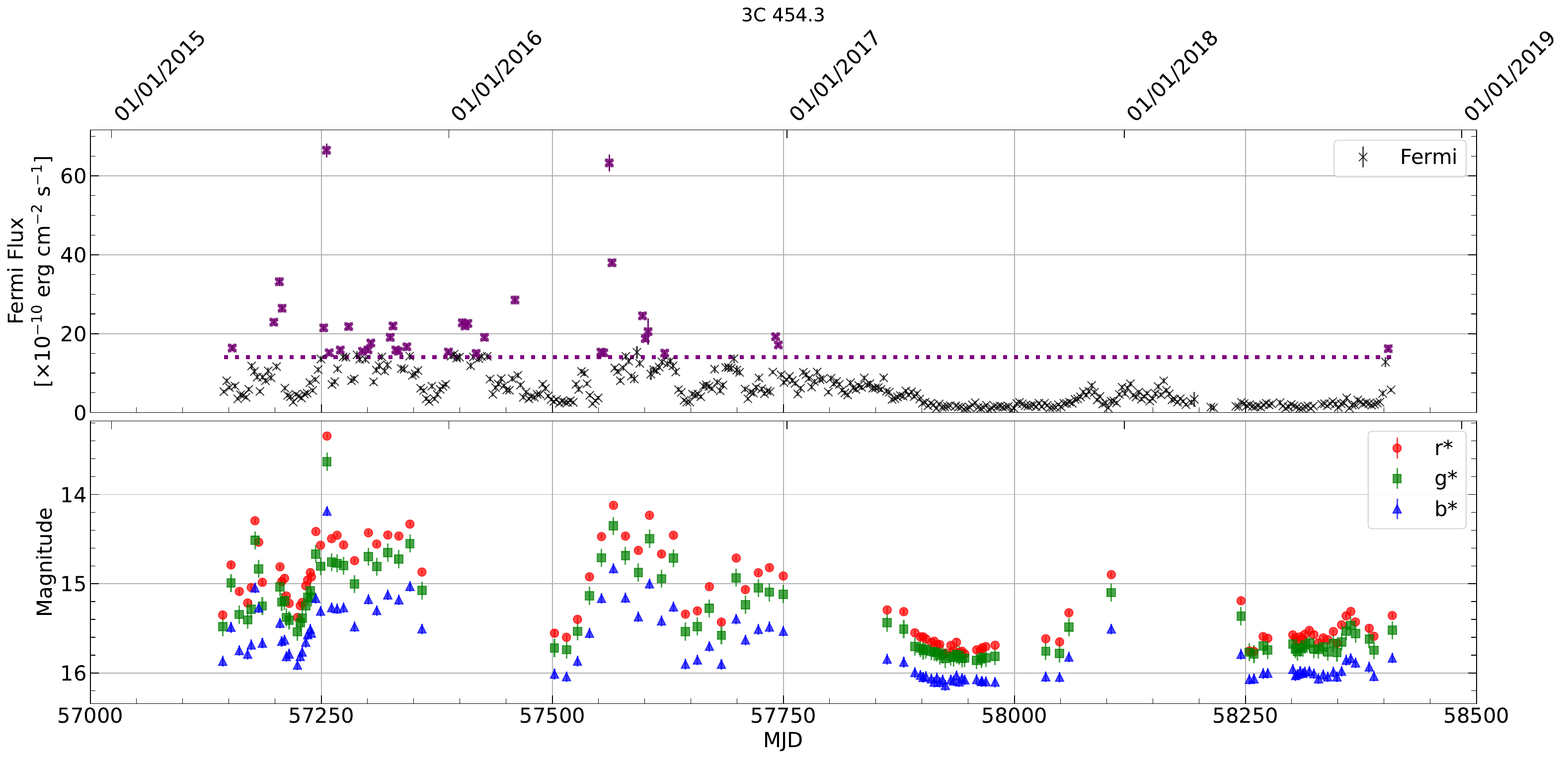}
\caption{As Fig. \ref{fig:1ES1011+496_lc} but for 3C 454.3.}
\label{fig:3C454.3_lc}
\end{figure*}

\bsp
\label{lastpage}
\end{document}